\documentclass{emulateapj}
\usepackage{hyperref}
\usepackage{amssymb,amsmath,natbib,threeparttable,rotating,color,float}
\usepackage[english]{babel}
\usepackage[graphicx]{realboxes}
\newcommand{\ha}{H$\alpha$}
\newcommand{\A}{\mathrm{\AA}}
\newcommand{\msun}{$\mathrm{M_{\odot}}$}

\newcommand*\pasa{PASA}
\newcommand*\na{New Astron.}

\newcommand{\tf}{Tully-Fisher}

\newcommand{\bc}{$(2.18\pm0.051)$}%
\newcommand{\afree}{$(0.193\pm0.108)$}

\newcommand{\bfixc}{$(2.17\pm0.047)$}
\newcommand{\ev}{$\Delta \mathrm{M}/$\msun$=-0.25\pm0.16$ dex}

\newcommand{\evbell}{$\Delta \mathrm{M}/$\msun$=-0.39\pm0.21$ dex}

\newcommand{\bsfree}{$(2.06\pm0.032)$}
\newcommand{\asfree}{$(0.211\pm0.086)$}

\newcommand{\bsfix}{($2.03\pm0.032$)}
\newcommand{\sev}{$\Delta \mathrm{M}/$\msun$=-0.45\pm0.13$ dex}

\newcommand{\sevnoc}{$\Delta \mathrm{M}/$\msun$=-0.26\pm0.12$ dex}

\newcommand{\evbright}{$\Delta \mathrm{M}/$\msun$=-0.27\pm0.17$ dex}
\newcommand{\evbrightbell}{$\Delta \mathrm{M}/$\msun$=-0.43\pm0.23$ dex}

\newcommand{\evbigsample}{$\Delta \mathrm{M}/$\msun$=-0.18\pm0.16$ dex}
\newcommand{\evbigsamplebell}{$\Delta \mathrm{M}/$\msun$=-0.29\pm0.21$ dex}

\begin{document}

\title{ZFIRE: the evolution of the stellar mass Tully-Fisher relation to redshift $2.0<z<2.5$ with MOSFIRE}

\author{Caroline M. S. Straatman\altaffilmark{1}, Karl Glazebrook\altaffilmark{2}, Glenn G. Kacprzak\altaffilmark{2}, Ivo Labb\'e\altaffilmark{3}, Themiya Nanayakkara\altaffilmark{2}, Leo Alcorn\altaffilmark{4}, Michael Cowley\altaffilmark{5,6}, Lisa J. Kewley\altaffilmark{7}, Lee R. Spitler\altaffilmark{5,6}, Kim-Vy H. Tran\altaffilmark{4}, Tiantian Yuan\altaffilmark{7}}

\altaffiltext{1}{Max-Planck Institut f\"ur Astronomie, K\"onigstuhl 17, D-69117, Heidelberg, Germany; straatman@mpia.de}
\altaffiltext{2}{Centre for Astrophysics and Supercomputing, Swinburne University, Hawthorn, VIC 3122, Australia}
\altaffiltext{3}{Leiden Observatory, Leiden University, PO Box 9513, 2300 RA Leiden, The Netherlands}
\altaffiltext{4}{George P. and Cynthia W. Mitchell Institute for Fundamental Physics and Astronomy, Department of Physics and Astronomy, Texas A\& M University, College Station, TX 77843}
\altaffiltext{5}{Australian Astronomical Observatory, PO Box 915, North Ryde, NSW 1670, Australia}      
\altaffiltext{6}{Department of Physics \& Astronomy, Macquarie University, Sydney, NSW 2109, Australia}
\altaffiltext{7}{Research School of Astronomy and Astrophysics, The Australian National University, Cotter Road, Weston Creek, ACT 2611, Australia}

\begin{abstract}
Using observations made with MOSFIRE on Keck I as part of the ZFIRE survey, we present the stellar mass \tf\ relation at $2.0 < z < 2.5$. The sample was drawn from a stellar mass limited, $K_s-$band selected catalog from ZFOURGE over the CANDELS area in the COSMOS field. We model the shear of the \ha\ emission line to derive rotational velocities at $2.2\times$ the scale radius of an exponential disk ($V_{2.2}$). We correct for the blurring effect of a two-dimensional PSF and the fact that the MOSFIRE PSF is better approximated by a Moffat than a Gaussian, which is more typically assumed for natural seeing. We find for the \tf\ relation at $2.0<z<2.5$ that $\mathrm{log}V_{2.2}=$\bc$+$\afree$(\mathrm{log}M/M_{\odot}-10)$ and infer an evolution of the zeropoint of \ev\ {or \evbell} compared to $z=0$ when adopting a fixed slope {of 0.29 or 1/4.5, respectively}. We also derive the alternative kinematic estimator $S_{0.5}$, with a best-fit relation $\mathrm{log}S_{0.5}=$\bsfree$+$\asfree$(\mathrm{log}M/M_{\odot}-10)$, and infer an evolution of \sev\ compared to $z<1.2$ if we adopt a fixed slope. We investigate and review various systematics, ranging from PSF effects, projection effects, systematics related to stellar mass derivation, selection biases {and slope}. We find that discrepancies between the various literature values are reduced when taking these into account. {Our observations correspond} well with {the gradual evolution predicted by} semi-analytic model{s}.  
\end{abstract}

\section{Introduction}

A major goal for galaxy evolution models is to understand the interplay between dark matter and baryons. In the current $\Lambda$CDM paradigm, galaxies are formed as gas cools and accretes into the centers of dark matter haloes. The gas maintains its angular momentum, settling in a disk at the center of the gravitational potential well \citep{Fall80} where it forms stars. This process can be disrupted by galaxy mergers, gas inflows, AGN and star formation feedback, which can affect the shape, star-formation history and kinematics of galaxies \citep[e.g.][]{Hammer05}.

From studies at $z=0$ of the kinematic properties of disk galaxies a correlation has emerged between disk rotational velocity and, initially, luminosity. This relation is now named the \tf\ relation, first reported by \citet{Tully77}, and originally used as a distance indicator. At $z=0$ the \tf\ relation is especially tight if expressed in terms of stellar mass instead of luminosity \citep{Bell01}. If studied at high redshift, it can be an important test of the mass assembly of galaxies over time, as it describes the relation between angular momentum and stellar mass, and the conversion of gas into stars versus the growth of dark matter haloes by accretion \citep[e.g.][]{Fall80,Mo98,Sales10}. With the increasing success of multiwavelength photometric surveys to study galaxy evolution, much insight has already been obtained into the structural evolution of galaxies to high redshift \citep[e.g.][]{Franx08,vanderWel14a,Straatman15}, and their stellar mass growth and star-formation rate histories \citep[e.g.][]{Whitaker12,Tomczak14,Tomczak15}. The study of galaxy kinematics at $z>1$ has been lagging behind, because of the faint magnitudes of high redshift galaxies and the on-going development of sensitive near-IR multiobject spectrographs needed for efficient follow-up observations.

In the past few years, studies of the \tf\ relation at $0 < z < 1$ were performed with the multiplexing optical spectrographs DEIMOS on Keck I \citep{Kassin07,Miller11} and LRIS on Keck II \citep{Miller12}, and optical Integral Field Unit (IFU) spectrographs such as VLT/GIRAFFE \citep{Puech08}, but beyond $z>1$ progress has been comparatively slow because of the reliance on mostly single-object integral field spectrographs, such as SINFONI \citep{Cresci09,Gnerucci11,Vergani12} on the VLT. These studies resulted in contra{ry} estimates of a potential evolution of the stellar mass zeropoint of the \tf\ relation with redshift {\citep{Glazebrook13}}. For example, studies by \citet{Puech08,Vergani12,Cresci09,Gnerucci11} and \citet{Simons16} indicate evolution already {at $z\geqslant0.6$}. At $z=0.6$ this amounts to $\Delta \mathrm{M}/$\msun$\sim-0.3$ dex \citep{Puech08}. At {$z\sim2$} $\Delta \mathrm{M}/$\msun$\sim-0.4$ dex  {\citep{Cresci09,Simons16}} and at $z=3$ $\Delta \mathrm{M}/$\msun$\sim-1.3$ dex \citep{Gnerucci11}. At the same time e.g., \citet{Miller11,Miller12} find no significant evolution up to $z=1.7$. 

Part of the inferred evolution however, or lack thereof, could be explained by selection bias, for example by preferentially selecting the most dynamically evolved galaxies at each redshift. This acts as a progenitor bias, \citep{vanDokkum01}, where the high redshift sample is an increasingly biased subset of the true distribution, leading to an underestimate of the evolution. Dynamically evolved galaxies could make up only a small fraction of the total population at high redshift, as irregular, dusty and dispersion dominated galaxies become more common towards higher redshifts \citep[e.g.][]{Abraham01,Kassin12,Spitler14} {and in a recent publication, \citet{Tiley16} showed that the inferred evolution is indeed larger for more rotationally supported galaxies}. Similarly, previous surveys at the highest redshift at $z>2$ tend to be biased towards the less dust-obscured or blue star-forming galaxies, such as Lyman Break galaxies, and often required previous rest-frame UV selection or a spectroscopic redshift from optical spectroscopy \citep[e.g.][]{ForsterSchreiber09,Gnerucci11}. As a consequence these samples may not be representative of massive galaxies at high redshift, which are more often reddened by dust-obscuration \citep[e.g.][]{Reddy05,Spitler14}. 

The different results between these studies could also be due to systematics arising from the different methodologies used to derive stellar mass, rotational velocity, and the different types of spectral data (one-dimensional long-slit spectra versus two-dimensional IFU data). As \citet{Miller12} note, a striking discrepancy exists between their long-slit results (no evolution) and IFU studies by \citet{Puech08,Vergani12} and \citet{Cresci09} ($\Delta \mathrm{M}/$\msun$=0.3-0.4$ dex). Sample size may also play a role: the highest redshift studies are based on small samples of only 14 galaxies at $z=2.2$ \citep{Cresci09} and 11 galaxies at $z=3$ \citep{Gnerucci11}. 

{At face value, a} non-evolving \tf\ relation would be a puzzling result{. In the framework of hierarchical clustering at fixed velocity the mass of a disk which is a fixed fraction of the total mass of an isothermal halo is predicted to change proportionately to the inverse of the Hubble constant \citep{Mo98,Glazebrook13}. }{T}he average properties of galaxies {also} evolve strongly with redshift. For example, the average star-formation rate of star-forming galaxies at fixed stellar mass tends to increase with redshift \citep[e.g.][]{Tomczak15}, as does their gas fraction \citep[e.g.][]{Papovich14}. At the same time their average size tends to be smaller \citep[e.g.][]{vanderWel14a}, which would by itself imply higher velocities at fixed stellar mass. {Yet semi-analytic models predict only a weak change in the stellar mass zeropoint, with most of the evolution occuring along the \tf\ relation \citep[e.g.,][]{Somerville08, Dutton11,Benson12}.}

It is clear that more studies with larger numbers of galaxies are needed to shed light on the observationally key epoch at $z\sim2$. In this study we use new spectra of galaxies at $2.0<z<2.5$ from the ZFIRE survey \citep{Nanayakkara16}. These were obtained from the newly installed MOSFIRE instrument on Keck I, a sensitive near-IR spectrograph {whose multiplexing capability} allows batch observations of large numbers of galaxies at the same time {to great depth}. The primary aim of ZFIRE is to spectroscopically confirm and study galaxies in two high redshift cluster clusters, one in the UDS field \citep{Lawrence07} at $z=1.62$ \citep{Papovich10} and one in the COSMOS field \citep{Scoville07} at $z=2.095$ \citep{Spitler12,Yuan14}. However, ZFIRE also targets many foreground and background galaxies at redshifts $1.5<z< 4.0$. {With ZFIRE the \ha\ (rest-frame vacuum $6564.614\mathrm{\AA}$) emission line is observed for a large number of galaxies at $z\sim2$, which can be used for studies of galaxy kinematics. In a recent paper \citet{Alcorn16} derived velocity dispersions and virial masses and investigated environmental dependence.} In this paper, we use the rich data set over the COSMOS field to study the \tf\ relation at $2.0 < z < 2.5$. Our aim is to provide improved constraints on the evolution of the stellar mass \tf\ relation with redshift.

In Section \ref{sec:data} we describe our data and sample of galaxies, in Section \ref{sec:analysis} we describe our analysis, in Section \ref{sec:tf22} we derive the \tf\ relation at $2<z<2.5$ and in Section \ref{sec:discussion} we discuss our results in an evolutionary context. Throughout, we use a standard cosmology with $\mathrm{\Omega_{\Lambda}=0.7}$, $\mathrm{\Omega_{m}=0.3}$ and $H_0=70$ km/s/Mpc. At {$2<z<2.5$}, {$1\arcsec$} corresponds to {$\sim 8$} kpc.

\section{Observations and selections}
\label{sec:data}

\subsection{Observations}

\subsubsection{Spectroscopic data}

This study makes use of data obtained with the Multi-Object Spectrometer for InfraRed Exploration \citep[MOSFIRE;][]{Mclean10} on Keck-I on Mauna Kea in Hawaii. The observations over COSMOS were carried out in 6 pointings with a $6.1\arcmin\times6.1\arcmin$ field of view. The observations were conducted on December 24-25, 2013 and February 10-13, 2014. Galaxies were observed in 8 masks in the $K-$band, which covers $1.93-2.45\micron$, and can be used to measure \ha\ and [NII] emission lines for galaxies at $z\sim2$. Two $H-$band masks were also included in the observations. The $H-$band coverage is $1.46-1.81\micron$, covering $H\beta$ and [OIII]. For this work, we limit the analysis to the \ha\ emission line data in the $K-$band. Further details on the $H-$band masks can be found in \citet{Nanayakkara16}. 

The total exposure time was 2 hours for each $K-$band mask. A $0.7\arcsec$ slit width was used, yielding {a} spectral resolution {of $R=3610$}. At $z=2.2$, the median redshift of the sample of galaxies in this study, this corresponds to {$\sim26$} km/s per pixel. The seeing conditions were $0.65-1.10\arcsec$, with a median of $0.7\arcsec$. We used a standard two-position dither pattern (ABBA). Before and after science target exposures, we measured the spectrum of an A0V type standard star in 0.7\arcsec\ slits to be used for telluric corrections and standard stars to be used for flux calibration in a slit of width 3\arcsec to minimize slit loss. Each individual mask also contained a star for monitoring purposes, such as measuring the seeing conditions.

The raw data were reduced using {a slightly modified version of} the publicly available {2015A} data reduction pipeline\footnote{The modified version is available at \url{https:// github.com/themiyan/MosfireDRP_Themiyan}.} developed by the MOSFIRE instrument team, resulting in two-dimensional spectra that were background subtracted, rectified and wavelength calibrated to vacuum wavelengths, with a typical residual of $<0.1\mathrm{\AA}$ \citep{Nanayakkara16}. To make up for the lack of skylines at the red end of the $K-$band, we used both night sky lines and a Neon arc lamp for wavelength calibration.

Based on the standard star, we applied a telluric correction and flux calibration to the two-dimensional spectra, similar to the procedure used by \citep{Steidel14}, and using our own custom IDL routines. {We subsequently scaled the flux values to agree with the photometric $K_s-$band magnitudes from the FourStar \citep{Persson13} Galaxy Evolution Survey \citep[ZFOURGE;][]{Straatman16}, resulting in a final median uncertainty of 0.08 magnitude \citep[see also][]{Nanayakkara16}}.

\begin{figure*}
\includegraphics[width=\textwidth]{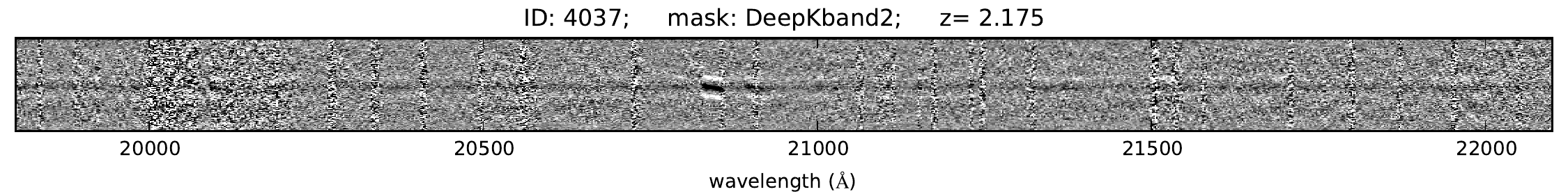}
\includegraphics[width=\textwidth]{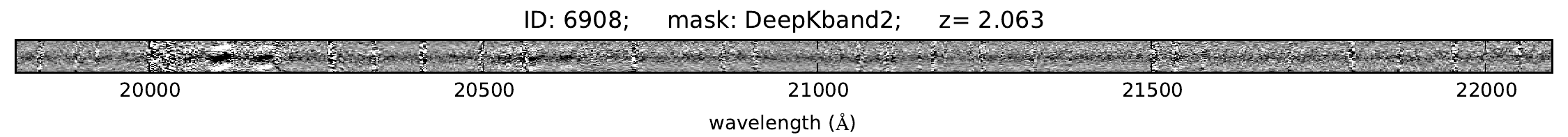}
\caption{Two example Keck MOSFIRE spectra (inverted grayscale) at $z=2.175$ and $z=2.063$, with \ha$\lambda6565$ clearly visible at $\lambda=20843.2\mathrm{\AA}$ (top) and $\lambda=20109.6\mathrm{\AA}$ (bottom). Other lines are visible as well, most notably [NII]$\lambda\lambda 6550,6585$ and [SII]$\lambda\lambda 6718,6733$.}
\label{fig:spectra}
\end{figure*}

In Figure \ref{fig:spectra} we show two example spectra at $z=2.175$ and $z=2.063$, with strong \ha\ emission at observed frame $\lambda=20843.2\mathrm{\AA}$ and $\lambda=20109.6\mathrm{\AA}$, respectively. Other lines are visible in the spectrum as well, most notably [NII]$\lambda\lambda 6550,6585$ and [SII]$\lambda\lambda 6718,6733$.

\subsubsection{Continuum subtraction}
From each two-dimensional spectrum we extracted spectral image stamps of $300\A$ wide (46 pixels) centered on the \ha\ emission lines. Night sky emission was masked using the publicly available night sky spectra taken during May 2012 engineering, at wavelengths where the sky spectrum exceeds $10^{-24}$ ergs/s/$\mathrm{cm^2/arcsec^2}$. We also masked $40\A$ wide boxes centered on the \ha\ line and the [NII] doublet. We subtracted the continuum using the following method: for each pixel row (one row corresponding to a one-dimensional spectrum with a length of $300\A$) we determined the median flux and the standard deviation. Next we iteratively rejected pixels at $>2.5\sigma$ from the median and recalculated both values. We repeated this a total of three times. The final median flux was our estimate of the continuum in that particular pixel row, which was then subtracted accordingly.

\subsubsection{PSF determination}
\label{sec:psf}
{The galaxies in this study are} small {($<0.7\arcsec$, see Section \ref{sec:tfsample})}, so the PSF needs to be properly characterized. Not only the FWHM of the PSF needs to be tracked, but even the detailed shape of the PSF can have a noticeable effect on the smoothing of the \ha\ line and its rotation profile. A simple Gaussian is often assumed, but this leads to underestimating the shear of the emission line -- and hence the velocity -- if the true PSF has stronger wings. Because the \tf\ relation is very steep \citep[e.g.][]{Bell01,Reyes11}, a small change in velocity could lead to significant offsets.

\begin{figure*}
\begin{center}
\includegraphics[width=0.8\textwidth]{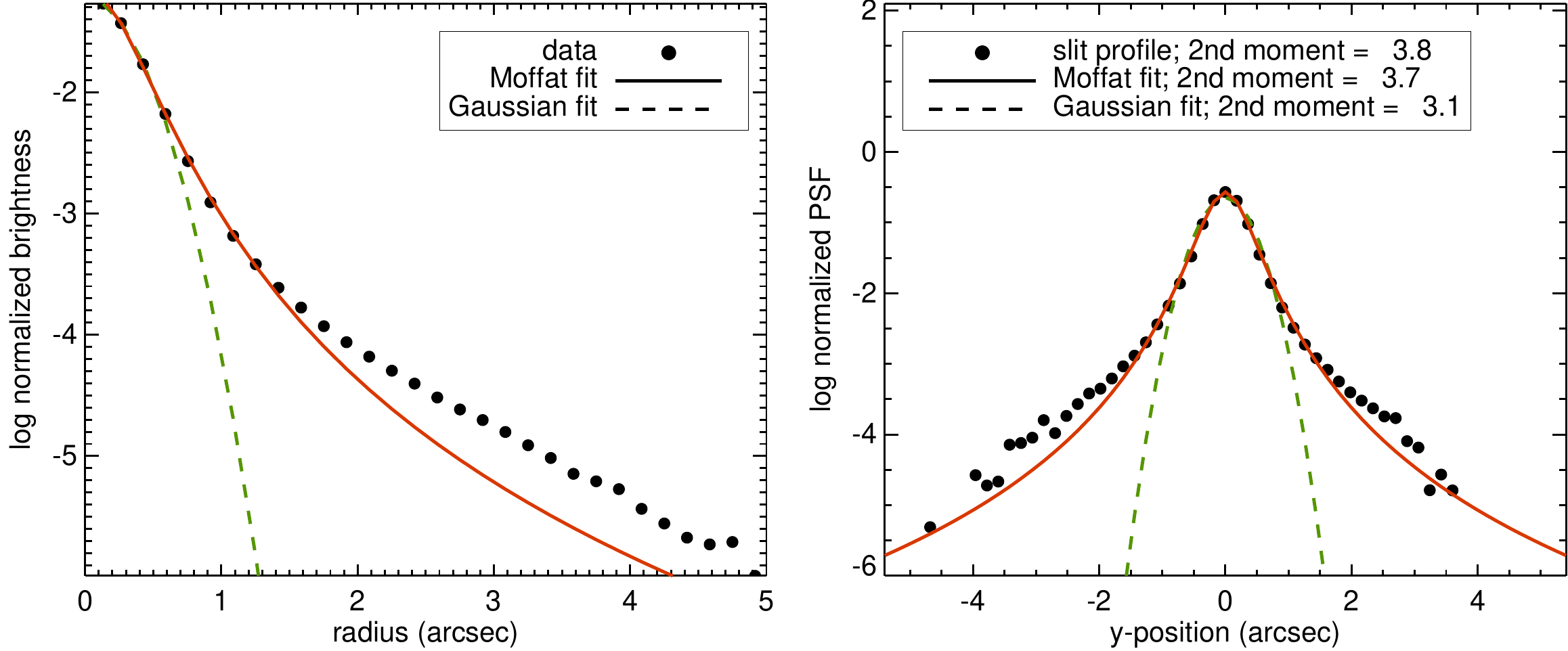}
\caption{Left: surface brightness profile of the two-dimensional $K_s-$band image PSF (dots) as a function of radius, with the best-fit Moffat (solid red line) and Gaussian (dashed green line). The Gaussian is quite steep, whereas the Moffat gives a better approximation of the flux at large radii. Right: a simulated one-dimensional spectral PSF, obtained from integrating the two-dimensional $K_s-$band PSF and the best-fits in a 0.7\arcsec\ virtual slit. The second order moment of the Moffat is close to that of the actual PSF, but that of the Gaussian is much smaller.} 
\label{fig:moment}
\end{center}
\end{figure*}

We first attempted to derive the PSF from the collapsed spectra of the monitor stars, which received the same exposure as the galaxies in the masks. The collapsed spectra were obtained by averaging over the flux in the wavelength direction, after masking skylines. The intensity profile had a very steep profile, which was {superficially} well fit by a Gaussian profile. Although adopting a Gaussian profile is common \citep[e.g.][]{Kriek15}, this was unexpected, because the MOSFIRE PSF in deep $K_s-$band imaging (Marchesini; private communication) clearly has strong wings, which {generally} are better fit with a Moffat profile (see Figure \ref{fig:moment}). Even small wings are important, because the effect of the PSF on convolution does not scale with the amount of flux in the wings, but with the second order moment of the PSF \citep{Franx89}. Even a few percent flux in the wings can have a significant effect, due to the $r^2$ weighting. For illustration, we calculate the second moment for a simulated spectral PSF derived from a deep MOSFIRE image at FWHM$=0.6\arcsec$ seeing. The image PSF was created by median stacking 5 unsaturated bright stars, after background subtraction and normalization. {First w}e measured the brightness profile of th{is two-dimensional} PSF as a function of radius and fitted {both} a Moffat and a Gaussian function, as shown in the left panel of Figure \ref{fig:moment}. To reproduce the one-dimensional spectral PSFs, we integrated the two-dimensional image PSF and its two model fits within a 0.7\arcsec\ virtual slit. Finally, we calculated the second order moments{, $F_2$ for the PSF, $G_2$ for the Gaussian model and $M_2$ for the Moffat model}. As shown in the right panel of Figure \ref{fig:moment}, the true PSF ($F_2 = 3.8$) is severely underestimated by a Gaussian approximation ($G_2 = 3.1$), whereas a Moffat fit produces good correspondence ($M_2 = 3.7$).\footnote{Note, to avoid noise amplification at large radii due to the $r^2$ weighting, we evaluate the second order moment at $r<2.6\arcsec$. The Gaussian is scaled up by 12\% for a consistent comparison to a Moffat in one dimension.}

\begin{figure*}
\includegraphics[width=0.32\textwidth]{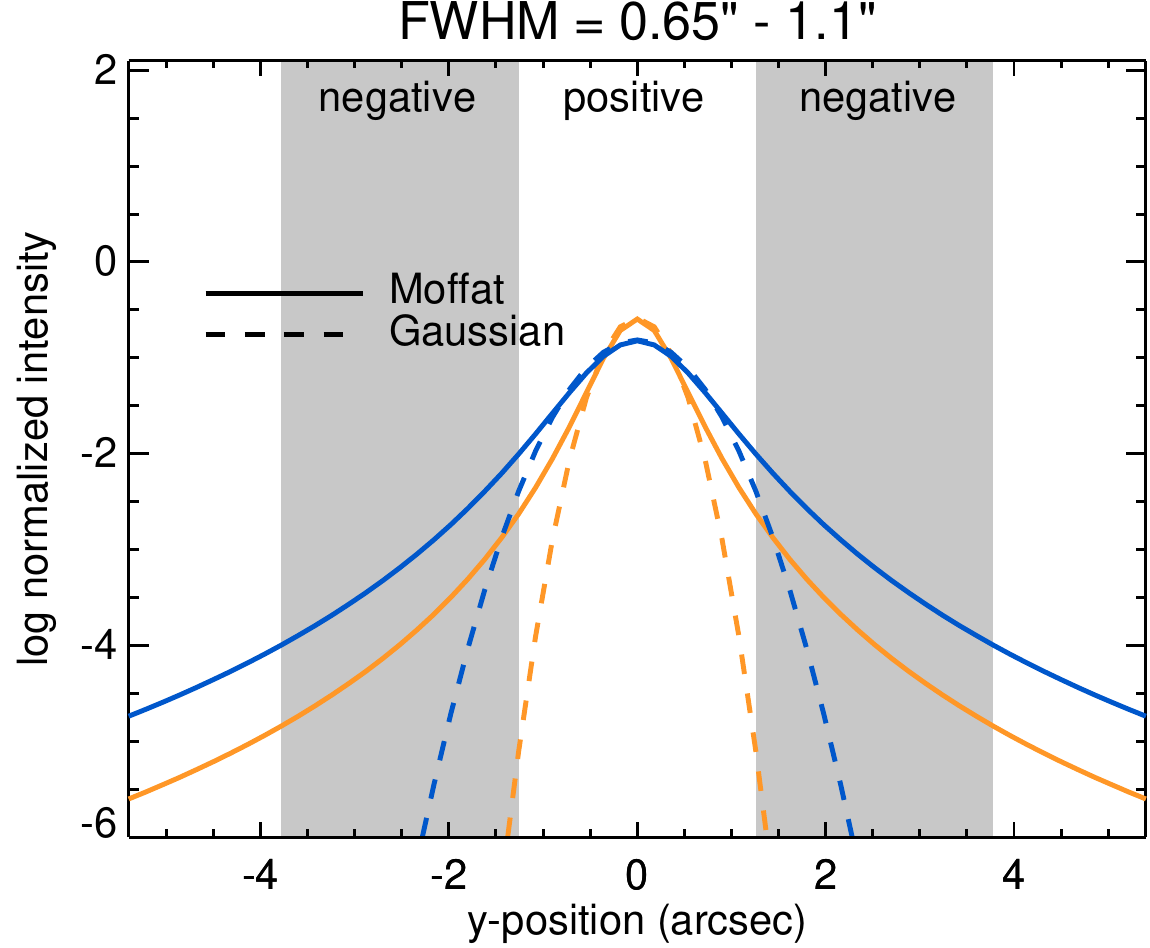}	
\includegraphics[width=0.32\textwidth]{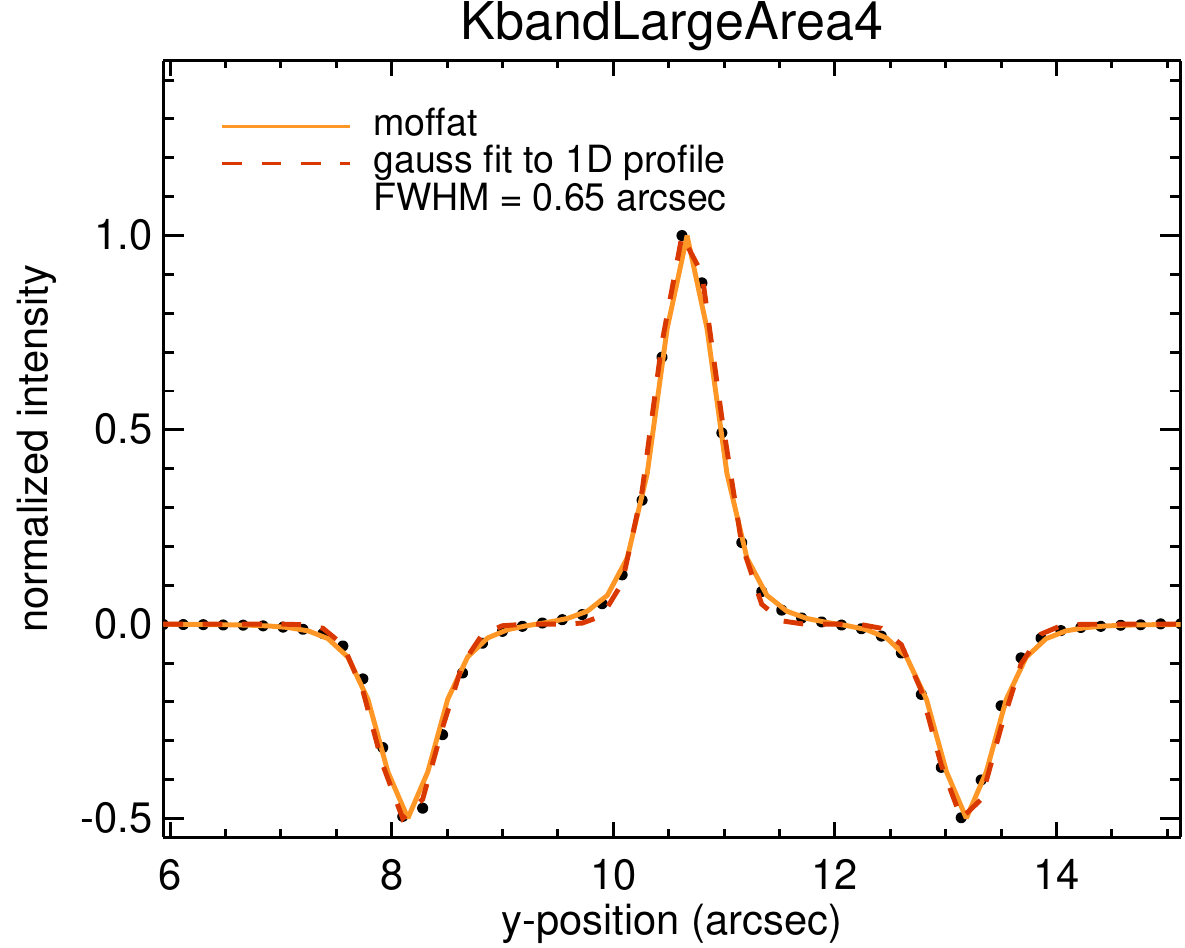}
\includegraphics[width=0.32\textwidth]{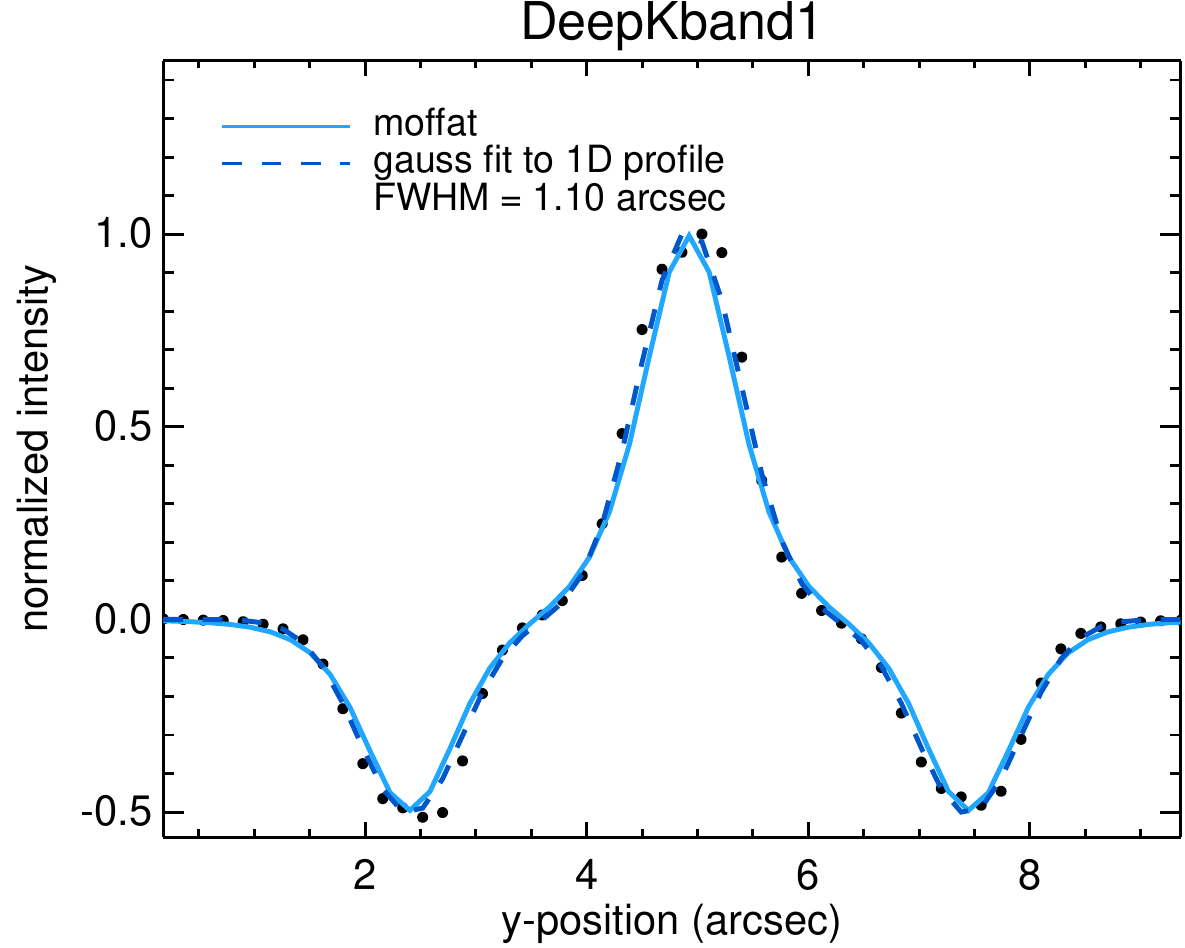}
\caption{Examples of spatial profiles of MOSFIRE PSFs. The solid and dashed curves are theoretically derived Moffat and Gaussian intensity profiles, respectively. They are shown at logarithmic scale in the left panel. A Moffat is a good representation of the original MOSFIRE PSF, but sky subtraction in the reduction process leaves negative imprints on each sides, which will subtract the strong wings. This makes the reduced PSF appear Gaussian. This is illustrated by the two examples of spatial profiles of monitor stars in the middle and right panels, with best and worst seeing, respectively. The black datapoints represent the star spectra collapsed in the wavelength direction. {The solid and dashed lines} are the {reconstructed Moffat} PSFs {and the one-dimensional Gaussian fits}, showing that {they} are nearly indistinguishable {in one dimension}.}
\label{fig:wings}
\end{figure*}	

Clearly it is important to account for the flux in the wings of the PSF. However, it turns out to be rather difficult to reconstruct the true shape of the PSF accurately from the spatial profile of a monitor star spectrum. The reason is that standard reduction of the ABBA dither pattern results in one positive and two negative imprints each $2.52\arcsec$ apart, meaning the PSF wings are largely subtracted out and the resulting profile is too steep. The problem is seeing dependent and becomes worse if the seeing is larger. We therefore proceeded to reconstruct the true PSF separately for every mask (with seeing varying from 0.65 to 1.1\arcsec). 
As the central regions of the PSFs are still well approximated by a Gaussian, we used Gaussian fits to the collapsed spectra of the monitor stars to characterize the seeing FWHM for each of the 8 $K-$band masks. We then reconstructed the approximate true PSF by first integrating a two-dimensional Moffat ($\beta=2.5$) PSF over the width of a $0.7\arcsec$ wide virtual slit and subtracting 1/2 times the intensity offset by $2.52\arcsec$ on either side to simulate the reduction process. Because the FWHM of a Gaussian fit to the resulting spectral PSF is 12\% broader than the original Moffat FWHM, we scaled the FWHM of the two-dimensional Moffat to match the simulated spectral PSF to the observations. Figure \ref{fig:wings} illustrates two extreme cases of best and worst seeing {for our data}.

We verified the effect of using a Gaussian or Moffat profile in our modelling by calculating rotational velocities using either the Moffat PSFs derived above, or Gaussian fits to the collapsed star spectra. The mean velocity is 4\% smaller if a Gaussian is assumed, with up to 15\% effects for some individual cases. 

\subsection{Target sample selection}

The primary ZFIRE sample was designed to spectroscopically confirm a large cluster of galaxies at $z=2.095$ \citep{Spitler12, Yuan14}  within the COSMOS field \citep{Scoville07}. The sample was optimized by focusing mostly on near-IR star-forming galaxies, with strong expected signatures such as \ha\ emission. Star-forming galaxies as part of the cluster were selected based on their rest-frame $U-V$ and $V-J$ colors, with photometric redshifts between $2.0<z<2.2$. $K$-band magnitudes of $K<24$ were priority sources, but fainter sources could be included as well. Non star-forming galaxies were prioritized next and lastly, field galaxies (not necessarily at the cluster redshift) could be used as fillers for the mask.  In total 187 unique sources were listed for $K-$band observations. 36 of these were observed in two different masks and 2 in three different masks, leading to a total of {224} spectra. 

Spectroscopic targets were originally obtained from the photometric redshift catalogs of ZFOURGE. The{se} were derived from ultra-deep near-IR $K_s-$band imaging ($\sim25.5$ mag). FourStar has a total of 6 near-IR medium bandwidth filters ($J_1,J_2,J_3,H_s,H_l$), that accurately sample the rest-frame $4000\mathrm{\AA}$/Balmer break at redshifts $1.5<z<4$. We combined these with a wealth of already public multiwavelength data at $0.3-24\micron$ to derive photometric redshifts, using the EAZY software \citep{Brammer08}. These redshifts were used as a prior for the MOSFIRE masks. The typical redshift uncertainty is $1-2\%$ for galaxies at $1.0<z<2.5$ \citep{Straatman16}. 

For this work we make use of the ZFOURGE stellar masses. These were calculated by fitting \citet{Bruzual03} stellar template models, using the software FAST \citep{Kriek09}, assuming a \citep{Chabrier03} initial mass function, exponentially declining star formation histories, solar metallicities and a \citet{Calzetti00} dust law. Galaxy sizes, axis-ratios and position angles are obtained from the size catalog of galaxies from the 3D-HST/CANDELS survey \citep{vanderWel14a,Skelton14}. These were crossmatched to ZFOURGE by looking for matches within $<0.7\arcsec$. The sizes were derived by fitting two-dimensional S\'ersic \citep{Sersic68} surface brightness profiles to HST/WFC3/F160W images, using the software GALFIT \citep{Peng10}.

From the original {$N=224$} ZFIRE $K_s$-band sample, we first selected 15{1} unique galaxies with $2.0<z_{spec}<2.5$, where we used spectroscopic redshifts derived from one-dimensional collapsed spectra \citep{Nanayakkara16}. Using the F160W position angles, we determined offsets with respect to the MOSFIRE masks: $\Delta\alpha=PA-\alpha_{mask}$, with PA the position angle of the major axis of the galaxy and $\alpha_{mask}$ the slit angle from the mask. We refined the sample by selecting only galaxies with $|\Delta\alpha|<40^{\circ}$ {to minimize slit angle corrections}, resulting in a sample of {81} galaxies. Some were included in more than one mask, and we have {102} spectra in total that follow these criteria. The \ha\ emission was inspected by eye for contamination from sky lines, and we only kept those instances that were largely free from skylines, removing {25}. Out of the remaining {77} spectra, {29} have very low SNR and were also omitted. We also looked for signs of AGN, by crossmatching with radio and X-ray catalogs \citep{Cowley16}. This revealed one AGN, which we removed. {Finally, we removed 5 spectra without corresponding HST/WFC3/F160W imaging.} The final high quality sample contains {42} spectra of {38} galaxies, and these form the basis for the kinematic analysis which we discuss next.

\section{Analysis}
\label{sec:analysis}

\subsection{\ha\ rotation model}

We modeled the rotation curves by fitting two-dimensional ($\lambda,r$) intensity models. We used the empirically motivated arctan function to model the velocity curve \citep{Courteau97,Willick97,Miller11}:

\begin{equation}
v(r)=V_0+\frac{2}{\pi}V_a \mathrm{arctan}\left(\frac{r-r_0}{r_t}\right)
\label{eq:vat}
\end{equation}

with $v(r)$ the velocity at radius $r$, $V_0$ the central velocity, $V_a$ the asymptotic velocity, $r_0$ the dynamic center and $r_t$ the turnover, or kinematic, scale radius. $r_t$ is a transitional point between the rising and flattening of the arctan curve. 

For relatively small proper motion if viewed on a cosmological scale, we can express the velocity as function of the wavelength difference with respect to the central wavelength $\lambda_0$ as:

\begin{equation}
\frac{v}{c}=\frac{\Delta\lambda}{\lambda_0}=\frac{\lambda-\lambda_0}{\lambda_0}
\label{eq:lv}
\end{equation}

Therefore we initially fit our model in wavelength space, and then afterwards convert the offset in $\lambda$ to velocity. In terms of wavelength, Equation \ref{eq:vat} becomes:

\begin{equation}
\lambda(r)=\lambda_0+\frac{2}{\pi}\lambda_a \mathrm{arctan}\left(\frac{r-r_0}{r_t}\right)
\label{eq:lat}
\end{equation}

We also model the spatial intensity of the emission, assuming an exponential disk:

\begin{equation}
I(r)=I_0\mathrm{exp}\left[\frac{-(r-r_0)}{R_s}\right]
\label{eq:exp}
\end{equation}

with $I(r)$ the intensity at radius $r$ and $I_0$ the central intensity. $r_0$ is the same in Equations \ref{eq:vat}, \ref{eq:lat} and \ref{eq:exp}, and the coordinates $(\lambda_0,r_0)$ represent the velocity centroid of the galaxy in \ha. $R_s$ is the scalelength of an exponential disk. At a given $r$, the intensity as a function of wavelength is modelled by a Gaussian profile, centered on $\lambda(r)$:

\begin{equation}
I(\lambda,r)=I(r)\mathrm{exp}\left[-\frac{(\lambda-\lambda(r))^2}{2(\sigma^2+\sigma_{instr}^2)}\right]
\label{eq:gf}
\end{equation}

with $\sigma$ the velocity dispersion and $\sigma_{instr}$ the instrumental broadening. $\sigma_{instr}=2.4\mathrm{\AA}$ was obtained from a Gaussian fit to a skyline. We allowed $\sigma$ to vary in the fit, but assumed it to be independent of radius. 

With Equations \ref{eq:lat} to \ref{eq:gf} we built a two-dimensional model of the \ha\ emission line, which was then smoothed with the PSF derived in Section \ref{sec:psf}. To avoid undersampling effects, we built the initial model on a grid with $3\times$ the spatial and wavelength resolution of the spectra. We also used a $3\times$ refined PSF. After convolving we rebinned the model by a factor $1/3$. We also subtracted half the intensity of the model at $\pm 14$ pixels to reproduce the dithering pattern. Parameters that can vary in the model are $\lambda_0,\lambda_a,r_0,r_t,I_0,R_s$ and $\sigma$. 

\subsection{Fitting procedure}
\label{sec:proc}

We fit the intensity model to $100\A$ wide spectral image stamps, centered on the \ha\ emission line. We used the Python scipy {\tt optimize.curve\_fit} algorithm, which is based on the Levenberg-Marquardt algorithm. This algorithm can be used to solve non-linear least squares minimization problems. The Levenberg-Marquardt algorithm can find local minima, but these are not necessarily the global minima, i.e. the best fits, that we are looking for. Therefore, we assessed each galaxy's spectral image stamp individually and we chose initial parameters for the model to be a reasonable match to the observed \ha\ emission.

In addition to the \ha\ stamps, we extracted corresponding images from the error spectra that are available for each observation. The error spectra represent standard errors on the flux in each pixel. The error stamps were matched by wavelength location to the \ha\ spectral image stamps, and we included these as weight arrays in the fit. We did not mask sky lines or pixels with low SNR, but simply used the (much) smaller weights from the error images at those locations.

\begin{figure*}
\begin{center}
\includegraphics[width=0.49\textwidth]{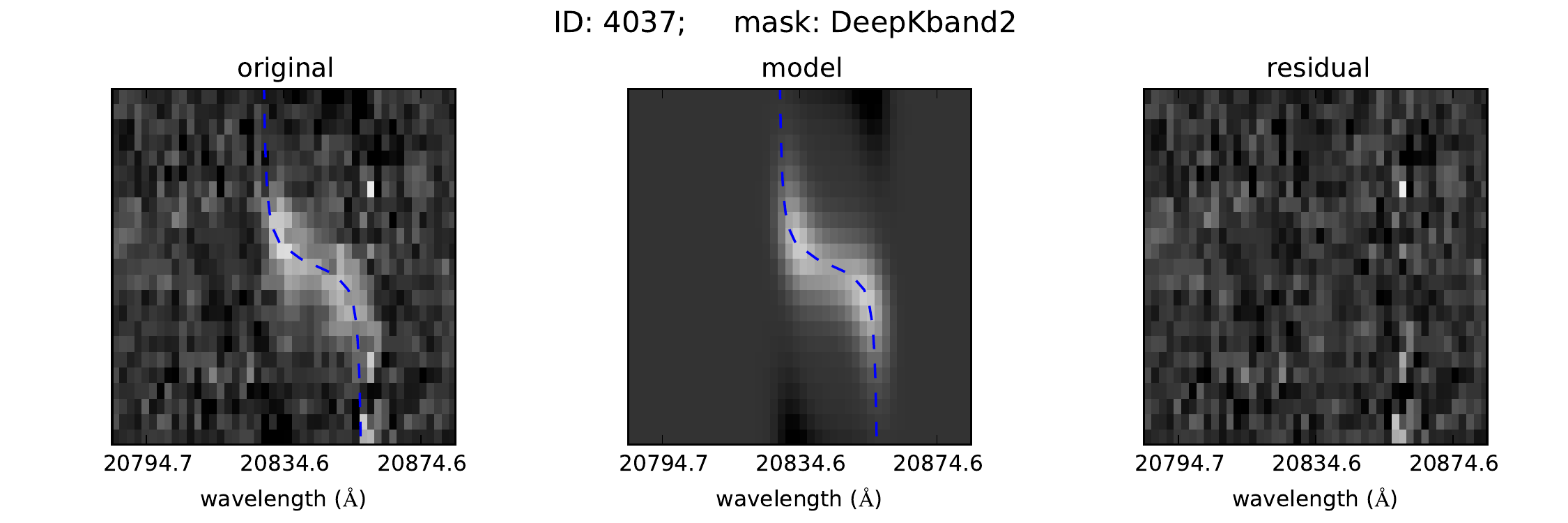}
\includegraphics[width=0.49\textwidth]{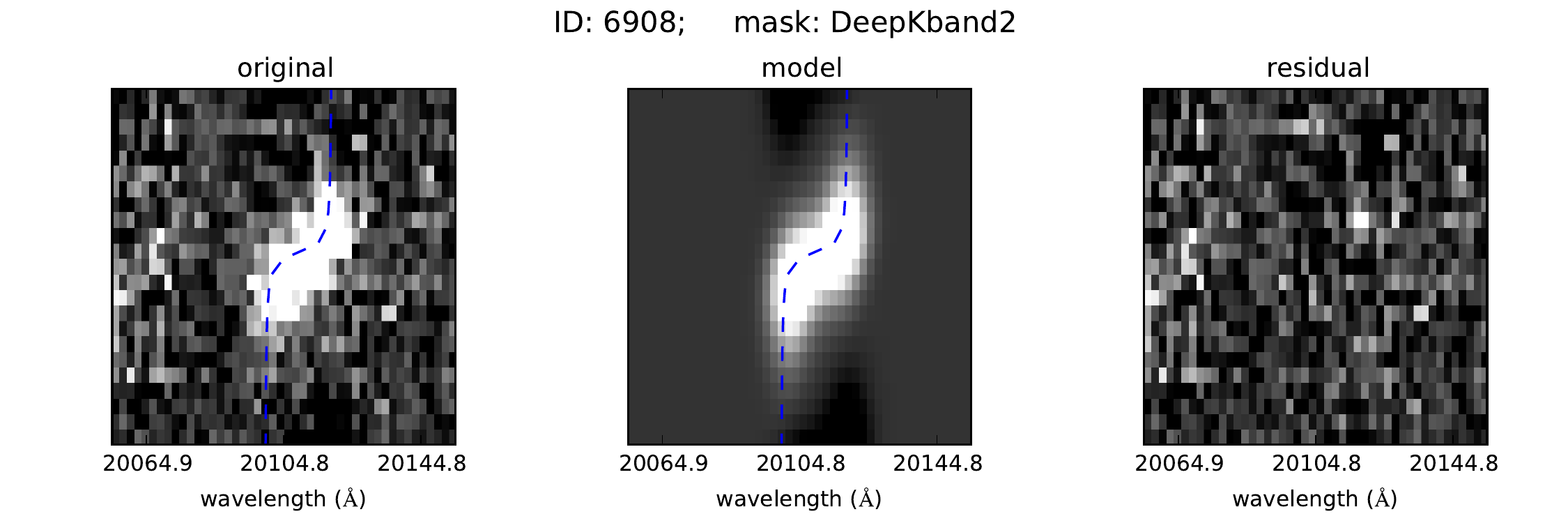}
\includegraphics[width=0.49\textwidth]{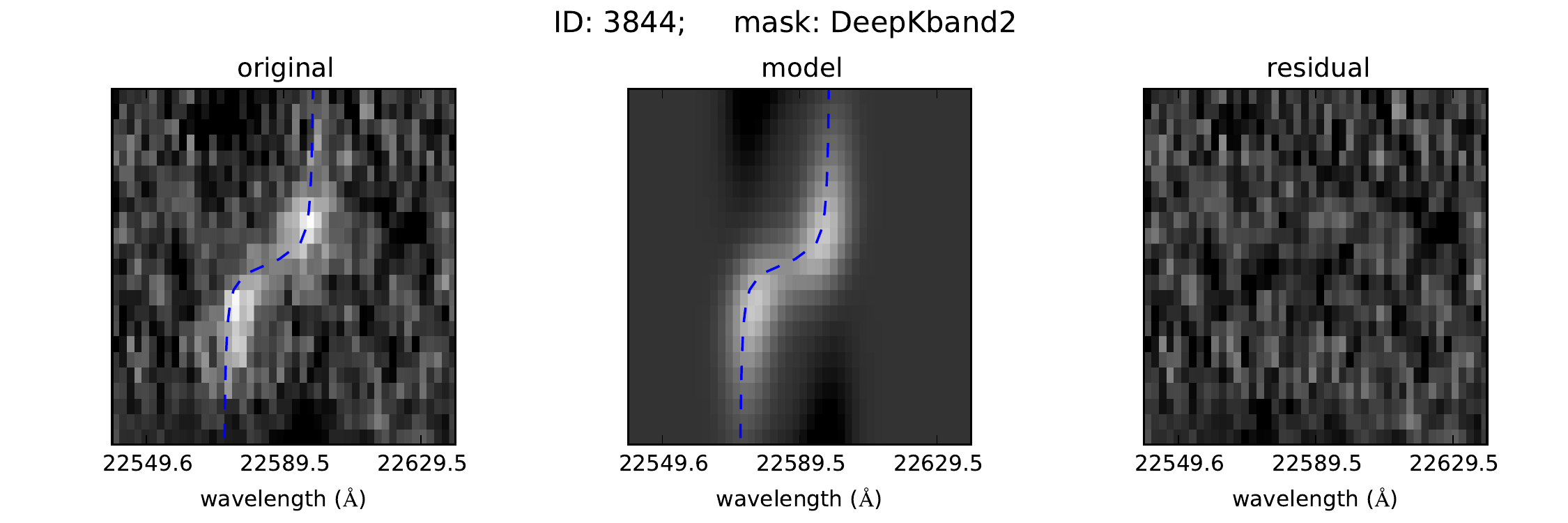}
\includegraphics[width=0.49\textwidth]{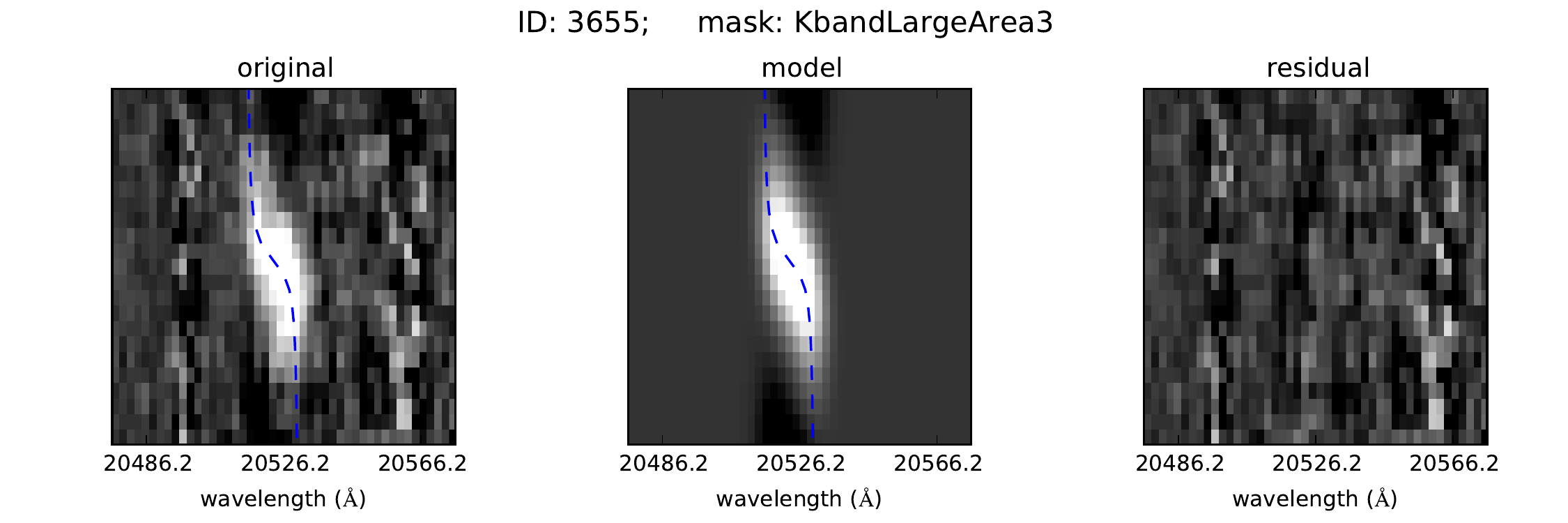}
\caption{Best-fit models. For each subsequent panel from left to right: the spectral image stamps, the best-fit model, the residual after subtracting the best-fit model. The blue dashed curves are the model arctan functions.}
\label{fig:bestfits}
\end{center}
\end{figure*}

In Figure \ref{fig:bestfits} we show the initial guesses and best-fit models for four example galaxies. The best-fit models are good representations of the \ha\ emission, with small residuals. {We also show the unconvolved arctangent functions constructed from the best-fit parameters using Equation \ref{eq:lat}. In Appendix \ref{app:collage} we show the spectral image stamps and best fits for the whole sample with additional radial velocity profiles derived from the emission in individual rows of each spectrum.} 

We estimated uncertainties on the parameters $\lambda_0,\lambda_a,r_0,r_t,I_0,R_s$ and $\sigma$, by applying a Monte Carlo procedure. For every source, we subtracted the best-fit two-dimensional model from the spectral image stamp, obtaining the residual images shown Figure \ref{fig:bestfits}. We then shifted the residual pixels by a random number of rows and columns, preserving local pixel-to-pixel correlations. The magnitude of the shift was drawn from a Gaussian distribution centered on zero, allowing negative values, i.e., shifting in the opposite direction, and with a standard deviation of two pixels. The number of rows and columns to be shifted were generated independently from each other. We then added the best-fit model back to the shifted residual and re-ran our fitting procedure. We repeated this process 200 times, obtaining for each parameter a distribution of values. We calculated the standard deviations for each parameter and used these as the uncertainties. 

\subsection{Velocities}
\label{sec:vel}

We measured the velocities from Equation \ref{eq:vat} at 2.2 times the scale radius ($R_s$) of the exponential brightness profile. We chose $r=2.2R_s$ as this is the radius where the rotation curve of a self-gravitating ideal exponential disc peaks \citep{Freeman70}. It is also a commonly adopted parameter in literature \citep[e.g.][]{Miller11}. Its main advantage is that it gives a consistent approximation of the rotational velocity across the sample, while avoiding extrapolations towards large radii and low SNR regions of the spectrum.

We corrected the velocities for inclination {using}: 

\begin{equation}
v'_{2.2}=\frac{v_{2.2}}{\mathrm{sin}(i)}
\end{equation}

with

\begin{equation}
i=\mathrm{cos^{-1}}\sqrt{\frac{(b/a)^2-q_0^2}{1-q_0^2}}
\end{equation}

We adopt here the convention that $i=0^{\circ}$ for galaxies viewed face-on and $i=90^{\circ}$ for edge-on galaxies. We used the axis-ratio's {($b/a$)} derived with GALFIT from \citet{vanderWel14a}. {Uncertainties on the axis-ratio were propagated and added to the velocity uncertainty from the Monte Carlo procedure.} 

$q_0\simeq0.1-0.2$ represents the intrinsic flattening ratio of an edge-on galaxy. Following convention we adopt $q_0=0.19$  \citep{Pizagno07, Haynes84}. {It has been shown that galaxies with $9<\mathrm{log}M/M_{\odot}<10$ at $z>1$ have a higher fraction of more elongated systems \citep[e.g.][]{vanderWel14b}. We note that using the axis-ratio's to derive the inclination may therefore lead to underestimated corrections for some of the galaxies in our sample, as 9/21 have $\mathrm{log}M/M_{\odot}<10$.}

\subsection{Two-dimensional PSF and projection effects}
\label{sec:smooth}

When considering slit spectra, with one spatial dimension, we need to account for systematic effects due to the two-dimensional nature of the PSF smoothing {as well as any mismatch between the slit angle and kinematic angle, here assumed to be the F160W position angle}. The main effect is that two-dimensional smoothing will effectively lead to an underestimation of the line-of-sight motion captured in one-dimensional spectra, as a flux component from lower velocity regions is mixed in. The effect depends on the apparent size of the galaxy relative to the size of the PSF and the size of the slit, i.e., mixing occurs even for an infinitely thin slit if the seeing is significant, and vice versa. 

To assess this effect, we generated a suite of 500 emission-line models \citep{Bekiaris16} of infinitely thin galaxies with similar sizes and velocities to our sample. We projected these onto the MOSFIRE 2D space, using various inclination angles ($0^{\circ}<i<90^{\circ}$) and slit angles relative to the major axes ($0^{\circ}<|\Delta \alpha|<45^{\circ}$), a finite slit width of $0.7\arcsec$, and a 2D Moffat PSF with FWHM$=0.5\arcsec$ and $\beta=2.5$. This PSF results in a Gaussian approximation of the seeing of $0.6\arcsec$ in 1D. In the simulations $R_s$ was varied between 1 and 5 kpc, $\sigma$ between 20 and 100 km/s and $V_a$ between 100 and 400 km/s. $r_t$ was defined as $R_s/3${, typical of the galaxies in this study}. We added noise based on the noise spectrum of observed galaxy 4037 and scaled to match the typical \ha\ SNR of the galaxies in our sample with a median of SNR$=25$ (see Section \ref{sec:res} for details on how we derive SNR). {We included two negative imprints to simulate the ABBA pattern.} The \citet{Bekiaris16} models are part of a fitting code designed to diagnose IFU data and are therefore an excellent sanity check on methods used for single-slit modeling. 

\begin{figure*}
\begin{center}
\includegraphics[width=0.8\textwidth]{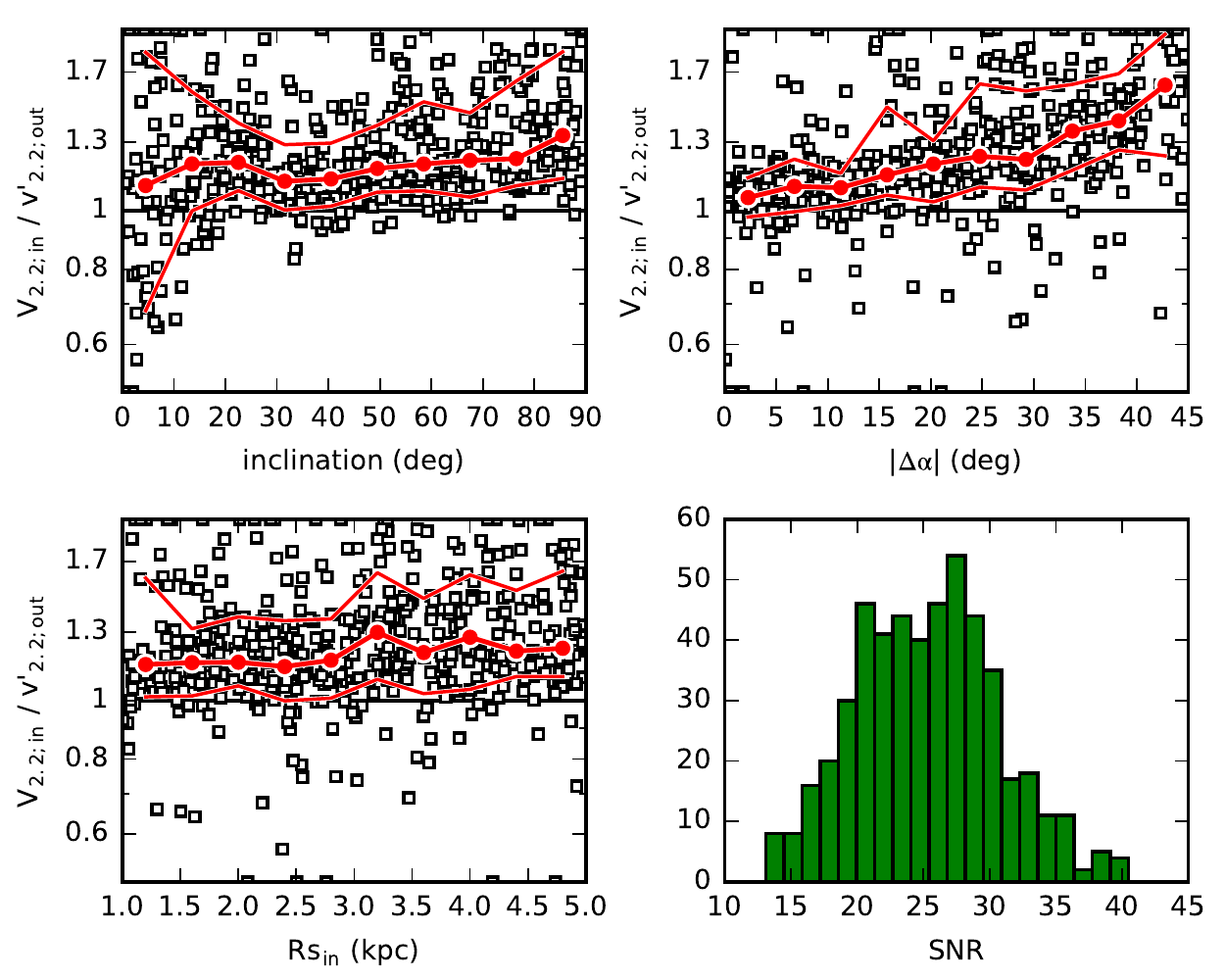}
\caption{{Results of simulating 500 MOSFIRE spectra, with a 0.6$\arcsec$ PSF and typical SNR. We show the offset between input and measured {velocity} as a function of inclination, $|\Delta \alpha|$ and input $R_s$. The red lines are the running median and $1\sigma$ percentiles. {There is a slight trend with inclination, indicating mixing of light plays a role, no trend with input $R_s$, and a strong trend with slit mismatch.}}}
\label{fig:simtest}
\end{center}
\end{figure*}

We measured the rotational velocity in the same way as for the observed spectra, including the correction for inclination angle. The {ratio between $V_{2.2;in}$ (the actual rotational velocity) and $v'_{2.2;out}$ (the measured rotational velocity) is} shown in Figure \ref{fig:simtest}. There is a slight trend with inclination, with on one side larger scatter for face-on galaxies, due to the general difficulties of measuring at small inclination. On the other side we find, as expected, an increase in both the scatter as well as the ratio $V_{2.2;in}/v'_{2.2;out}$ for very inclined galaxies, that suffer the most from 2D smoothing effects. {T}here is a {strong} trend of increasing $V_{2.2;in}/v'_{2.2;out}$ towards higher $|\Delta \alpha|$, {but we do not find a significant trend with} input radi{us}. {In general the result is that the recovered velocities are too small by a median factor of 1.19, depending on $|\Delta \alpha|$, with a scatter of 0.24}.

We found similar results for different seeing values {or a factor 2 lower SNR}. Given that {there is a clear trend with slit angle mismatch}, {we derived a $|\Delta \alpha|-$dependent correction} to our velocities{, using the median offset in $V_{2.2;in}/v'_{2.2;out}$ in a five degree bin around the value of $|\Delta \alpha|$ associated with each spectrum. We propagated the scatter around the median offset into the velocity uncertainty already derived from the Monte Carlo procedure. From hereon we use the symbol $V_{2.2}$ for the final slit angle and projection corrected velocities}.

\subsection{Results}
\label{sec:res}

\begin{figure*}
\begin{center}
\includegraphics[width=0.8\textwidth]{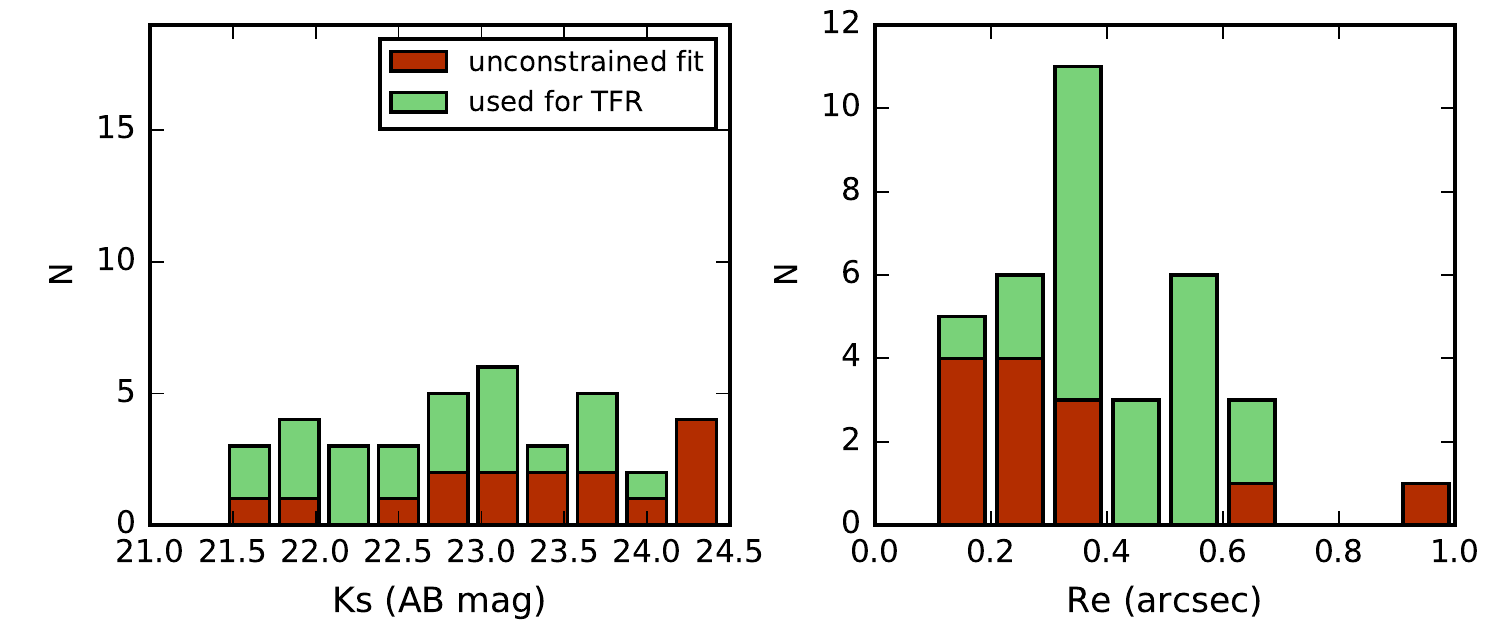}
\caption{$K_s$-band magnitude and effective F160W radius $R_e$ stacked histograms of {38} galaxies in the high quality sample. For {18} spectra of {16} galaxies the fits were poorly constrained {or could not be fit with an exponential brightness profile}. The remaining {22} galaxies were used to derive the \tf\ relation. The removed galaxies have fainter $K_s$-band magnitudes and smaller sizes on average, which resulted in a selection bias towards larger and/or brighter galaxies.}
\label{fig:nohist}
\end{center}
\end{figure*}

\begin{table*}
\begin{center}
\caption{Results}
\begin{scriptsize}
\begin{threeparttable}
\begin{tabular}{l l l l l l l l}
\hline
\hline
ID & mask & seeing & $\mathrm{z_{centroid}}$ & $V_a$ & $r_t$ & $R_s$ & $\mathrm{SNR_{H\alpha}}$\\
& & ($\arcsec$) & & (km/s) & ($\arcsec$) & ($\arcsec$) & \\
\hline
1814 & KbandLargeArea4 & 0.65 & 2.1704$\pm$0.00016 & 60.6$\pm$83.0 & 6.5e-10$\pm$1.3e-01 & 0.09$\pm$0.03 & 46\\
2715 & mask2 & 0.67 & 2.0824$\pm$0.00004 & 52.3$\pm$90.5 & 8.2e-07$\pm$1.6e-01 & 0.16$\pm$0.03 & 22\\
2723 & mask2 & 0.67 & 2.0851$\pm$0.00004 & 102.1$\pm$105.7 & 2.7e-06$\pm$1.6e-04 & 0.13$\pm$0.03 & 12\\
2765 & mask1 & 0.71 & 2.2279$\pm$0.00008 & 242.6$\pm$83.2 & 3.5e-01$\pm$1.7e-01 & 0.24$\pm$0.03 & 36\\
3074 & mask1 & 0.71 & 2.2267$\pm$0.00005 & 117.8$\pm$114.4 & 2.6e-02$\pm$2.6e-02 & 0.37$\pm$0.05 & 16\\
3527 & KbandLargeArea4 & 0.65 & 2.1890$\pm$0.00004 & 87.9$\pm$3.2 & 1.6e-03$\pm$3.4e-04 & 0.23$\pm$0.01 & 71\\
3598 & mask2 & 0.67 & 2.2279$\pm$0.00006 & 145.3$\pm$19.5 & 1.3e-01$\pm$3.6e-02 & 0.34$\pm$0.05 & 11\\
3633 & DeepKband2 & 0.80 & 2.0991$\pm$0.00008 & 170.7$\pm$22.2 & 7.9e-02$\pm$2.8e-02 & 0.44$\pm$0.09 & 20\\
3655 & KbandLargeArea3 & 1.09 & 2.1263$\pm$0.00003 & 110.0$\pm$31.5 & 2.2e-01$\pm$1.7e-01 & 0.35$\pm$0.06 & 41\\
3680 & mask3 & 0.68 & 2.1753$\pm$0.00005 & 132.7$\pm$31.7 & 1.2e-01$\pm$6.4e-02 & 0.19$\pm$0.04 & 10\\
3714 & mask3 & 0.68 & 2.1761$\pm$0.00005 & 81.1$\pm$76.7 & 2.1e-02$\pm$1.2e-02 & 0.22$\pm$0.01 & 33\\
3844 & DeepKband2 & 0.80 & 2.4404$\pm$0.00001 & 177.5$\pm$181.8 & 1.2e-01$\pm$1.3e-01 & 0.70$\pm$0.02 & 16\\
4010 & KbandLargeArea4 & 0.65 & 2.2216$\pm$0.00011 & 56.2$\pm$24.9 & 6.4e-09$\pm$2.3e-01 & 0.21$\pm$0.05 & 18\\
4037 & DeepKband2 & 0.80 & 2.1750$\pm$0.00004 & 212.2$\pm$212.9 & 1.5e-01$\pm$1.5e-01 & 0.33$\pm$0.04 & 16\\
$-$ & mask2 & 0.67 & 2.1747$\pm$0.00005 & 190.8$\pm$23.5 & 8.9e-02$\pm$3.0e-02 & 0.34$\pm$0.03 & 20\\
4099 & mask3 & 0.68 & 2.4391$\pm$0.00002 & 51.1$\pm$52.0 & 1.3e-02$\pm$1.2e-02 & 0.28$\pm$0.03 & 15\\
4645 & DeepKband1 & 1.10 & 2.1011$\pm$0.00005 & 184.7$\pm$27.7 & 1.7e-01$\pm$4.6e-02 & 0.26$\pm$0.03 & 3\\
4930 & DeepKband2 & 0.80 & 2.0974$\pm$0.00002 & 70.4$\pm$27.6 & 1.5e-01$\pm$1.9e-01 & 0.24$\pm$0.03 & 14\\
5630 & KbandLargeArea4 & 0.65 & 2.2427$\pm$0.00003 & 152.1$\pm$61.9 & 2.2e-01$\pm$2.2e-01 & 0.27$\pm$0.02 & 26\\
5870 & mask4 & 0.66 & 2.1036$\pm$0.00006 & 51.7$\pm$8.6 & 3.2e-06$\pm$1.0e-02 & 0.20$\pm$0.05 & 18\\
6908 & DeepKband2 & 0.80 & 2.0633$\pm$0.00002 & 145.8$\pm$17.1 & 8.3e-02$\pm$3.4e-02 & 0.30$\pm$0.02 & 19\\
$-$ & mask1 & 0.71 & 2.0632$\pm$0.00004 & 144.4$\pm$142.5 & 4.1e-02$\pm$3.6e-02 & 0.33$\pm$0.05 & 26\\
8108 & mask2 & 0.67 & 2.1622$\pm$0.00005 & 230.9$\pm$54.0 & 1.5e-01$\pm$5.8e-02 & 0.15$\pm$0.03 & 21\\
9420 & mask3 & 0.68 & 2.0633$\pm$0.00016 & 344.0$\pm$337.2 & 4.2e-01$\pm$4.0e-01 & 0.28$\pm$0.05 & 5\\
\hline
ID & mask & $\sigma$ & $v_{2.2}$ & $V_{2.2}$ & $V_{2.2;in}/v'_{2.2;out}$ & $\mathrm{sin}(i)$ & $\alpha_{mask}$\\
& & (km/s) & (km/s) & (km/s) & & (deg) & \\
\hline
1814 & KbandLargeArea4 & 47.2$\pm$10.6 & 60.6$\pm$15.7 & 115.5$\pm$29.0 & 1.20$\pm$0.14 & 0.63 & 2.0\\
2715 & mask2 & 57.0$\pm$5.6 & 52.3$\pm$11.5 & 72.2$\pm$15.3 & 1.11$\pm$0.08 & 0.81 & -47.3\\
2723 & mask2 & 69.4$\pm$6.1 & 102.1$\pm$7.4 & 283.4$\pm$46.2 & 1.36$\pm$0.23 & 0.49 & -47.3\\
2765 & mask1 & 83.6$\pm$11.0 & 150.2$\pm$29.4 & 242.0$\pm$68.0 & 1.26$\pm$0.31 & 0.78 & 134.0\\
3074 & mask1 & 79.4$\pm$11.7 & 115.4$\pm$14.3 & 144.9$\pm$19.1 & 1.10$\pm$0.07 & 0.87 & 134.0\\
3527 & KbandLargeArea4 & 68.4$\pm$4.1 & 87.7$\pm$3.3 & 107.2$\pm$7.5 & 1.09$\pm$0.07 & 0.89 & 2.0\\
3598 & mask2 & 75.8$\pm$12.8 & 128.9$\pm$19.4 & 184.8$\pm$36.1 & 1.39$\pm$0.20 & 0.97 & -47.3\\
3633 & DeepKband2 & 52.2$\pm$4.9 & 161.9$\pm$19.3 & 202.9$\pm$29.1 & 1.19$\pm$0.12 & 0.95 & -62.0\\
3655 & KbandLargeArea3 & 27.3$\pm$10.5 & 90.5$\pm$10.9 & 288.2$\pm$39.6 & 1.10$\pm$0.09 & 0.35 & 59.0\\
3680 & mask3 & 0.0$\pm$17.4 & 108.9$\pm$19.5 & 154.9$\pm$38.6 & 1.24$\pm$0.26 & 0.87 & 14.8\\
3714 & mask3 & 70.2$\pm$4.4 & 78.9$\pm$7.4 & 130.9$\pm$15.6 & 1.19$\pm$0.11 & 0.72 & 14.8\\
3844 & DeepKband2 & 52.8$\pm$4.5 & 168.6$\pm$3.1 & 251.5$\pm$22.8 & 1.10$\pm$0.11 & 0.73 & -62.0\\
4010 & KbandLargeArea4 & 104.6$\pm$9.8 & 56.2$\pm$14.1 & 72.8$\pm$18.1 & 1.10$\pm$0.11 & 0.85 & 2.0\\
4037 & DeepKband2 & 21.9$\pm$8.9 & 185.1$\pm$9.2 & 283.0$\pm$29.4 & 1.10$\pm$0.11 & 0.72 & -62.0\\
$-$ & mask2 & 61.0$\pm$12.8 & 176.5$\pm$12.2 & 270.4$\pm$30.8 & 1.10$\pm$0.11 & $-$ & -47.3\\
4099 & mask3 & 25.8$\pm$5.2 & 50.4$\pm$7.7 & 86.8$\pm$20.3 & 1.24$\pm$0.26 & 0.72 & 14.8\\
4645 & DeepKband1 & 0.0$\pm$7.8 & 151.1$\pm$16.3 & 164.8$\pm$21.4 & 1.05$\pm$0.08 & 0.97 & 2.0\\
4930 & DeepKband2 & 50.0$\pm$5.1 & 57.8$\pm$12.0 & 70.9$\pm$21.9 & 1.23$\pm$0.26 & 1.00 & -62.0\\
5630 & KbandLargeArea4 & 65.9$\pm$10.4 & 117.8$\pm$21.8 & 163.3$\pm$34.9 & 1.37$\pm$0.19 & 0.99 & 2.0\\
5870 & mask4 & 21.1$\pm$10.1 & 51.7$\pm$6.9 & 74.9$\pm$11.6 & 1.10$\pm$0.11 & 0.76 & -63.0\\
6908 & DeepKband2 & 46.1$\pm$7.8 & 134.0$\pm$11.0 & 324.9$\pm$46.7 & 1.36$\pm$0.18 & 0.56 & -62.0\\
$-$ & mask1 & 86.9$\pm$11.3 & 139.3$\pm$22.0 & 298.0$\pm$53.7 & 1.20$\pm$0.15 & $-$ & 134.0\\
8108 & mask2 & 45.6$\pm$16.8 & 168.8$\pm$21.8 & 190.9$\pm$27.8 & 1.11$\pm$0.08 & 0.98 & -47.3\\
9420 & mask3 & 36.7$\pm$23.8 & 212.9$\pm$40.7 & 257.2$\pm$51.6 & 1.19$\pm$0.11 & 0.99 & 14.8\\
\hline
\end{tabular}
\begin{tablenotes}
\item {Columns explained from left to right.}
\item {ID: galaxy ID; mask: observing mask; seeing: measured gaussian seeing; $\mathrm{z_{cenroid}}$: redshift based on kinematic center; $V_a$: best-fit $V_a$; $r_t$ best-fit $r_t$; $R_s$: best-fit $R_s$; $\mathrm{SNR_{H\alpha}}$: signal-to-noise of the \ha\ emission line; $\sigma$: best-fit intrinsic velocity dispersion; $v_{2.2}$: velocity derived at $2.2R_s$; $V_{2.2}$: velocity after correcting for inclination, projection effects and slit misalignment; $V_{2.2;in}/v'_{2.2;out}$: median input versus output ratio of emission line models for the given $|\Delta \alpha|$; $\mathrm{sin}(i)$: inclination correction; $\alpha_{mask}$: slit angle.}
\item {The final velocity was derived from $v_{2.2}$ as $V_{2.2}=(v_{2.2}/\mathrm{sin}(i))(V_{2.2;in}/v'_{2.2;out})$.}
\end{tablenotes}
\end{threeparttable}
\end{scriptsize}
\label{t:tr}
\end{center}
\end{table*}

The best-fit parameters of the rotation model and their uncertainties, along with $v_{2.2}$ and $V_{2.2}$, are shown in Table \ref{t:tr}. Of the {42} spectra in the high quality sample, we obtained good fits for 24 (of 22 galaxies), while for {18} spectra we obtained poorly constrained fits{, with large random uncertainties ($>30\%$) on the velocities}. We therefore removed these {18} spectra (of {16} galaxies) from the sample. To evaluate if removing the failed fits introduces biases relative to the target sample we show the distribution of the $K_s-$band magnitudes and sizes in Figure \ref{fig:nohist}. The $K_s-$band magnitudes for the good fits are brighter than those of the full target sample (median {$K_s= 22.8$} versus median {$K_s=23.5$}) and the galaxies are slightly larger (median {$R_e= 0.40\arcsec$} versus {$R_e= 0.26\arcsec$}). So removing these galaxies does bias the sample to somewhat brighter and larger galaxies. 

The {22} galaxies for which we will derive the \tf\ relation have high velocities and velocity dispersions, with a median {$V_{2.2}=164$ km/s}, $\sigma=53$ km/s and $V/\sigma=3${$.5$}. We note that these dispersions could be slightly overestimated, e.g., the dispersion reflects mixing of velocity gradients on scales smaller than the seeing.

{In case of high SNR the uncertainties are not dominated by random errors and the Monte Carlo procedure would result in relatively small errors. Because it is unlikely that we can derive velocities with more than 10\% accuracy, we impose a minimum error of 10\% of the measured velocities for all sources.}

\begin{figure*}
\begin{center}
\includegraphics[width=0.8\textwidth]{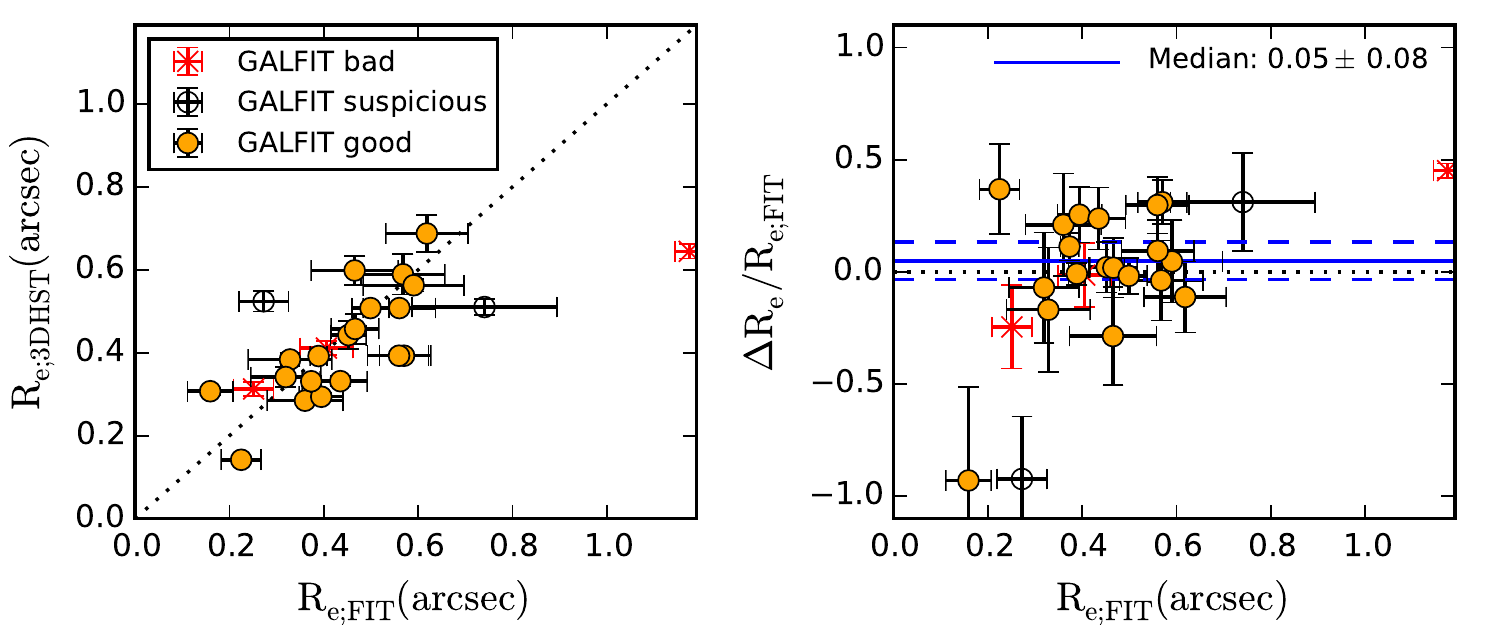}
\caption{Left: $R_e=1.678*R_s$ in the $K-$band from our fits versus $R_e$ in the HST/WFC3/F160W-band measured by \citet{vanderWel14a}, for {25} spectra. The dotted line indicates the one-to-one relation. Right: $\Delta R_e/R_{e;FIT}=(R_{e;FIT}-R_{e;3DHST})/R_{e;FIT}$ as a function of $R_{e,FIT}$. The bootstrapped median and $1\sigma$ error on the median {(excluding any GALFIT bad fits indicated by red crosses)} are shown as the solid and dashed {blue} lines, respectively.}
\label{fig:rscmp}
\end{center}
\end{figure*}

At high redshift measuring the kinematic profile of a galaxy is more difficult due to the smaller angular scales for distant galaxies, and seeing effects and SNR play a larger role. We therefore verified our size measurements. We converted the best-fit $R_s$ derived from the $K-$band spectra to effective radius ($R_e$), using $R_e=1.678R_s$ (valid for exponential disks), and compared this with the effective radii from HST/WFC3/F160W image reported by \citet{vanderWel14a}. On average we find good agreement, with some scatter, and we derived a bootstrapped median {$\Delta R_e/R_e=0.05\pm0.08$} (Figure \ref{fig:rscmp}). The most prominent outliers, with $|\Delta R_e/R_e|>0.5$, occur for two small galaxies{, that have $R_s<25\%$ of the seeing}. {One additionally} has a very irregular morphology, and was flagged by \citet{vanderWel14a} as a suspicious GALFIT result. 

We note that {7/25} fits resulted in very small $r_t$, with $r_t<0.02\arcsec$. This is clearly much less than the resolution of a pixel: $0.18\arcsec$. To investigate the potential impact of small $r_t$ on the velocities, we re-fit the{se} spectra limiting $r_t$ to $r_t > 0.02\arcsec$, 	obtaining a median velocity that is 10\% higher. This may indicate that the velocities are underestimated for sources with small $r_t$, but without knowing the true $r_t$, the effect is difficult to quantify. 

The total SNR is included in Table \ref{t:tr}. We measured the SNR within $5R_s$ above and below the center of the line, but never beyond $1.26\arcsec$ to avoid the negative imprints of the emission line in the spectrum. We also defined a wavelength region within which to measure SNR, defined by the maximum shear of the line, plus a buffer of $3FWHM_{\lambda}=3(2\sqrt{2\mathrm{ln}2})\sqrt{\sigma^2+\sigma_{instr}^2}\A$. The SNR within these limits was calculated by summing the flux and summing the squares of the equivalent pixels in the noise spectrum, and dividing the first by the square root of the latter. 

{Two} galaxies were included in two masks (4037 and 6908 in Table \ref{t:tr}). As they were observed under different seeing conditions and have different SNR and slit-angle, they provide a useful check on consistency. Encouragingly, we find that these galaxies have velocities, redshifts and scale parameters that agree between masks within their uncertainties. We averaged their velocities to derive the \tf\ relation in the next Section.

\section{The Tully-Fisher relation at $2.0<z<2.5$}\label{sec:tf22}

\subsection{Tully-Fisher sample}\label{sec:tfsample}
 
\begin{figure*}
\begin{center}
\vspace{20pt}
\includegraphics[width=0.8\textwidth]{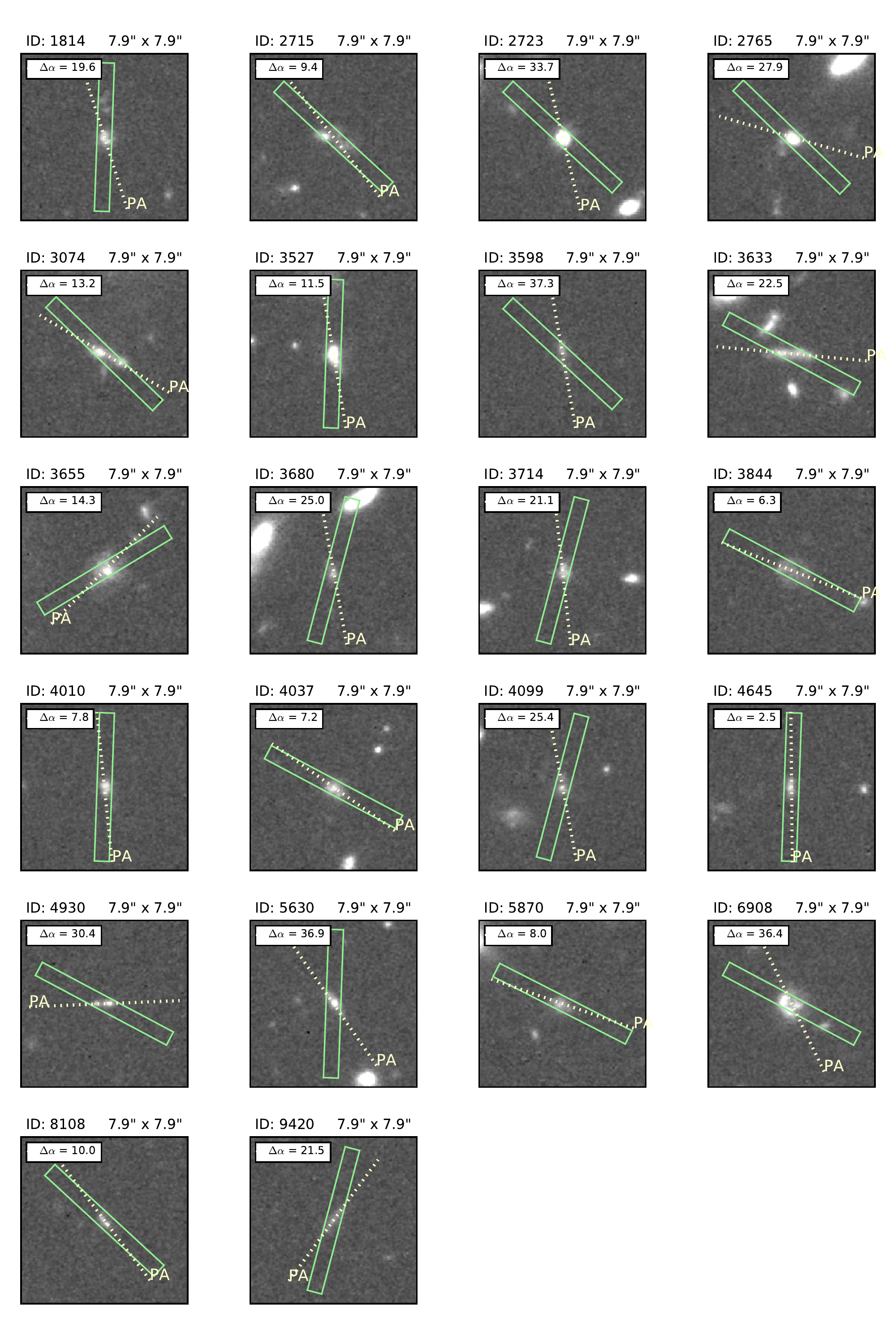}
\caption{{HST/WFC3/}F160W images of the galaxies in our \tf\ sample. {The green box shows the dimensions and orientation of the slit compared to the galaxies. The dotted line indicates the PA of the major axis.}}
\label{fig:stamps}
\end{center}
\end{figure*}

\begin{table*}
\begin{center}
\caption{Full sample}
\begin{scriptsize}
\begin{threeparttable}
\begin{tabular}{l r r l l l l l l l l l}
\hline
\hline
ID & R.A. & Decl & $\mathrm{K_{s}}$ & F160W & $\mathrm{M/10^{10}}$ & SFR & $R_{e}$ & GALFIT & $b/a$ & $n_{sersic}$ & P.A.\\
& (deg) & (deg) & (AB mag) & (AB mag) & $\mathrm{(M_{\odot})}$ & $\mathrm{(M_{\odot}/yr)}$ & ($\arcsec$) & flag\tnote{a} & & & (deg) \\
\hline
1814 & 150.1680908 & 2.2112861 & 23.0 & 23.3 & 0.6 & 17.5 & 0.31$\pm$0.01 & 0 & 0.79$\pm$0.02 & 0.6 & -17.6\\
2715 & 150.0895386 & 2.2235634 & 22.6 & 23.0 & 0.8 & 34.4 & 0.52$\pm$0.02 & 1 & 0.61$\pm$0.03 & 0.6 & -37.9\\
2723 & 150.1172638 & 2.2238791 & 21.5 & 22.0 & 7.8 & 66.4 & 0.14$\pm$0.00 & 0 & 0.88$\pm$0.02 & 4.5 & -13.6\\
2765 & 150.119339 & 2.2241209 & 21.9 & 22.5 & 2.4 & 101.1 & 0.29$\pm$0.01 & 0 & 0.64$\pm$0.01 & 1.8 & -73.9\\
3074 & 150.1209106 & 2.2288201 & 22.2 & 22.6 & 1.4 & 30.9 & 0.69$\pm$0.04 & 0 & 0.51$\pm$0.02 & 2.3 & -59.2\\
3527 & 150.1825714 & 2.2358665 & 21.9 & 22.5 & 2.1 & 168.8 & 0.39$\pm$0.01 & 0 & 0.48$\pm$0.01 & 1.0 & -9.5\\
3598 & 150.1120911 & 2.2368469 & 22.9 & 23.7 & 2.3 & 69.9 & 0.59$\pm$0.05 & 0 & 0.30$\pm$0.03 & 1.1 & -10.0\\
3633 & 150.1249237 & 2.236979 & 22.1 & 22.7 & 2.5 & 150.5 & 0.51$\pm$0.02 & 1 & 0.37$\pm$0.02 & 0.9 & -84.5\\
3655 & 150.1691284 & 2.2383816 & 21.9 & 22.4 & 2.1 & 127.3 & 0.56$\pm$0.01 & 0 & 0.94$\pm$0.01 & 1.1 & 44.7\\
3680 & 150.063446 & 2.237031 & 24.0 & 24.1 & 0.2 & 16.9 & 0.34$\pm$0.02 & 0 & 0.51$\pm$0.04 & 0.9 & -10.2\\
3714 & 150.0707703 & 2.2381561 & 22.7 & 23.3 & 1.5 & 23.1 & 0.33$\pm$0.01 & 0 & 0.71$\pm$0.02 & 0.7 & -6.3\\
3844 & 150.1094666 & 2.2400432 & 22.4 & 23.0 & 1.5 & 65.2 & 0.64$\pm$0.02 & 2 & 0.69$\pm$0.02 & 0.2 & -68.3\\
4010 & 150.1798706 & 2.2423265 & 23.0 & 23.5 & 1.1 & 131.2 & 0.28$\pm$0.01 & 0 & 0.55$\pm$0.02 & 0.4 & -5.8\\
4037 & 150.0981293 & 2.2428052 & 22.2 & 23.1 & 4.9 & 95.7 & 0.39$\pm$0.01 & 0 & 0.71$\pm$0.02 & 0.6 & -54.8\\
4099 & 150.0718231 & 2.243396 & 23.1 & 23.7 & 2.2 & 57.5 & 0.46$\pm$0.04 & 0 & 0.71$\pm$0.03 & 1.7 & -10.6\\
4645 & 150.0743256 & 2.2516196 & 23.6 & 23.8 & 0.3 & 14.1 & 0.33$\pm$0.01 & 0 & 0.32$\pm$0.03 & 0.5 & -0.5\\
4930 & 150.0559387 & 2.2557058 & 23.5 & 23.8 & 0.3 & 15.3 & 0.41$\pm$0.02 & 2 & 0.09$\pm$0.02 & 0.2 & 87.6\\
5630 & 150.2009735 & 2.2665324 & 22.9 & 23.3 & 0.9 & 111.3 & 0.44$\pm$0.03 & 0 & 0.24$\pm$0.02 & 2.4 & -34.9\\
5870 & 150.0609436 & 2.2696433 & 23.1 & 23.5 & 0.8 & 15.2 & 0.38$\pm$0.02 & 0 & 0.67$\pm$0.02 & 0.7 & -71.0\\
6908 & 150.0834198 & 2.2857671 & 21.7 & 22.2 & 3.0 & 138.9 & 0.51$\pm$0.01 & 0 & 0.84$\pm$0.01 & 0.4 & -25.6\\
8108 & 150.0622711 & 2.3044007 & 23.8 & 24.1 & 0.5 & 16.3 & 0.31$\pm$0.02 & 2 & 0.27$\pm$0.03 & 0.2 & -37.3\\
9420 & 150.0947418 & 2.3236084 & 23.6 & 24.0 & 0.7 & 17.1 & 0.60$\pm$0.03 & 0 & 0.24$\pm$0.03 & 0.5 & 36.3\\
\hline
\end{tabular}
\begin{tablenotes}
\item {Columns explained from left to right.}
\item {ID: galaxy ID; R.A.: right ascension; Decl: declination; K$_s$: total FourStar/K$_s-$band magnitude; F160W: total HST/WFC3/F160W magnitude; M: stellar mass; SFR: star-formation rate; $R_e$: effective radius from \citet{vanderWel14a}; GALFIT flag: quality flag provided by \citet{vanderWel14a}; $b/a$: axis ratio: $n_{sersic}$: sersic index; P.A.; position angle of the major axis.}
\item[a] 0: good fit; 1: suspicious fit; 2: bad fit \citep{vanderWel12}.
\end{tablenotes}
\end{threeparttable}
\end{scriptsize}
\label{t:tg}
\end{center}
\end{table*}

\begin{figure*}
\begin{center}
\includegraphics[width=\textwidth]{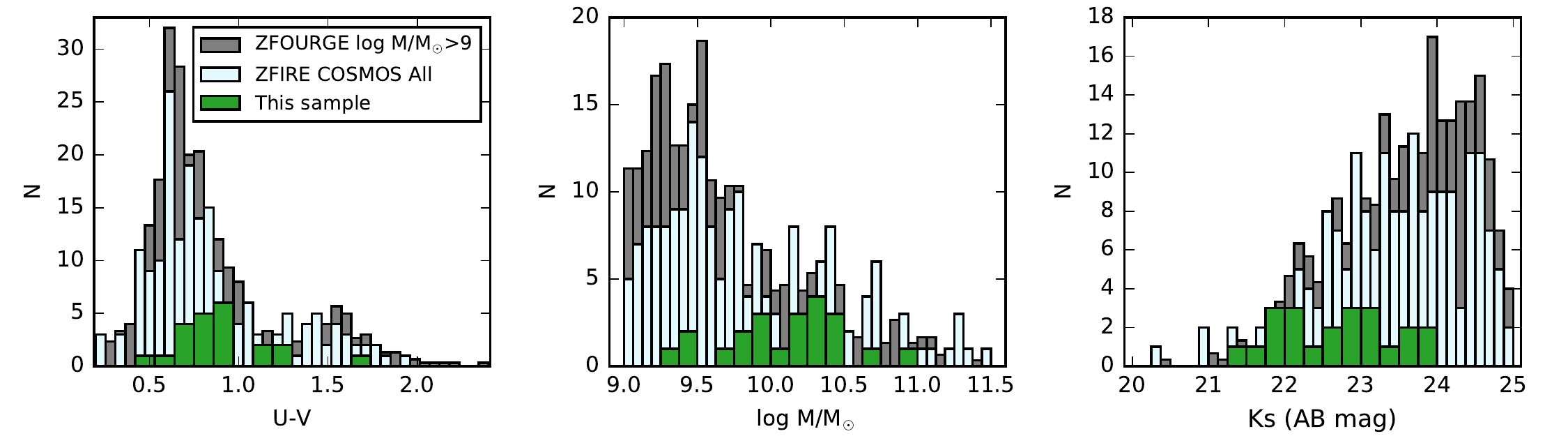}
\caption{Rest-frame $U-V$ colours, stellar masses and ZFOURGE $K_s$-band magnitudes for the {22} galaxies used here to derive the \tf\ relation (green), the ZFIRE target sample (lightblue), and a parent sample drawn from ZFOURGE with $2<z<2.5$ and $M/M_{\odot}>10^9$ (gray). The gray histograms were reduced by a factor of three for reasons of visibility. The {22} galaxies of this study have a large range in $U-V$, stellar mass and brightness.}
\label{fig:samplehist}
\end{center}
\end{figure*}

\begin{figure*}
\begin{center}
\includegraphics[width=\textwidth]{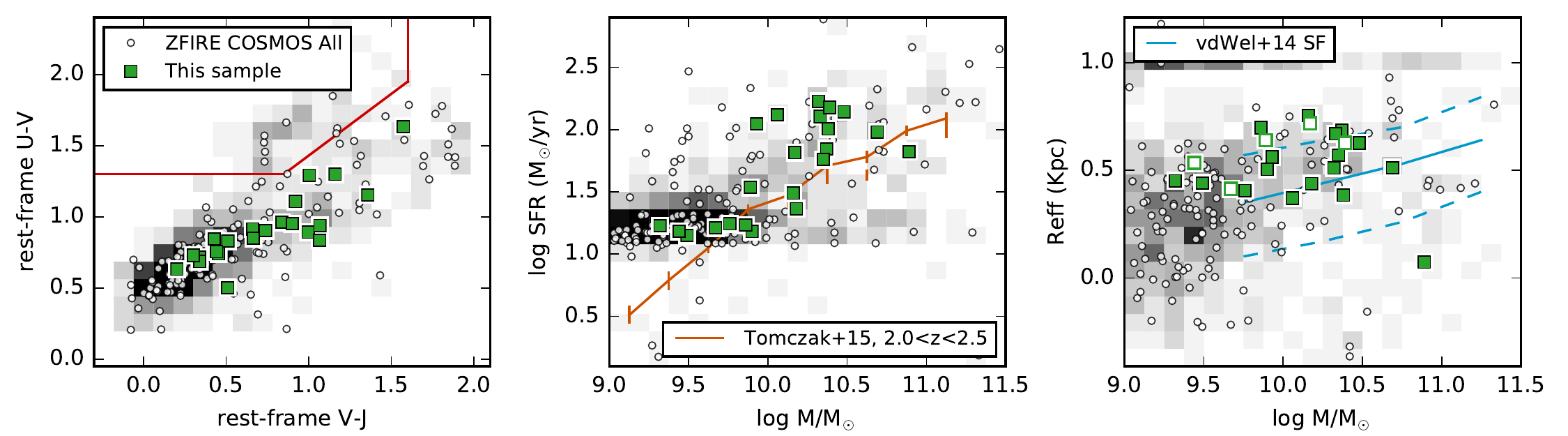}
\caption{Left: UVJ diagram of the ZFIRE sample (open symbols) and the {22} galaxies studied in this work (squares). The underlying histogram is the full distribution of $2.0<z<2.5$ galaxies with $M>10^9$\ \msun from ZFOURGE. This diagram separates quiescent galaxies from star-forming galaxies based on their rest-frame $U-V$ and $V-J$ colors, obtained from ZFOURGE photometry. The {22} galaxies in the sample span the full range in color typical of star-forming galaxies (region below the red line). Middle: stellar mass versus the logarithm of SFR. The orange line shows the median SFR as function of stellar mass of star-forming galaxies at $2.0<z<2.5$ \citep{Tomczak15}. Most of the galaxies in the sample are above the SFR-stellar mass relation at that redshift \citep{Tomczak15}. Right: stellar mass versus effective radius, with the size-mass relation at $2.0<z<2.5$ for star-forming galaxies shown as a blue line. The dashed lines are the corresponding 16th and 84th percentiles. Open square datapoints are flagged as suspicious or bad fits in the catalogs of \citet{vanderWel14a}. }
\label{fig:sample}
\end{center}
\end{figure*}

We show F160W images of the remaining {22} galaxies in the \tf\ sample in Figure \ref{fig:stamps}, and illustrate the orientation of their major axis and the MOSFIRE slits. Physical properties of the sample are shown in Table \ref{t:tg}, Figure \ref{fig:samplehist} and  Figure \ref{fig:sample}. We also compare with the primary 187 ZFIRE targets as well as with the general population of galaxies at this redshift obtained from ZFOURGE. For the ZFOURGE sample we selected galaxies with stellar mass $M/M_{\odot}>10^9$. The 19 galaxies in our sample cover the full range of the star-forming region of the UVJ diagram (below the red line), but they have higher star-formation rates (SFRs) compared to the SFR-stellar mass relation for star-forming galaxies at $2.0<z<2.5$ \citep{Tomczak15}. They lie at the bright, high-mass end of the general galaxy population. They have a large spread in size (Figure \ref{fig:sample}), including even a massive compact galaxy with effective size $R_e=0.14\arcsec$, but on average they are larger than predicted by the size-mass relation at $2.0<z<2.5$ \citep{vanderWel14a}. 

Of the {22} galaxies, 6 are spectroscopically confirmed to be part of the $z=2.095$ galaxy cluster. However, due to the small number of cluster galaxies {in this sample}, a study of the effects of environment on the evolution of the \tf\ relation is not feasible here. \citet{Alcorn16} measured the velocity dispersions of a larger sample of ZFIRE cluster galaxies and found no evidence for environmental effects at this redshift.

\subsection{The \tf\ relation}
\label{sec:result}

\begin{table}
\begin{center}
\caption{\tf\ variables}
\begin{footnotesize}
\begin{threeparttable}
\begin{tabular}{l l l l}
\hline
\hline
ID & $V_{2.2;TF}$ & $S_{05;TF}$ & $\sigma_{TF}$\\
& (km/s) & (km/s) & (km/s)\\
\hline
1814& 115.5$\pm$29.0 & 94.4$\pm$18.5 & 47.2$\pm$10.6\\
2715& 72.2$\pm$15.3 & 76.5$\pm$8.3 & 57.0$\pm$5.6\\
2723& 283.4$\pm$46.2 & 212.1$\pm$31.0 & 69.4$\pm$6.1\\
2765& 242.0$\pm$68.0 & 190.4$\pm$43.5 & 83.6$\pm$11.0\\
3074& 144.9$\pm$19.1 & 129.6$\pm$12.9 & 79.4$\pm$11.7\\
3527& 107.2$\pm$10.7 & 102.1$\pm$6.3 & 68.4$\pm$4.1\\
3598& 184.8$\pm$36.1 & 151.1$\pm$23.0 & 75.8$\pm$12.8\\
3633& 202.9$\pm$29.1 & 152.7$\pm$19.4 & 52.2$\pm$4.9\\
3655& 288.2$\pm$39.6 & 205.6$\pm$27.8 & 27.3$\pm$10.5\\
3680& 154.9$\pm$38.6 & 109.5$\pm$27.3 & 0.0$\pm$17.4\\
3714& 130.9$\pm$15.6 & 116.2$\pm$9.2 & 70.2$\pm$4.4\\
3844& 251.5$\pm$25.2 & 185.5$\pm$17.1 & 52.8$\pm$4.5\\
4010& 72.8$\pm$18.1 & 116.6$\pm$10.5 & 104.6$\pm$9.8\\
4037& 276.7$\pm$42.6 & 200.0$\pm$29.7 & 41.5$\pm$15.6\\
4099& 86.8$\pm$20.3 & 66.6$\pm$13.4 & 25.8$\pm$5.2\\
4645& 164.8$\pm$21.4 & 116.5$\pm$15.1 & 0.0$\pm$7.8\\
4930& 70.9$\pm$21.9 & 70.8$\pm$11.5 & 50.0$\pm$5.1\\
5630& 163.3$\pm$34.9 & 132.9$\pm$22.1 & 65.9$\pm$10.4\\
5870& 74.9$\pm$11.6 & 57.0$\pm$8.5 & 21.1$\pm$10.1\\
6908& 311.5$\pm$71.2 & 230.1$\pm$48.4 & 66.5$\pm$13.7\\
8108& 190.9$\pm$27.8 & 142.5$\pm$19.4 & 45.6$\pm$16.8\\
9420& 257.2$\pm$51.6 & 185.5$\pm$36.1 & 36.7$\pm$23.8\\
\hline
\end{tabular}
\begin{tablenotes}
\item {$V_{2.2;TF}$, $S_{05;TF}$ and $\sigma_{TF}$ were used to generate Figure \ref{fig:tf}. For sources with observations in multiple masks, they are average values.}
\end{tablenotes}
\end{threeparttable}
\end{footnotesize}
\label{t:tf}
\end{center}
\end{table}

\begin{figure*}
\begin{center}
\includegraphics[width=0.49\textwidth]{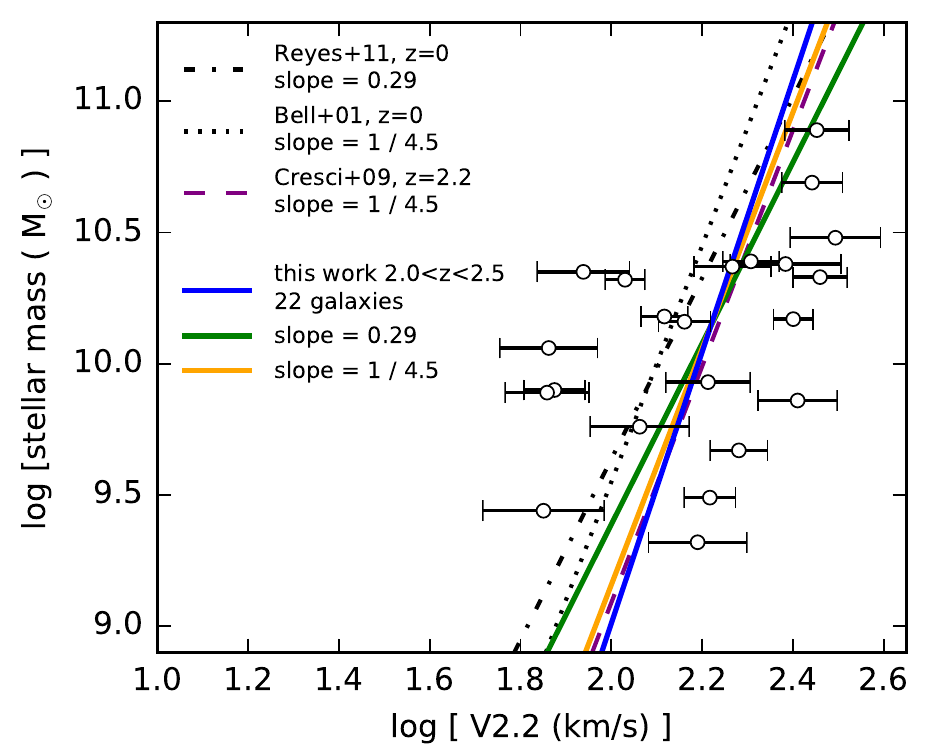}
\includegraphics[width=0.49\textwidth]{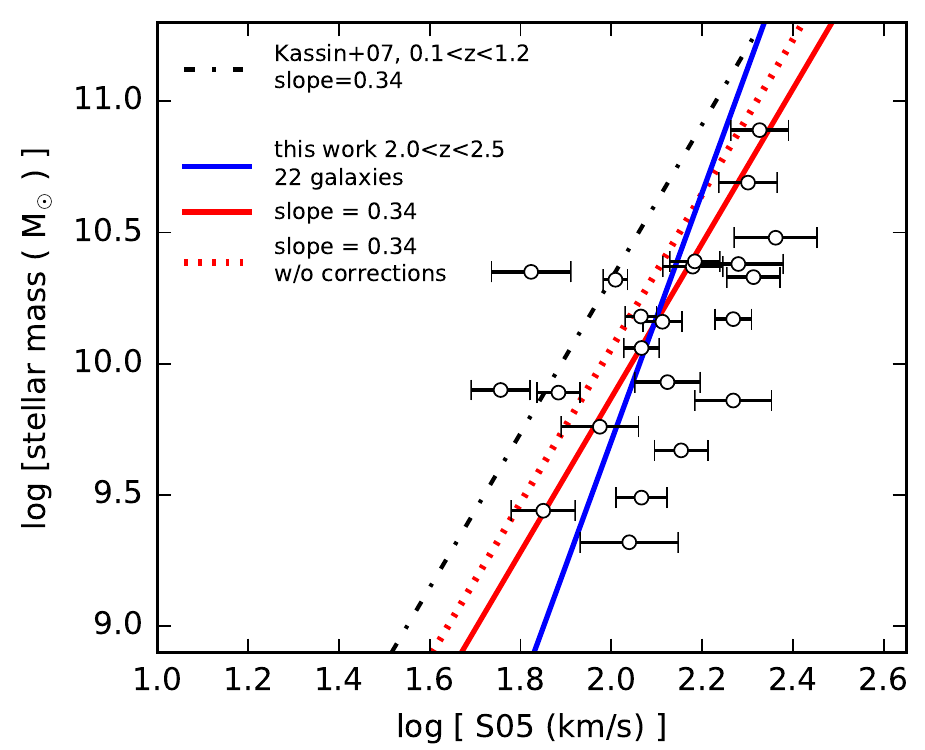}
\caption{Left: Stellar mass versus velocity and the best-fit \tf\ relation for galaxies at $2.0<z<2.5$: $\mathrm{log}V_{2.2}=$\bc$+$\afree$(\mathrm{log}M/M_{\odot}-10)$ (blue line). {For comparison, the dashed purple line is the $z\sim2.2$ result from \citep{Cresci09}.} The {dash-}dotted {and dotted} line{s are} the $z=0$ result{s} from \citet{Reyes11} {and \citet{Bell01}, respectively}. The green {and orange} line{s} {are} the best-fit result{s} with the slope fixed to th{e respective slopes} at $z=0$. The scatter in velocity {around the relation} is {0.19} dex. Right: Stellar mass versus $S_{0.5}=\sqrt{0.5V_{2.2}^2+\sigma^2}$. The scatter in $S_{0.5}$ is smaller than in velocity: 0.16 dex. We derive a best-fit relation $\mathrm{log}S_{0.5}=(2.07\pm0.03)+(0.224\pm0.060)(\mathrm{log}M/M_{\odot}-10)$ ({red} line), with a steeper slope than at $0.1<z<1.2$ \citep[{dash-}dotted line;][]{Kassin07}. The solid {red} line is the best-fit with the slope fixed to match the relation of \citet{Kassin07}. The {red} dotted line shows the same result if we remove the smoothing {and slit misalignment} correction{s} and assume a Gaussian PSF (velocities 4\% smaller).}
\label{fig:tf}
\end{center}
\end{figure*}

The Tully-Fisher relation is the relation between rotational velocity and stellar mass. We show our rotation measurements (also shown in Table \ref{t:tf}) versus stellar mass in the left panel of Figure \ref{fig:tf}, using the stellar masses taken from the ZFOURGE catalogs {and averaging values if galaxies were observed in two masks}. {We} performed a linear regression to the data following:

\begin{equation}
\mathrm{log}V_{2.2}=B+A(\mathrm{log}M/M_{\odot}-10)
\label{eq:tf}
\end{equation}

The Tully-Fisher relation is by convention shown in diagrams with stellar mass on the y-axis. However, the dominant uncertainty here is that in velocity and therefore we performed regression with $V_{2.2}$ as the dependent variable. This is also a method very commonly used in literature which acts against Malmquist bias \citep{Bamford06,Weiner06b,Kelly07}.

We obtain from the fit $B=$\bc\ and $A=$\afree. We derived the uncertainties by bootstrapping the sample 1000 times, and taking the standard deviation from the bootstrapped distributions of $B$ and $A$. The slope of the \tf\ relation, \afree, is consistent with previous results at $z=0$. For example, \citet{Reyes11} find $A=0.29$ and \citet{Bell01} find $A=1/4.5=0.22$. Our study has too few numbers to significantly constrain evolution in the slope between $z=0$ and $2<z<2.5$, but if we fix the slope to that at lower redshift we can study the evolution of the zeropoint. Setting $A=0.29$, we find $B=$\bfixc. Compared to $z=0$ \citep{Reyes11}, this implies an evolution of the zeropoint (in stellar mass) of \ev. We included here a small correction of -0.05 dex in stellar mass to account for the \citet{Kroupa01} IMF used by \citet{Reyes11} instead of the \citet{Chabrier03} IMF used here. {Similarly, we can compare to the $z=0$ result of \citet{Bell01}, by setting the slope to 1/4.5. This results in an observed evolution of \evbell.} The{se} offset{s} in steller mass {are} consistent with the findings of \citet{Cresci09} {and \citet{Simons16}} {who} derived $\mathrm{\Delta M/M_{\Sun}}=-0.41\pm0.11$ dex{ and $\mathrm{\Delta M/M_{\Sun}}=-0.44\pm0.16$, respectively}. 

As an additional consistency check, {we investigated the effects of sample selection. First we refined the sample even more} and we fitted the \tf\ relation only to the galaxies with highest SNR, fixing the slope to $A=0.29$ {or 1/4.5}. {W}e obtained consistent result{s} with \evbright\ and {\evbrightbell, respectively,} for the {11} galaxies with spectra with SNR$>20$. Then we tested applying a less severe sample selection, including spectra with velocity errors $<50\%$ instead of $<30\%$. This resulted in \evbigsample\ {and \evbigsamplebell, respectively,} for a sample of 2{7} galaxies. This is a rather large difference and we take note that it may be a potential caveat if one selects the brightest (and therefore easiest to fit) galaxies.

\begin{figure*}
\includegraphics[width=\textwidth]{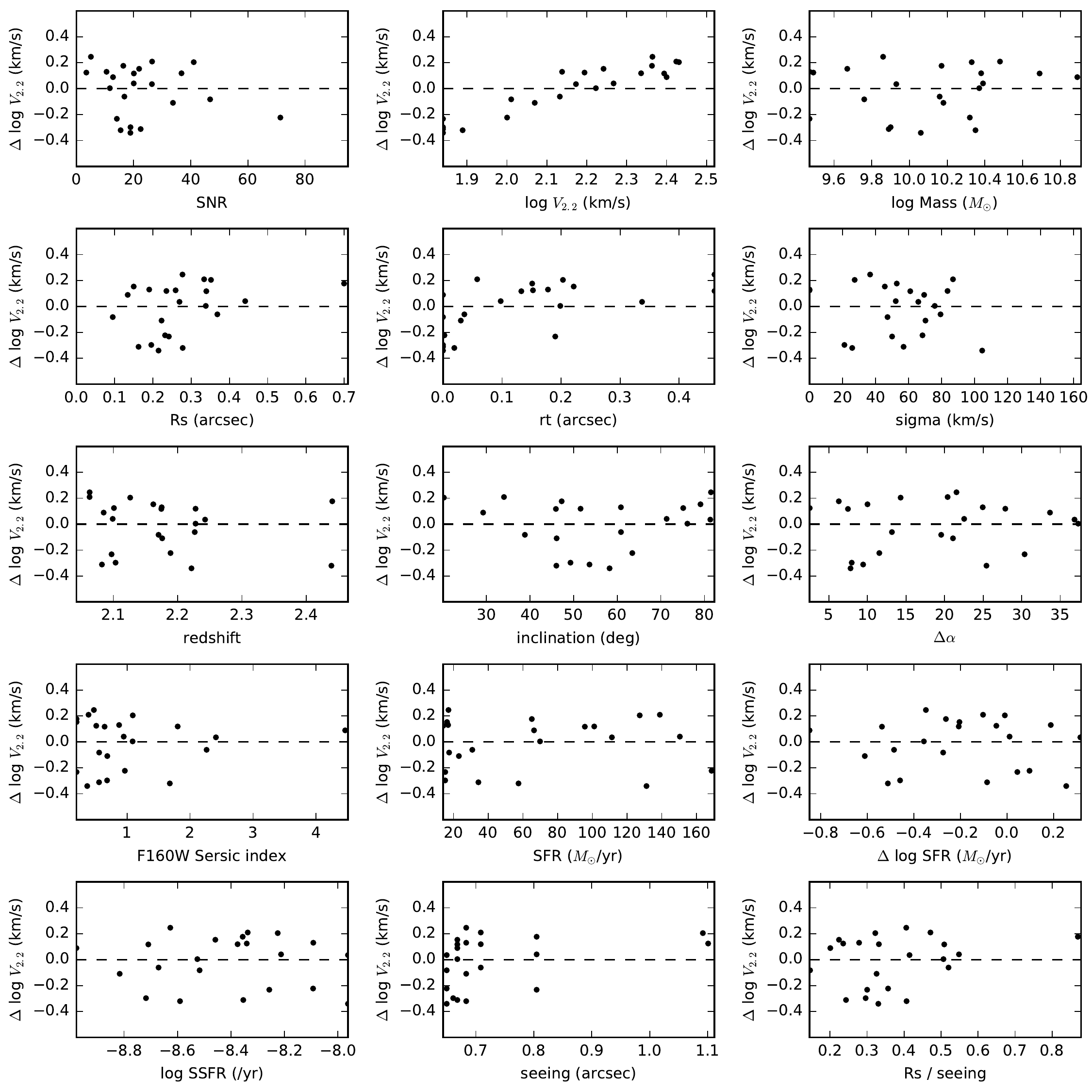}
\caption{Difference between the observed velocity and the velocity predicted by the best-fit \tf\ relation. We plot against model and observational parameters. From left to right, top to bottom: SNR, log $V_{2.2}$, log stellar mass, best-fit $R_s$, best-fit $r_t$,  best-fit $\sigma$, redshift, inclination, $\Delta\alpha$, Sersic index, SFR, SFR minus predicted SFR at $2.0<z<2.5$ \citep{Tomczak15}, SSFR, seeing and $R_s$ relative to the seeing. We find almost no correlations, except for best-fit $r_t$, with a negative offset in $\Delta$log$V_{2.2}$ for fits with $r_t\approx0$.}
\label{fig:devs}
\end{figure*}

To investigate if there are remaining systematic trends we show in Figure \ref{fig:devs} the velocity residuals of the best-fit Tully-Fisher relation with respect to various parameters and properties of the galaxies (such as SFR). We define the residual as $\Delta \mathrm{log} V_{2.2} = \mathrm{log} V_{2.2} - \mathrm{log} V_{TFR}$ with $V_{TFR}$ the rotational velocity predicted from the fit for a specific stellar mass. There are no systematic effects related to Sersic Index, $R_s$, stellar mass, SFR, specific SFR (SSFR), or offset from the SFR-stellar mass relation at $2.0<z<2.5$ \citep{Tomczak15}. In addition there is no clear relation with inclination, PA, or seeing. A few prominent outliers have a negative $\Delta \mathrm{log} V_{2.2}$, i.e., they are located to the left of the \tf\ relation in Figure \ref{fig:tf}. These have average values of very small $r_t$, the kinematic scale radius. As we have shown in Section \ref{sec:vel}, resolution effects may play a role in determining $r_t$, and we derive somewhat higher velocities {if we limit $r_t$ to $r_t > 0.02\arcsec$ for these spectra, resulting in $\Delta \mathrm{log}M/M_{\odot}=-0.31\pm0.14$ dex {and $\Delta \mathrm{log}M/M_{\odot}=-0.47\pm0.17$ dex for slopes of 0.29 and 1/4.5, respectively, for the whole sample. We note that a value of zero for $r_t$ is possible in the presence of non-circular motion, for example if the galaxy has a bar \citep{Franx92}. We inspected the F160W images, but found no indications of a bar-like morphology.}

The scatter of the residual velocities with respect to the \tf\ relation is significant, with {$\sigma=0.19$} dex. This is more than at $z=0$, and {could partly be due to the low $r_t$ outliers. It} may {also} be related to star-forming galaxies at high redshift showing more variety in kinematics, and the increase of non-rotationally supported galaxies \citep[e.g.][]{Kassin07}. An alternative to the stellar mass-velocity relation is the stellar mass-$S_{0.5}$ relation, with $S_{0.5}=\sqrt{0.5V_{2.2}^2+\sigma^2}$. This relation was first coined by \citet{Weiner06a}, and \citet{Kassin07} showed that the scatter decreases significantly if $S_{0.5}$ is used. They also found that it does not evolve significantly between $z=0.1$ and $z=1.2$.

We calculated $S_{0.5}$ for the galaxies in our sample (right panel in Figure \ref{fig:tf}) and find the scatter is indeed smaller: {$\sigma_{rms}=0.15$} dex{, similar to what \citet{Kassin07} derived at $0.1<z<1.2$ (0.16 dex) and to a recent study by \citet{Price15}, using MOSFIRE at $1.4<z<2.6$ (0.17 dex)}. We derived the best-fit relation to the data with $A$ and $B$ free in the fit and found $\mathrm{log}S_{0.5}=$\bsfree$+$\asfree$(\mathrm{log}M/M_{\odot}-10)$. Here the slope is steeper than at $0.1<z<1.2$, where \citet{Kassin07} found that $A=0.34$. This is in agreement with the previous study by \citet{Cresci09}, who did not derive a best-fit to their data, but they do find higher $S_{0.5}$ values towards smaller stellar mass compared to the $0.2<z<1.2$ relation. Keeping the slope fixed at $A=0.34$, we found $B=$\bsfix. This implies a zeropoint evolution of \sev\ compared to $0.1<z<1.2$. \citet{Price15} also find an offset {in $\Delta \mathrm{M}/$\msun} for galaxies at high redshift, implying $\Delta \mathrm{M}/$\msun$\sim-0.3$ dex, and their data does not indicate a steeper slope. Their offset from $0.1<z<1.2$ is inconsistent with and less than what we find, which could be due to the inclusion of galaxies at $z<2$, their assumption of a Gaussian PSF, and our correction for two-dimensional PSF effects. If we account for the Gauss-Moffat difference and remove the correction for smoothing {and slit misalignment}, the inferred evolution {compared with} $z<1.2$ is less: \sevnoc. Both the findings from \citet{Price15} and ours point towards evolution of the zeropoint of the stellar mass-$S_{0.5}$ relation between $z<1.2$ and $z\gtrsim2$, but no evolution for the scatter in $S_{0.5}$.

{One caveat could be the possible misclassification of mergers as rotating disks in our sample. As \citet{Hung15} showed using artificially redshifted IFU data, a large fraction of high redshift ($z>1.5$) interacting galaxies would still be kinematically classified as single rotating disks. Inspection of Figure \ref{fig:stamps} indicates that some of the galaxies here have multiple components with small angular separations, e.g., \#1814, \#4930, \#6908. Such components may contribute differently to the kinematics of the system, but to investigate this  in detail is beyond the scope of this paper.}

\section{Discussion}
\label{sec:discussion}

\subsection{Comparison to literature}
\label{sec:discuss1}

\begin{figure*}
\begin{center}
\includegraphics[width=0.49\textwidth]{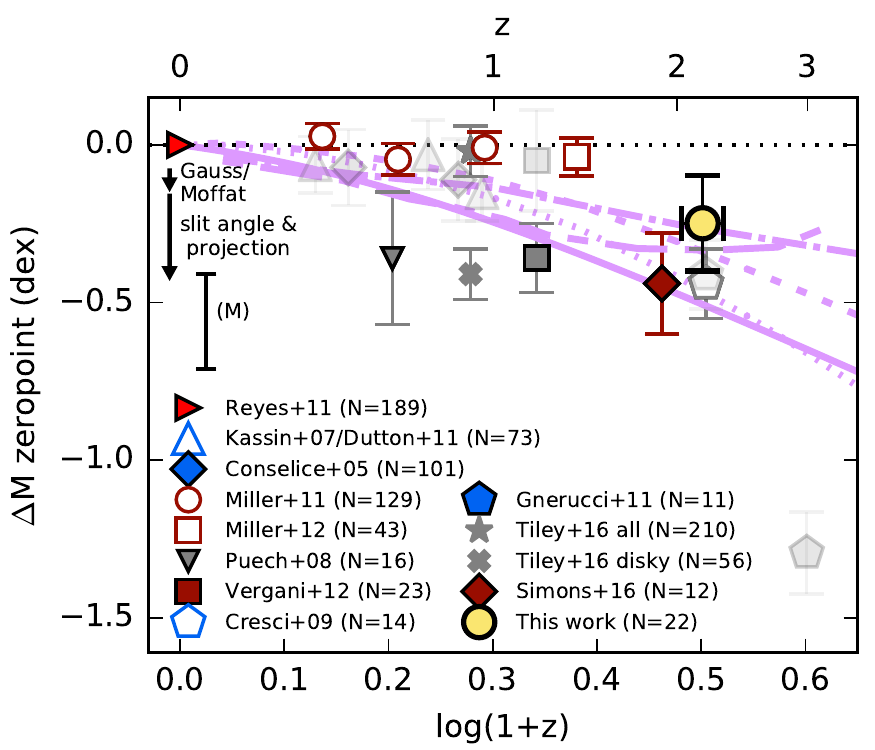}
\includegraphics[width=0.49\textwidth]{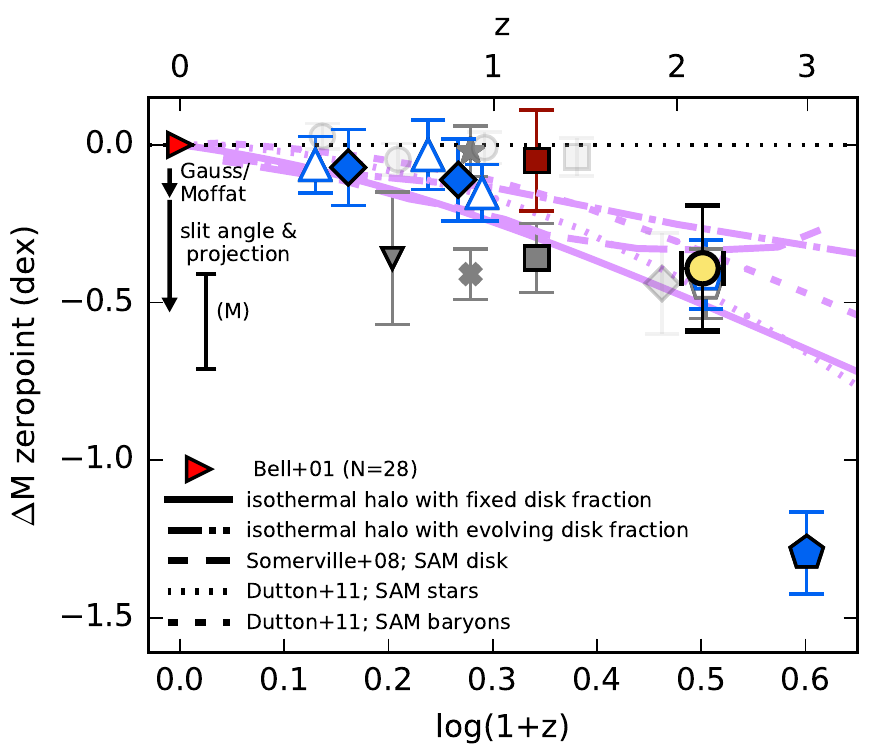}
\caption{{T}he evolution of the stellar mass zeropoint with redshift. The {yellow} datapoint{s represent the observed evolution} from our survey at $2.0<z<2.5$ {for fixed slopes of 0.29 (left panel) or 1/4.5 (right panel)}. {Uncertainties are derived from bootstrap resampling and} {t}he horizontal error bar indicates the standard deviation of redshift in our sample. Results from other surveys (as quoted in the corresponding papers) are shown with symbols as indicated in the legend. {For studies that do not compare directly with \citet{Bell01} or \citet{Reyes11} we use gray symbols.} The magnitude of the systematic effects that we have corrected for are indicated by arrows, and we have also indicated {the magnitude of a factor 2 uncertainty in stellar mass with a vertical errorbar (M)}. We also show the predictions from semi-analytical models ({pink} lines).}
\label{fig:lit}
\end{center}
\end{figure*}

To put our result into context, we show the evolution of the stellar mass zeropoint in Figure \ref{fig:lit}, and include previous results from literature. These were all derived from the stellar mass-velocity relation, with quite strong discrepancies between different studies\footnote{We show the stellar mass zeropoint offsets as quoted by the authors, which were carefully derived and corrected for the different IMFs used in $z=0$ studies. We verified the corrections applied to each datapoint, but could not confirm the IMF-correction by \citet{Conselice05}. The correction from \citet{Vergani12} was unclear.} {at $z>0.5$}. However, before comparing with other studies at different redshifts, several major caveats have to be taken into account: studies use different galaxy selections, different methodologies to derive velocity and stellar mass, and different types of spectroscopic observations. We will discuss these first and then review and compare the studies.

The first is selection bias. At $z>2$ star-forming galaxies have different properties on average than at $z=0$. For example, they have higher SFRs, higher gas masses and smaller sizes \citep[e.g.][]{Papovich14,vanderWel14a}. At $z>2$ dust-obscured galaxies are more common, and for these galaxies the \ha\ luminosity is attenuated \citep[e.g.][]{Reddy05,Spitler14}. Samples that are UV or \ha\ selected may therefore not be a complete distribution of star-forming galaxies at high redshift and changes in incompleteness may mimick evolution with redshift. Mergers and galaxies with irregular morphologies are also more common than at $z=0$ \citep[e.g.][]{Abraham01,Mortlock13}. These galaxies have less ordered velocity fields \citep[e.g.][]{Kassin07} higher velocity dispersions relative to circular velocities, and are often excluded from \tf\ samples because it is difficult to describe these galaxies with smooth rotating models \citep{Cresci09,Gnerucci11}. At high redshift the angular extent of galaxies is often small compared to the seeing, which may give the appearance that the galaxy is dispersion dominated if the velocity gradient is unresolved \citep[e.g.][]{Miller12}. If the selection requires ordered rotation, this leads to biases towards larger galaxies.

Here we have attempted to introduce as little selection bias as possible, but it could not be entirely avoided. As {described} in Section \ref{sec:res}, we have excluded galaxies with poor fits, which tended to be galaxies with smaller sizes and fainter magnitudes than the overall photometric sample. {Despite the large uncertainties on the velocities of these poor fits, in Section \ref{sec:result} we have shown that such a selection may indeed bias the result towards larger average velocities and hence a stronger evolution of the stellar mass zeropoint.} 

Another caveat when comparing different results from literature is methodology. In many studies the PSF is assumed to be Gaussian, but for our MOSFIRE data a Moffat profile is a better approximation. The difference between using a Gaussian and a Moffat in our modelling leads to a 0.06{$-0.08$} dex shift in the stellar mass zeropoint of the \tf\ relation{, depending on the slope}. 
In addition, several different possibilities exist to model the velocity field, e.g. the one-dimensional arctan model we used here \citep[and also used by e.g.][]{Miller11,Miller12} or a two-dimensional integrated mass model \citep{Cresci09,Gnerucci11}. Different choices for the radius at which to evaluate velocity exist as well. In some cases $R_{80}$ is used, encapsulating 80\% of the optical light \citep{Reyes11}. In other cases $V_{max}$ is used, or the asymptotic velocity $V_{a}$ in the arctan model, which is often extrapolated at a radius beyond the optically observed extent of the galaxy \citep[e.g.][]{Weiner06b}. Most studies in Figure {\ref{fig:lit}} employ $V_{max}$. We prefer $V_{2.2}$, because it is more robust, and it is used in several other studies \citep[e.g.][]{Miller11,Miller12}. The relation \citet{Reyes11} derived for $V_{80}$, which is close to $V_{max}$, implies a 4\% increase in velocities relative to $V_{2.2}$, or a $\sim0.06${$-0.08$} dex effect on the inferred stellar mass zeropoint. Lastly, uncertainty on the stellar mass has to be taken into account. We derived our stellar mass from fitting to SEDs obtained from photometry, which depends on several assumptions of the stellar population models. Differences between a \citet{Salpeter55}, \citet{Kroupa01}, Diet Salpeter \citep{Bell03} and \citet{Chabrier03} IMF are $0.05-0.3$ dex. In addition, different stellar populations models can produce stellar masses different by a factor of 2 \citep[e.g. the review of][]{Conroy13}. Also, fitting models to SEDs versus applying M/L ratios based on ($g-r$) colors \citep{Bell03,Puech08} can amount to up to a factor of 2 differences \citep{Reyes11}. 

Another important issue is simply that the datasets between surveys are of a different kind, such as single-slit data versus integral field spectroscopy. For example, \citet{Cresci09}, who use IFS, employ a three-dimensional method, by modelling a datacube with x,y and $\lambda$ dimensions. This kind of modelling already includes effects from the two-dimensional PSF and projection, whereas (y,$\lambda$) modelling of single-slit data using a one-dimensional PSF (as performed in this study and by \citet{Conselice05,Kassin07,Miller11,Miller12} at high redshift) results in systematically underestimating the velocity. 

In summary, methodology and data sets can introduce significant {velocity} offsets. Reviewing the studies at different redshifts with this in mind we can try to understand these discrepancies. For example, there exist clear differences between \citet{Puech08} and other studies \citep[e.g.][]{Conselice05,Kassin07,Miller11} at $z\sim0.5$ of $0.3-0.4$ dex. {There is also a strong apparent evolution between $z=1.7$ \citep[][based on one-dimensional modeling of single-slit data]{Miller12} and our results at $z=2.2$. This can at least in part be explained if we take into account our slit misalignment and projection corrections (see Section \ref{sec:smooth}). For example, \citet{Weiner06a} already speculated velocities may become underestimated for larger $|\Delta \alpha|$ and based on this \citet{Kassin07} select sources with $|\Delta \alpha|<40^{\circ}$, but do not correct for slit misalignment. In Section \ref{sec:smooth} however we showed that the effects of slit misalignment are still strong even for $|\Delta \alpha|<40^{\circ}$. \citet{Miller11} and \citet{Miller12} apply a correction for slit misalignment \citep[][Equation 4]{Miller11}, but if we apply the same to our simulations, we still find a residual $V_{2.2;in}/v'_{2.2;out}=1.1$, independent of $|\Delta \alpha|$. We show the median correction applied to our data in both panels of Figure \ref{fig:lit}. It is likely that a similar correction is applicable to all single-slit studies based on one-dimensional modeling, but we also note that such corrections are dependent on slitwidth as well as the angular size of a galaxy. The effect may thus be smaller at $z<2$. Encouragingly, \citet{Harrison17} derive a similar median correction of 7\% on circular velocites measured in virtual single slits projected on IFU data of $z\sim0.9$ galaxies.} {The differences at $z=0.6$ with the result from} \citet{Puech08} {may further be explained by a different conversion from light to stellar mass (M/L ratios versus SED fitting).}

{The effects of sample selection on the observed evolution have been shown by \citet{Tiley16}, who measured the stellar mass \tf\ relation both for a parent sample of more general properties, e.g., \ha\ detected, non-zero rotation with $<30\%$ uncertainties, as well as for a subsample of rotation dominated galaxies with $v/\sigma>3$. They found a $\Delta \mathrm{M}/$\msun$=0.39$ dex difference, with stronger evolution for the rotation dominated sample. The median $V_{2.2}/\sigma$ of our sample is 3.{5} and we must keep in mind that our sample may trace the evolution of the more rotationally supported disks.}  The differences between $0 < z < 2$ studies and the $z=3$ result of \citet{Gnerucci11} could be related to their rest-frame UV selection and requirement of optical (rest-UV) spectroscopic redshifts, which tends to be much bluer than near-IR selected samples at $1 < z < 2$.

Different studies also compare to different $z=0$ relations. The most common references are the results from \citet{Bell01}, \citet{Pizagno05}, and \citet{Reyes11}. These are based on different IMF, choice of velocity indicator, and method to derive stellar mass. In most high redshift studies, estimates of the evolution of the \tf\ relation are derived very carefully, but a major factor of uncertainty is the derived slope of the relation at $z=0$. For example, \citet{Bell01} derive a much steeper slope, with $A=1/4.5$, than \citet{Pizagno05} and \citet{Reyes11}. This is illustrated by \citet{Vergani12}, who find a $-0.36\pm0.11$ dex evolution compared to \citet{Pizagno07} and only $-0.05\pm 0.16$ dex compared to \citet{Bell01}. Other studies that compare to \citet{Bell01} are those by \citet{Conselice05}, \citet[][based on the results of \citet{Kassin07}]{Dutton11}, \citet{Cresci09} and \citet{Gnerucci11}. \citet{Miller11}, \citet{Miller12} and \citet{Simons16} use the $z=0$ relation from \citet{Reyes11}, while \citet{Puech08} {and \citet{Tiley16}} derive the stellar mass \tf\ relation both at $z=0$ and {the redshift of their study} and compare internally.

{Despite potential selection effects and differences in methodogy, our results are consistent with the two other studies at $z\sim2$ \citep{Cresci09,Simons16}, and suggest that the stellar mass \tf\ relation has evolved since $z\sim2.2$.}

\subsection{Interpretation of the evolution of the \tf\ relation}

Taking into account the various systematic differences between studies at high redshift, {w}e now discuss the observations in a framework of {dark matter halo physics and }semi-analytical models. {First we calculate the expected disk mass evolution assuming an isothermal dark matter halo and a constant fraction of the total halo mass corresponding to the disk \citep[see also Equation 4 of][]{Mo98}. Here the evolution scales with the inverse of the Hubble constant: $\sim 1/H(z)$ (solid pink line). In reality it is likely the disk mass fraction evolves over time \citep[e.g.,][]{Papovich14}. We therefore matched the predicted mass growth of a halo of $10^{13}$\msun\ at $z=0$ with the results from \citet{Behroozi10}, who analysed the stellar mass-halo mass relation for galaxies between $z=0.1$ and $z=4$. Here we assumed a zero gas fraction for the disk. This resulted in softer evolution (dash-dotted pink line).}

{An even more detailed approximation for the evolution of a disk is explored with the semi-analytic models of \citet{Somerville08} (long dashes) and \citet{Dutton11} (dotted line for the stellar mass and dashed line for the baryonic disk mass evolution). The key differences between these models is that \citet{Somerville08} allow for halo contraction and haloes with a mass distribution following the prescription of \citet{Navarro97} and assume purely stellar disks, whereas \citet{Dutton11} assume isothermal haloes without contraction and include gas in their models.} The semi-analytic models also predict a softer evolution of the stellar mass zeropoint than for a simple isothermal halo. It is also worth noting that \citet{Dutton11} find a more gradual evolution for the full baryonic disk mass than for the stellar mass only at fixed velocity. The stellar mass growth of a galaxy will be affected by internal feedback processes, which are not addressed here. However, these could further dampen the predicted evolution \citep{Sales10}.

Our result at $z=2.2$ is consistent with these predictions{, if we assume either the slope of \citet{Bell01} or \citet{Reyes11}}. There is good agreement between the models and {most of} the observations{, given the uncertainties due to mass derivation and sample selection}, with a gradual zeropoint evolution. {We note that our median corrections as indicated by the downwards pointing arrows in Figure \ref{fig:lit}, if applied to single-slit observations at $z<1.7$, could potentially move these datapoints towards a more negative $\Delta \mathrm{M}/$\msun. This would bring earlier studies into better agreement with each other, but would also lead to a stronger observed evolution than predicted at this redshift. At the same time, if such corrections become less severe due to the increasing median size of galaxies towards lower redshift \citep[e.g.,][]{vanderWel14a} it may be possible to reconcile the apparent strong observed evolution between $z=1.7$ \citep{Miller12} and $z\sim2.2$, while lower redshift results still remain consistent with the models.}
The datapoint from \citet{Gnerucci11} is still an outlier and if representative it may point to non-self similar evolution at high redshift ($z>3$).

{As a final note, the increase in stellar mass at fixed velocity over time could simply reflect the conversion of gas into stellar mass.} The median $V/\sigma\sim3.5$ for galaxies in our sample is larger than the median of, e.g., \citet{Price15} at $1.4<z<2.6$ of $V/\sigma=2.1$, and above the treshold of what \citet{Kassin12} consider a kinematically settled disk. This could {be due to} our selection of bright galaxies with clear rotation. Nevertheless it confirms the emerging picture that at high redshift disk galaxies are more often pressure supported {and} the evolution {that we observe} in {the} stellar mass {zeropoint} could {partly} reflect the conversion from gas to stars over time{, a scenario that was also suggested by \citet{Simons16}}.

 \section{Summary}
 
In this work we have derived the stellar mass-velocity and $S_{0.5}$-velocity scaling relations at $2.0<z<2.5$, making use of 24 MOSFIRE single-slit spectra of 22 star-forming galaxies, as part of the ZFIRE survey. The diagnostic used was the \ha\ emission line, and we fitted model spectral image stamps to the data to trace the rotational velocities and dispersions of the \ha\ gas in the galaxies.

We conducted a careful check of systematics and corrected our results where necessary, and subsequently fitted and interpreted the stellar mass \tf\ evolution to $0 < z < 2.5$. We found the following main results:
\begin{itemize}
\item The MOSFIRE PSF can be best approached by a Moffat function with $\beta=2.5$, instead of a Gaussian. Assuming a Gaussian PSF instead leads to 4\% underestimated velocities on average, implying a 0.06$-0.08$ dex effect on the stellar mass zeropoint of the \tf\ relation.
\item Two-dimensional PSF and slit projection effects cause flux from lower velocity regions of a galaxy to be mixed within the slit. {From simulations of emission line models we derive a bias towards} {on average 19}\% smaller velocities, {depending on how well the slit is aligned with the position angle of the galaxy. Depending on the slope of the \tf\ relation, this translates into a 0.26 dex to 0.34 dex effect on the stellar mas zeropoint}.
\item Taking this into account, we derived the stellar mass \tf\ relation $\mathrm{log}V_{2.2}=$\bc$+$\afree$(\mathrm{log}M/M_{\odot}-10)$ and inferred an evolution of \ev\ compared to $z=0${, assuming a fixed slope of 0.29 or \evbell\ assuming a slope of 1/4.5}.
\item The best-fit modified \tf\ relation, the $S_{0.5}$-velocity relation, is $\mathrm{log}S_{0.5}=$\bsfree$+$ \asfree$(\mathrm{log}M/M_{\odot}-10)$, with an inferred zeropoint evolution of \sev\ compared to $0.1<z<1.2$.
\item We reviewed previous results in literature, which have strong discrepancies between IFS and single-slit studies over a large redshift range. We give as an explanation for these discrepancies that single-slit results may suffer from PSF and projection effects{, uncertainties in stellar mass, and selection bias}. 
\item {T}he overall evolution of the stellar mass zeropoint at $0 < z < 2.5$ is reasonably well matched by {predictions from hierarchical cluster{ing} \citep{Mo98} and} the semi-analytic models of {\citet{Somerville08} and} \citet{Dutton11}. However, in detail some discrepancies with the models remain. Furthermore, our data confirm previous observations of increased contributions from non-rotationally supported galaxies, which are not included in the models. The increase of the average velocity dispersion towards higher redshift is related to the higher gas fractions in galaxies{, which could drive part of the evolution in $\Delta \mathrm{M}/$\msun}. It is possible that the evolution in $\Delta \mathrm{M}/$\msun\ {is softened by additional processes within a galaxy, such as stellar feedback}.
\end{itemize}

\section{Acknowledgements}
The data presented herein were obtained at the W.M. Keck Observatory, which is operated as a scientific partnership among the California Institute of Technology, the University of California, and the National Aeronautics and Space Administration. The Observatory was made possible by the generous financial support of the W.M. Keck Foundation. We wish to thank the W. M. Keck Observatory support staff for their enthusiastic support. We recognize and acknowledge the very significant cultural role and reverence that the summit of Mauna Kea has always had within the indigenous Hawaiian community. We thank the anonymous referee for a helpful report. We thank Gabe Brammer and Danilo Marchesini for kindly sharing a MOSFIRE $K_s$-band image, and George Bekiaris for help with software. We are grateful to {Marijn Franx and} Susan Kassin for in-depth discussions. CMSS gratefully acknowledges the support of the Australian Government through an Endeavour Research Fellowship. {TN, KG and GGK acknowledge Swinburne-Caltech collaborative Keck time.} GGK acknowledges the support of the Australian Research Council through the award of a Future Fellowship (FT140100933). {KG acknowledges the support of the Australian Research Council through Discovery Proposal awards DP1094370, DP130101460, and DP130101667. KT acknowledges the support of the National Science Foundation under Grant \#1410728.}


\begin{thebibliography}{}
\expandafter\ifx\csname natexlab\endcsname\relax\def\natexlab#1{#1}\fi

\bibitem[{{Abraham} \& {van den Bergh}(2001)}]{Abraham01}
{Abraham}, R.~G., \& {van den Bergh}, S. 2001, Science, 293, 1273

\bibitem[{{Alcorn} {et~al.}(2016){Alcorn}, {Tran}, {Kacprzak}, {Nanayakkara},
  {Straatman}, {Yuan}, {Allen}, {Cowley}, {Dav{\'e}}, {Glazebrook}, {Kewley},
  {Labb{\'e}}, {Quadri}, {Spitler}, \& {Tomczak}}]{Alcorn16}
{Alcorn}, L.~Y., {Tran}, K.-V.~H., {Kacprzak}, G.~G., {et~al.} 2016, \apjl,
  825, L2

\bibitem[{{Bamford} {et~al.}(2006){Bamford}, {Arag{\'o}n-Salamanca}, \&
  {Milvang-Jensen}}]{Bamford06}
{Bamford}, S.~P., {Arag{\'o}n-Salamanca}, A., \& {Milvang-Jensen}, B. 2006,
  \mnras, 366, 308

\bibitem[{{Behroozi} {et~al.}(2010){Behroozi}, {Conroy}, \&
  {Wechsler}}]{Behroozi10}
{Behroozi}, P.~S., {Conroy}, C., \& {Wechsler}, R.~H. 2010, \apj, 717, 379

\bibitem[{{Bekiaris} {et~al.}(2016){Bekiaris}, {Glazebrook}, {Fluke}, \&
  {Abraham}}]{Bekiaris16}
{Bekiaris}, G., {Glazebrook}, K., {Fluke}, C.~J., \& {Abraham}, R. 2016,
  \mnras, 455, 754

\bibitem[{{Bell} \& {de Jong}(2001)}]{Bell01}
{Bell}, E.~F., \& {de Jong}, R.~S. 2001, \apj, 550, 212

\bibitem[{{Bell} {et~al.}(2003){Bell}, {McIntosh}, {Katz}, \&
  {Weinberg}}]{Bell03}
{Bell}, E.~F., {McIntosh}, D.~H., {Katz}, N., \& {Weinberg}, M.~D. 2003, \apjs,
  149, 289

\bibitem[{{Benson}(2012)}]{Benson12}
{Benson}, A.~J. 2012, \na, 17, 175

\bibitem[{{Brammer} {et~al.}(2008){Brammer}, {van Dokkum}, \&
  {Coppi}}]{Brammer08}
{Brammer}, G.~B., {van Dokkum}, P.~G., \& {Coppi}, P. 2008, \apj, 686, 1503

\bibitem[{{Bruzual} \& {Charlot}(2003)}]{Bruzual03}
{Bruzual}, G., \& {Charlot}, S. 2003, \mnras, 344, 1000

\bibitem[{{Calzetti} {et~al.}(2000){Calzetti}, {Armus}, {Bohlin}, {Kinney},
  {Koornneef}, \& {Storchi-Bergmann}}]{Calzetti00}
{Calzetti}, D., {Armus}, L., {Bohlin}, R.~C., {et~al.} 2000, \apj, 533, 682

\bibitem[{Chabrier(2003)}]{Chabrier03}
Chabrier, G. 2003, Publications of the Astronomical Society of the Pacific,
  115, 763

\bibitem[{{Conroy}(2013)}]{Conroy13}
{Conroy}, C. 2013, \araa, 51, 393

\bibitem[{{Conselice} {et~al.}(2005){Conselice}, {Bundy}, {Ellis}, {Brichmann},
  {Vogt}, \& {Phillips}}]{Conselice05}
{Conselice}, C.~J., {Bundy}, K., {Ellis}, R.~S., {et~al.} 2005, \apj, 628, 160

\bibitem[{{Courteau}(1997)}]{Courteau97}
{Courteau}, S. 1997, \aj, 114, 2402

\bibitem[{{Cowley} {et~al.}(2016){Cowley}, {Spitler}, {Tran}, {Rees},
  {Labb{\'e}}, {Allen}, {Brammer}, {Glazebrook}, {Hopkins}, {Juneau},
  {Kacprzak}, {Mullaney}, {Nanayakkara}, {Papovich}, {Quadri}, {Straatman},
  {Tomczak}, \& {van Dokkum}}]{Cowley16}
{Cowley}, M.~J., {Spitler}, L.~R., {Tran}, K.-V.~H., {et~al.} 2016, \mnras,
  457, 629

\bibitem[{{Cresci} {et~al.}(2009){Cresci}, {Hicks}, {Genzel}, {Schreiber},
  {Davies}, {Bouch{\'e}}, {Buschkamp}, {Genel}, {Shapiro}, {Tacconi},
  {Sommer-Larsen}, {Burkert}, {Eisenhauer}, {Gerhard}, {Lutz}, {Naab},
  {Sternberg}, {Cimatti}, {Daddi}, {Erb}, {Kurk}, {Lilly}, {Renzini},
  {Shapley}, {Steidel}, \& {Caputi}}]{Cresci09}
{Cresci}, G., {Hicks}, E.~K.~S., {Genzel}, R., {et~al.} 2009, \apj, 697, 115

\bibitem[{{Dutton} {et~al.}(2011){Dutton}, {van den Bosch}, {Faber}, {Simard},
  {Kassin}, {Koo}, {Bundy}, {Huang}, {Weiner}, {Cooper}, {Newman}, {Mozena}, \&
  {Koekemoer}}]{Dutton11}
{Dutton}, A.~A., {van den Bosch}, F.~C., {Faber}, S.~M., {et~al.} 2011, \mnras,
  410, 1660

\bibitem[{{Fall} \& {Efstathiou}(1980)}]{Fall80}
{Fall}, S.~M., \& {Efstathiou}, G. 1980, \mnras, 193, 189

\bibitem[{{F{\"o}rster Schreiber} {et~al.}(2009){F{\"o}rster Schreiber},
  {Genzel}, {Bouch{\'e}}, {Cresci}, {Davies}, {Buschkamp}, {Shapiro},
  {Tacconi}, {Hicks}, {Genel}, {Shapley}, {Erb}, {Steidel}, {Lutz},
  {Eisenhauer}, {Gillessen}, {Sternberg}, {Renzini}, {Cimatti}, {Daddi},
  {Kurk}, {Lilly}, {Kong}, {Lehnert}, {Nesvadba}, {Verma}, {McCracken},
  {Arimoto}, {Mignoli}, \& {Onodera}}]{ForsterSchreiber09}
{F{\"o}rster Schreiber}, N.~M., {Genzel}, R., {Bouch{\'e}}, N., {et~al.} 2009,
  \apj, 706, 1364

\bibitem[{{Franx} \& {de Zeeuw}(1992)}]{Franx92}
{Franx}, M., \& {de Zeeuw}, T. 1992, \apjl, 392, L47

\bibitem[{{Franx} {et~al.}(1989){Franx}, {Illingworth}, \& {Heckman}}]{Franx89}
{Franx}, M., {Illingworth}, G., \& {Heckman}, T. 1989, \aj, 98, 538

\bibitem[{{Franx} {et~al.}(2008){Franx}, {van Dokkum}, {Schreiber}, {Wuyts},
  {Labb{\'e}}, \& {Toft}}]{Franx08}
{Franx}, M., {van Dokkum}, P.~G., {Schreiber}, N.~M.~F., {et~al.} 2008, \apj,
  688, 770

\bibitem[{{Freeman}(1970)}]{Freeman70}
{Freeman}, K.~C. 1970, \apj, 160, 811

\bibitem[{{Glazebrook}(2013)}]{Glazebrook13}
{Glazebrook}, K. 2013, \pasa, 30, 56

\bibitem[{{Gnerucci} {et~al.}(2011){Gnerucci}, {Marconi}, {Cresci}, {Maiolino},
  {Mannucci}, {Calura}, {Cimatti}, {Cocchia}, {Grazian}, {Matteucci}, {Nagao},
  {Pozzetti}, \& {Troncoso}}]{Gnerucci11}
{Gnerucci}, A., {Marconi}, A., {Cresci}, G., {et~al.} 2011, \aap, 528, A88

\bibitem[{{Hammer} {et~al.}(2005){Hammer}, {Flores}, {Elbaz}, {Zheng}, {Liang},
  \& {Cesarsky}}]{Hammer05}
{Hammer}, F., {Flores}, H., {Elbaz}, D., {et~al.} 2005, \aap, 430, 115

\bibitem[{{Harrison} {et~al.}(2017){Harrison}, {Johnson}, {Swinbank}, {Stott},
  {Bower}, {Smail}, {Tiley}, {Bunker}, {Cirasuolo}, {Sobral}, {Sharples},
  {Best}, {Bureau}, {Jarvis}, \& {Magdis}}]{Harrison17}
{Harrison}, C.~M., {Johnson}, H.~L., {Swinbank}, A.~M., {et~al.} 2017, ArXiv
  e-prints, arXiv:1701.05561

\bibitem[{{Haynes} \& {Giovanelli}(1984)}]{Haynes84}
{Haynes}, M.~P., \& {Giovanelli}, R. 1984, \aj, 89, 758

\bibitem[{{Hung} {et~al.}(2015){Hung}, {Rich}, {Yuan}, {Larson}, {Casey},
  {Smith}, {Sanders}, {Kewley}, \& {Hayward}}]{Hung15}
{Hung}, C.-L., {Rich}, J.~A., {Yuan}, T., {et~al.} 2015, \apj, 803, 62

\bibitem[{{Kassin} {et~al.}(2007){Kassin}, {Weiner}, {Faber}, {Koo}, {Lotz},
  {Diemand}, {Harker}, {Bundy}, {Metevier}, {Phillips}, {Cooper}, {Croton},
  {Konidaris}, {Noeske}, \& {Willmer}}]{Kassin07}
{Kassin}, S.~A., {Weiner}, B.~J., {Faber}, S.~M., {et~al.} 2007, \apjl, 660,
  L35

\bibitem[{{Kassin} {et~al.}(2012){Kassin}, {Weiner}, {Faber}, {Gardner},
  {Willmer}, {Coil}, {Cooper}, {Devriendt}, {Dutton}, {Guhathakurta}, {Koo},
  {Metevier}, {Noeske}, \& {Primack}}]{Kassin12}
---. 2012, \apj, 758, 106

\bibitem[{{Kelly}(2007)}]{Kelly07}
{Kelly}, B.~C. 2007, \apj, 665, 1489

\bibitem[{{Kriek} {et~al.}(2009){Kriek}, {van Dokkum}, {Labb{\'e}}, {Franx},
  {Illingworth}, {Marchesini}, \& {Quadri}}]{Kriek09}
{Kriek}, M., {van Dokkum}, P.~G., {Labb{\'e}}, I., {et~al.} 2009, \apj, 700,
  221

\bibitem[{{Kriek} {et~al.}(2015){Kriek}, {Shapley}, {Reddy}, {Siana}, {Coil},
  {Mobasher}, {Freeman}, {de Groot}, {Price}, {Sanders}, {Shivaei}, {Brammer},
  {Momcheva}, {Skelton}, {van Dokkum}, {Whitaker}, {Aird}, {Azadi}, {Kassis},
  {Bullock}, {Conroy}, {Dav{\'e}}, {Kere{\v s}}, \& {Krumholz}}]{Kriek15}
{Kriek}, M., {Shapley}, A.~E., {Reddy}, N.~A., {et~al.} 2015, \apjs, 218, 15

\bibitem[{{Kroupa}(2001)}]{Kroupa01}
{Kroupa}, P. 2001, \mnras, 322, 231

\bibitem[{{Lawrence} {et~al.}(2007){Lawrence}, {Warren}, {Almaini}, {Edge},
  {Hambly}, {Jameson}, {Lucas}, {Casali}, {Adamson}, {Dye}, {Emerson},
  {Foucaud}, {Hewett}, {Hirst}, {Hodgkin}, {Irwin}, {Lodieu}, {McMahon},
  {Simpson}, {Smail}, {Mortlock}, \& {Folger}}]{Lawrence07}
{Lawrence}, A., {Warren}, S.~J., {Almaini}, O., {et~al.} 2007, \mnras, 379,
  1599

\bibitem[{{McLean} {et~al.}(2010){McLean}, {Steidel}, {Epps}, {Matthews},
  {Adkins}, {Konidaris}, {Weber}, {Aliado}, {Brims}, {Canfield}, {Cromer},
  {Fucik}, {Kulas}, {Mace}, {Magnone}, {Rodriguez}, {Wang}, \&
  {Weiss}}]{Mclean10}
{McLean}, I.~S., {Steidel}, C.~C., {Epps}, H., {et~al.} 2010, in Society of
  Photo-Optical Instrumentation Engineers (SPIE) Conference Series, Vol. 7735,
  Society of Photo-Optical Instrumentation Engineers (SPIE) Conference Series,
  1

\bibitem[{{Miller} {et~al.}(2011){Miller}, {Bundy}, {Sullivan}, {Ellis}, \&
  {Treu}}]{Miller11}
{Miller}, S.~H., {Bundy}, K., {Sullivan}, M., {Ellis}, R.~S., \& {Treu}, T.
  2011, \apj, 741, 115

\bibitem[{{Miller} {et~al.}(2012){Miller}, {Ellis}, {Sullivan}, {Bundy},
  {Newman}, \& {Treu}}]{Miller12}
{Miller}, S.~H., {Ellis}, R.~S., {Sullivan}, M., {et~al.} 2012, \apj, 753, 74

\bibitem[{{Mo} {et~al.}(1998){Mo}, {Mao}, \& {White}}]{Mo98}
{Mo}, H.~J., {Mao}, S., \& {White}, S.~D.~M. 1998, \mnras, 295, 319

\bibitem[{{Mortlock} {et~al.}(2013){Mortlock}, {Conselice}, {Hartley},
  {Ownsworth}, {Lani}, {Bluck}, {Almaini}, {Duncan}, {van der Wel},
  {Koekemoer}, {Dekel}, {Dav{\'e}}, {Ferguson}, {de Mello}, {Newman}, {Faber},
  {Grogin}, {Kocevski}, \& {Lai}}]{Mortlock13}
{Mortlock}, A., {Conselice}, C.~J., {Hartley}, W.~G., {et~al.} 2013, \mnras,
  433, 1185

\bibitem[{{Nanayakkara} {et~al.}(2016){Nanayakkara}, {Glazebrook}, {Kacprzak},
  {Yuan}, {Tran}, {Spitler}, {Kewley}, {Straatman}, {Cowley}, {Fisher},
  {Labbe}, {Tomczak}, {Allen}, \& {Alcorn}}]{Nanayakkara16}
{Nanayakkara}, T., {Glazebrook}, K., {Kacprzak}, G.~G., {et~al.} 2016, \apj,
  828, 21

\bibitem[{{Navarro} {et~al.}(1997){Navarro}, {Frenk}, \& {White}}]{Navarro97}
{Navarro}, J.~F., {Frenk}, C.~S., \& {White}, S.~D.~M. 1997, \apj, 490, 493

\bibitem[{{Papovich} {et~al.}(2010){Papovich}, {Momcheva}, {Willmer},
  {Finkelstein}, {Finkelstein}, {Tran}, {Brodwin}, {Dunlop}, {Farrah}, {Khan},
  {Lotz}, {McCarthy}, {McLure}, {Rieke}, {Rudnick}, {Sivanandam}, {Pacaud}, \&
  {Pierre}}]{Papovich10}
{Papovich}, C., {Momcheva}, I., {Willmer}, C.~N.~A., {et~al.} 2010, \apj, 716,
  1503

\bibitem[{{Papovich} {et~al.}(2015){Papovich}, {Labb{\'e}}, {Quadri}, {Tilvi},
  {Behroozi}, {Bell}, {Glazebrook}, {Spitler}, {Straatman}, {Tran}, {Cowley},
  {Dav{\'e}}, {Dekel}, {Dickinson}, {Ferguson}, {Finkelstein}, {Gawiser},
  {Inami}, {Faber}, {Kacprzak}, {Kawinwanichakij}, {Kocevski}, {Koekemoer},
  {Koo}, {Kurczynski}, {Lotz}, {Lu}, {Lucas}, {McIntosh}, {Mehrtens},
  {Mobasher}, {Monson}, {Morrison}, {Nanayakkara}, {Persson}, {Salmon},
  {Simons}, {Tomczak}, {van Dokkum}, {Weiner}, \& {Willner}}]{Papovich14}
{Papovich}, C., {Labb{\'e}}, I., {Quadri}, R., {et~al.} 2015, \apj, 803, 26

\bibitem[{{Peng} {et~al.}(2010){Peng}, {Ho}, {Impey}, \& {Rix}}]{Peng10}
{Peng}, C.~Y., {Ho}, L.~C., {Impey}, C.~D., \& {Rix}, H.-W. 2010, \aj, 139,
  2097

\bibitem[{{Persson} {et~al.}(2013){Persson}, {Murphy}, {Smee}, {Birk},
  {Monson}, {Uomoto}, {Koch}, {Shectman}, {Barkhouser}, {Orndorff}, {Hammond},
  {Harding}, {Scharfstein}, {Kelson}, {Marshall}, \& {McCarthy}}]{Persson13}
{Persson}, S.~E., {Murphy}, D.~C., {Smee}, S., {et~al.} 2013, \pasp, 125, 654

\bibitem[{{Pizagno} {et~al.}(2005){Pizagno}, {Prada}, {Weinberg}, {Rix},
  {Harbeck}, {Grebel}, {Bell}, {Brinkmann}, {Holtzman}, \& {West}}]{Pizagno05}
{Pizagno}, J., {Prada}, F., {Weinberg}, D.~H., {et~al.} 2005, \apj, 633, 844

\bibitem[{{Pizagno} {et~al.}(2007){Pizagno}, {Prada}, {Weinberg}, {Rix},
  {Pogge}, {Grebel}, {Harbeck}, {Blanton}, {Brinkmann}, \& {Gunn}}]{Pizagno07}
---. 2007, \aj, 134, 945

\bibitem[{{Price} {et~al.}(2015){Price}, {Kriek}, {Shapley}, {Reddy},
  {Freeman}, {Coil}, {de Groot}, {Shivaei}, {Siana}, {Azadi}, {Barro},
  {Mobasher}, {Sanders}, \& {Zick}}]{Price15}
{Price}, S.~H., {Kriek}, M., {Shapley}, A.~E., {et~al.} 2015, ArXiv e-prints,
  arXiv:1511.03272

\bibitem[{{Puech} {et~al.}(2008){Puech}, {Flores}, {Hammer}, {Yang}, {Neichel},
  {Lehnert}, {Chemin}, {Nesvadba}, {Epinat}, {Amram}, {Balkowski}, {Cesarsky},
  {Dannerbauer}, {di Serego Alighieri}, {Fuentes-Carrera}, {Guiderdoni},
  {Kembhavi}, {Liang}, {{\"O}stlin}, {Pozzetti}, {Ravikumar}, {Rawat},
  {Vergani}, {Vernet}, \& {Wozniak}}]{Puech08}
{Puech}, M., {Flores}, H., {Hammer}, F., {et~al.} 2008, \aap, 484, 173

\bibitem[{{Reddy} {et~al.}(2005){Reddy}, {Erb}, {Steidel}, {Shapley},
  {Adelberger}, \& {Pettini}}]{Reddy05}
{Reddy}, N.~A., {Erb}, D.~K., {Steidel}, C.~C., {et~al.} 2005, \apj, 633, 748

\bibitem[{{Reyes} {et~al.}(2011){Reyes}, {Mandelbaum}, {Gunn}, {Pizagno}, \&
  {Lackner}}]{Reyes11}
{Reyes}, R., {Mandelbaum}, R., {Gunn}, J.~E., {Pizagno}, J., \& {Lackner},
  C.~N. 2011, \mnras, 417, 2347

\bibitem[{{Sales} {et~al.}(2010){Sales}, {Navarro}, {Schaye}, {Dalla Vecchia},
  {Springel}, \& {Booth}}]{Sales10}
{Sales}, L.~V., {Navarro}, J.~F., {Schaye}, J., {et~al.} 2010, \mnras, 409,
  1541

\bibitem[{{Salpeter}(1955)}]{Salpeter55}
{Salpeter}, E.~E. 1955, \apj, 121, 161

\bibitem[{{Scoville} {et~al.}(2007){Scoville}, {Aussel}, {Brusa}, {Capak},
  {Carollo}, {Elvis}, {Giavalisco}, {Guzzo}, {Hasinger}, {Impey}, {Kneib},
  {LeFevre}, {Lilly}, {Mobasher}, {Renzini}, {Rich}, {Sanders}, {Schinnerer},
  {Schminovich}, {Shopbell}, {Taniguchi}, \& {Tyson}}]{Scoville07}
{Scoville}, N., {Aussel}, H., {Brusa}, M., {et~al.} 2007, \apjs, 172, 1

\bibitem[{{Sersic}(1968)}]{Sersic68}
{Sersic}, J.~L. 1968, {Atlas de galaxias australes}

\bibitem[{{Simons} {et~al.}(2016){Simons}, {Kassin}, {Trump}, {Weiner},
  {Heckman}, {Barro}, {Koo}, {Guo}, {Pacifici}, {Koekemoer}, \&
  {Stephens}}]{Simons16}
{Simons}, R.~C., {Kassin}, S.~A., {Trump}, J.~R., {et~al.} 2016, \apj, 830, 14

\bibitem[{{Skelton} {et~al.}(2014){Skelton}, {Whitaker}, {Momcheva}, {Brammer},
  {van Dokkum}, {Labbe}, {Franx}, {van der Wel}, {Bezanson}, {Da Cunha},
  {Fumagalli}, {Foerster Schreiber}, {Kriek}, {Leja}, {Lundgren}, {Magee},
  {Marchesini}, {Maseda}, {Nelson}, {Oesch}, {Pacifici}, {Patel}, {Price},
  {Rix}, {Tal}, {Wake}, \& {Wuyts}}]{Skelton14}
{Skelton}, R.~E., {Whitaker}, K.~E., {Momcheva}, I.~G., {et~al.} 2014, ArXiv
  e-prints, arXiv:1403.3689

\bibitem[{{Somerville} {et~al.}(2008){Somerville}, {Barden}, {Rix}, {Bell},
  {Beckwith}, {Borch}, {Caldwell}, {H{\"a}u{\ss}ler}, {Heymans}, {Jahnke},
  {Jogee}, {McIntosh}, {Meisenheimer}, {Peng}, {S{\'a}nchez}, {Wisotzki}, \&
  {Wolf}}]{Somerville08}
{Somerville}, R.~S., {Barden}, M., {Rix}, H.-W., {et~al.} 2008, \apj, 672, 776

\bibitem[{{Spitler} {et~al.}(2012){Spitler}, {Labb{\'e}}, {Glazebrook},
  {Persson}, {Monson}, {Papovich}, {Tran}, {Poole}, {Quadri}, {van Dokkum},
  {Kelson}, {Kacprzak}, {McCarthy}, {Murphy}, {Straatman}, \&
  {Tilvi}}]{Spitler12}
{Spitler}, L.~R., {Labb{\'e}}, I., {Glazebrook}, K., {et~al.} 2012, \apjl, 748,
  L21

\bibitem[{{Spitler} {et~al.}(2014){Spitler}, {Straatman}, {Labb{\'e}},
  {Glazebrook}, {Tran}, {Kacprzak}, {Quadri}, {Papovich}, {Persson}, {van
  Dokkum}, {Allen}, {Kawinwanichakij}, {Kelson}, {McCarthy}, {Mehrtens},
  {Monson}, {Nanayakkara}, {Rees}, {Tilvi}, \& {Tomczak}}]{Spitler14}
{Spitler}, L.~R., {Straatman}, C.~M.~S., {Labb{\'e}}, I., {et~al.} 2014, \apjl,
  787, L36

\bibitem[{{Steidel} {et~al.}(2014){Steidel}, {Rudie}, {Strom}, {Pettini},
  {Reddy}, {Shapley}, {Trainor}, {Erb}, {Turner}, {Konidaris}, {Kulas}, {Mace},
  {Matthews}, \& {McLean}}]{Steidel14}
{Steidel}, C.~C., {Rudie}, G.~C., {Strom}, A.~L., {et~al.} 2014, \apj, 795, 165

\bibitem[{{Straatman} {et~al.}(2015){Straatman}, {Labb{\'e}}, {Spitler},
  {Glazebrook}, {Tomczak}, {Allen}, {Brammer}, {Cowley}, {van Dokkum},
  {Kacprzak}, {Kawinwanichakij}, {Mehrtens}, {Nanayakkara}, {Papovich},
  {Persson}, {Quadri}, {Rees}, {Tilvi}, {Tran}, \& {Whitaker}}]{Straatman15}
{Straatman}, C.~M.~S., {Labb{\'e}}, I., {Spitler}, L.~R., {et~al.} 2015, \apjl,
  808, L29

\bibitem[{{Straatman} {et~al.}(2016){Straatman}, {Spitler}, {Quadri}, {Labbe},
  {Glazebrook}, {Persson}, {Papovich}, {Tran}, {Brammer}, {Cowley}, {Tomczak},
  {Nanayakkara}, {Alcorn}, {Allen}, {Broussard}, {van Dokkum}, {Forrest}, {van
  Houdt}, {Kacprzak}, {Kawinwanichakij}, {Kelson}, {Lee}, {McCarthy},
  {Mehrtens}, {Monson}, {Murphy}, {Rees}, {Tilvi}, \& {Whitaker}}]{Straatman16}
{Straatman}, C.~M.~S., {Spitler}, L.~R., {Quadri}, R.~F., {et~al.} 2016, ArXiv
  e-prints, arXiv:1608.07579

\bibitem[{{Tiley} {et~al.}(2016){Tiley}, {Stott}, {Swinbank}, {Bureau},
  {Harrison}, {Bower}, {Johnson}, {Bunker}, {Jarvis}, {Magdis}, {Sharples},
  {Smail}, {Sobral}, \& {Best}}]{Tiley16}
{Tiley}, A.~L., {Stott}, J.~P., {Swinbank}, A.~M., {et~al.} 2016, \mnras, 460,
  103

\bibitem[{{Tomczak} {et~al.}(2014){Tomczak}, {Quadri}, {Tran}, {Labb{\'e}},
  {Straatman}, {Papovich}, {Glazebrook}, {Allen}, {Brammer}, {Kacprzak},
  {Kawinwanichakij}, {Kelson}, {McCarthy}, {Mehrtens}, {Monson}, {Persson},
  {Spitler}, {Tilvi}, \& {van Dokkum}}]{Tomczak14}
{Tomczak}, A.~R., {Quadri}, R.~F., {Tran}, K.-V.~H., {et~al.} 2014, \apj, 783,
  85

\bibitem[{{Tomczak} {et~al.}(2015){Tomczak}, {Quadri}, {Tran}, {Labbe},
  {Straatman}, {Papovich}, {Glazebrook}, {Allen}, {Brammer}, {Cowley},
  {Dickinson}, {Elbaz}, {Inami}, {Kacprzak}, {Morrison}, {Nanayakkara},
  {Persson}, {Rees}, {Salmon}, {Schreiber}, {Spitler}, \&
  {Whitaker}}]{Tomczak15}
---. 2015, ArXiv e-prints, arXiv:1510.06072

\bibitem[{{Tully} \& {Fisher}(1977)}]{Tully77}
{Tully}, R.~B., \& {Fisher}, J.~R. 1977, \aap, 54, 661

\bibitem[{{van der Wel} {et~al.}(2012){van der Wel}, {Bell}, {H{\"a}ussler},
  {McGrath}, {Chang}, {Guo}, {McIntosh}, {Rix}, {Barden}, {Cheung}, {Faber},
  {Ferguson}, {Galametz}, {Grogin}, {Hartley}, {Kartaltepe}, {Kocevski},
  {Koekemoer}, {Lotz}, {Mozena}, {Peth}, \& {Peng}}]{vanderWel12}
{van der Wel}, A., {Bell}, E.~F., {H{\"a}ussler}, B., {et~al.} 2012, \apjs,
  203, 24

\bibitem[{{van der Wel} {et~al.}(2014{\natexlab{a}}){van der Wel}, {Franx},
  {van Dokkum}, {Skelton}, {Momcheva}, {Whitaker}, {Brammer}, {Bell}, {Rix},
  {Wuyts}, {Ferguson}, {Holden}, {Barro}, {Koekemoer}, {Chang}, {McGrath},
  {H{\"a}ussler}, {Dekel}, {Behroozi}, {Fumagalli}, {Leja}, {Lundgren},
  {Maseda}, {Nelson}, {Wake}, {Patel}, {Labb{\'e}}, {Faber}, {Grogin}, \&
  {Kocevski}}]{vanderWel14a}
{van der Wel}, A., {Franx}, M., {van Dokkum}, P.~G., {et~al.}
  2014{\natexlab{a}}, \apj, 788, 28

\bibitem[{{van der Wel} {et~al.}(2014{\natexlab{b}}){van der Wel}, {Chang},
  {Bell}, {Holden}, {Ferguson}, {Giavalisco}, {Rix}, {Skelton}, {Whitaker},
  {Momcheva}, {Brammer}, {Kassin}, {Martig}, {Dekel}, {Ceverino}, {Koo},
  {Mozena}, {van Dokkum}, {Franx}, {Faber}, \& {Primack}}]{vanderWel14b}
{van der Wel}, A., {Chang}, Y.-Y., {Bell}, E.~F., {et~al.} 2014{\natexlab{b}},
  \apjl, 792, L6

\bibitem[{{van Dokkum} \& {Franx}(2001)}]{vanDokkum01}
{van Dokkum}, P.~G., \& {Franx}, M. 2001, \apj, 553, 90

\bibitem[{{Vergani} {et~al.}(2012){Vergani}, {Epinat}, {Contini}, {Tasca},
  {Tresse}, {Amram}, {Garilli}, {Kissler-Patig}, {Le F{\`e}vre}, {Moultaka},
  {Paioro}, {Queyrel}, \& {L{\'o}pez-Sanjuan}}]{Vergani12}
{Vergani}, D., {Epinat}, B., {Contini}, T., {et~al.} 2012, \aap, 546, A118

\bibitem[{{Weiner} {et~al.}(2006{\natexlab{a}}){Weiner}, {Willmer}, {Faber},
  {Melbourne}, {Kassin}, {Phillips}, {Harker}, {Metevier}, {Vogt}, \&
  {Koo}}]{Weiner06a}
{Weiner}, B.~J., {Willmer}, C.~N.~A., {Faber}, S.~M., {et~al.}
  2006{\natexlab{a}}, \apj, 653, 1027

\bibitem[{{Weiner} {et~al.}(2006{\natexlab{b}}){Weiner}, {Willmer}, {Faber},
  {Harker}, {Kassin}, {Phillips}, {Melbourne}, {Metevier}, {Vogt}, \&
  {Koo}}]{Weiner06b}
---. 2006{\natexlab{b}}, \apj, 653, 1049

\bibitem[{{Whitaker} {et~al.}(2012){Whitaker}, {van Dokkum}, {Brammer}, \&
  {Franx}}]{Whitaker12}
{Whitaker}, K.~E., {van Dokkum}, P.~G., {Brammer}, G., \& {Franx}, M. 2012,
  \apjl, 754, L29

\bibitem[{{Willick}(1999)}]{Willick97}
{Willick}, J.~A. 1999, \apj, 516, 47

\bibitem[{{Yuan} {et~al.}(2014){Yuan}, {Nanayakkara}, {Kacprzak}, {Tran},
  {Glazebrook}, {Kewley}, {Spitler}, {Poole}, {Labb{\'e}}, {Straatman}, \&
  {Tomczak}}]{Yuan14}
{Yuan}, T., {Nanayakkara}, T., {Kacprzak}, G.~G., {et~al.} 2014, \apjl, 795,
  L20

\end{thebibliography}

\appendix
\section{Individual fits}
\label{app:collage}

\begin{figure*}
\begin{center}
\includegraphics[width=0.49\textwidth]{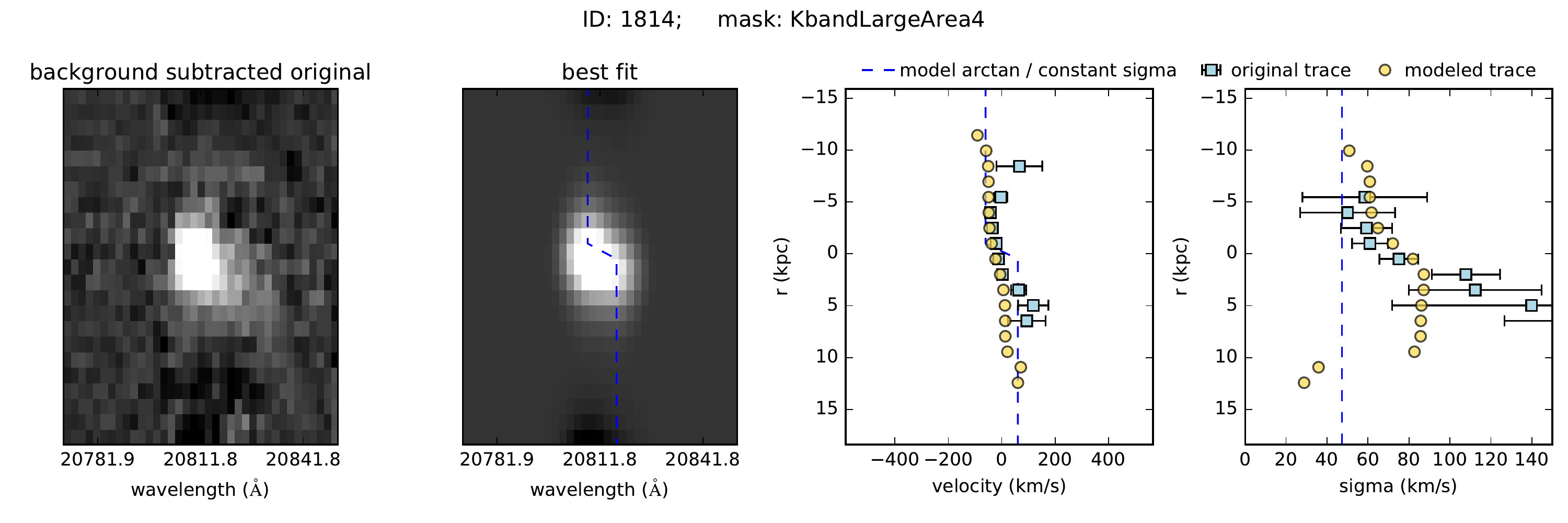}
\includegraphics[width=0.49\textwidth]{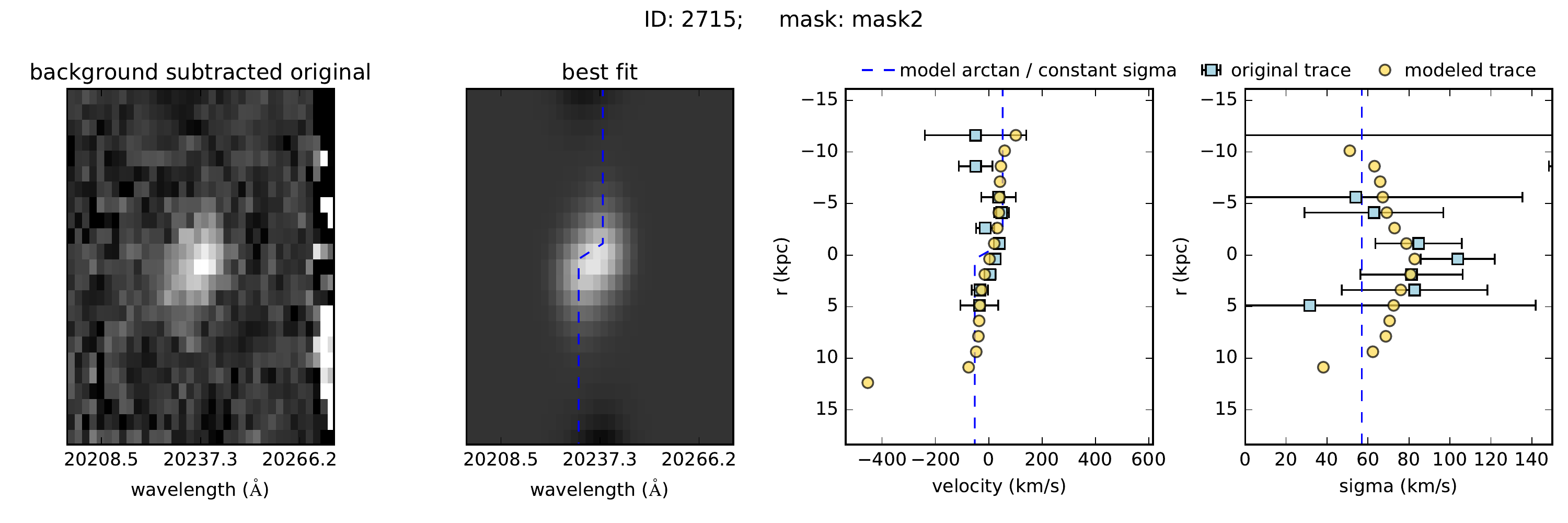}\\
\includegraphics[width=0.49\textwidth]{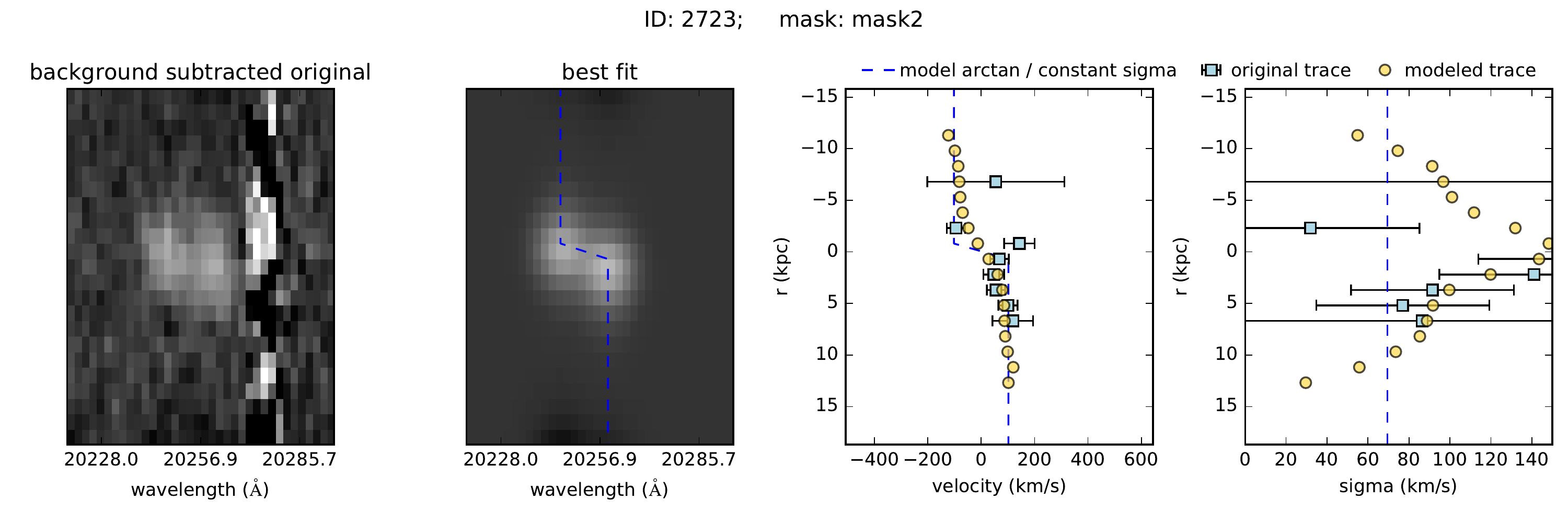}
\includegraphics[width=0.49\textwidth]{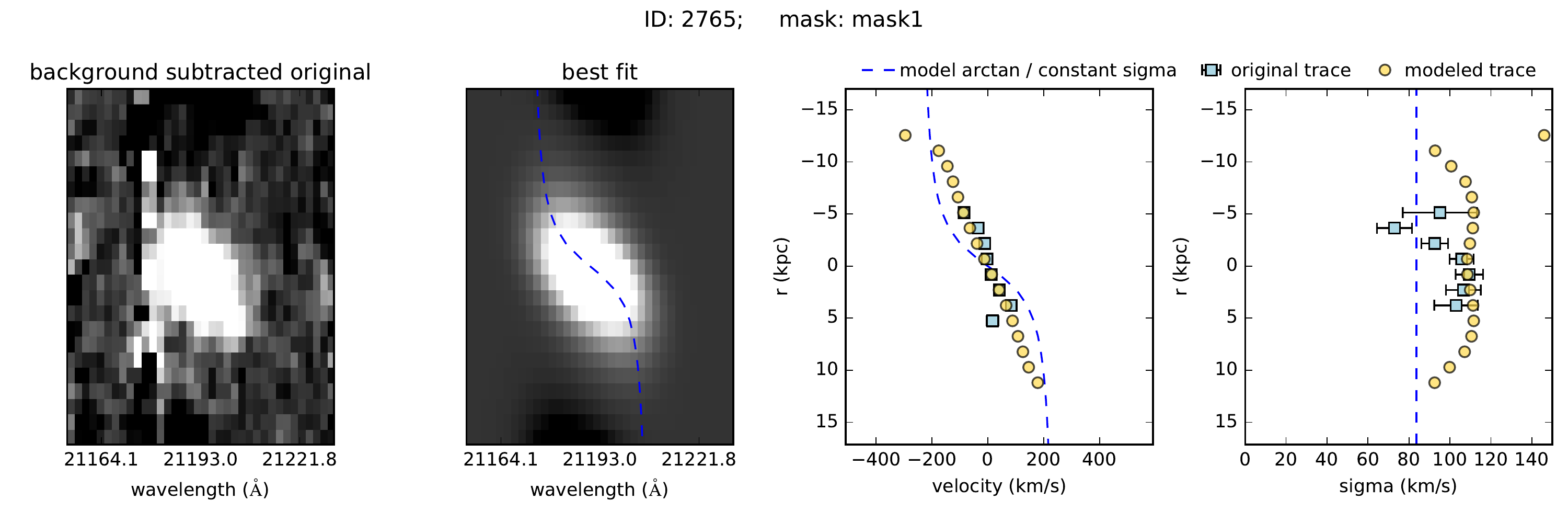}\\
\includegraphics[width=0.49\textwidth]{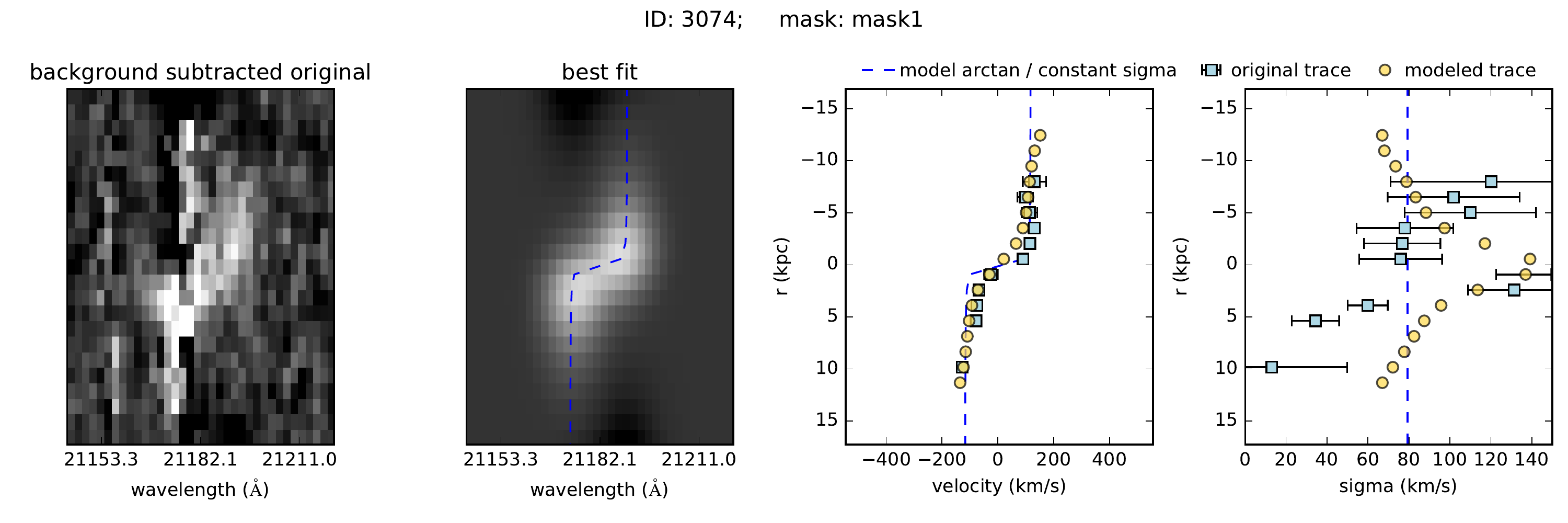}
\includegraphics[width=0.49\textwidth]{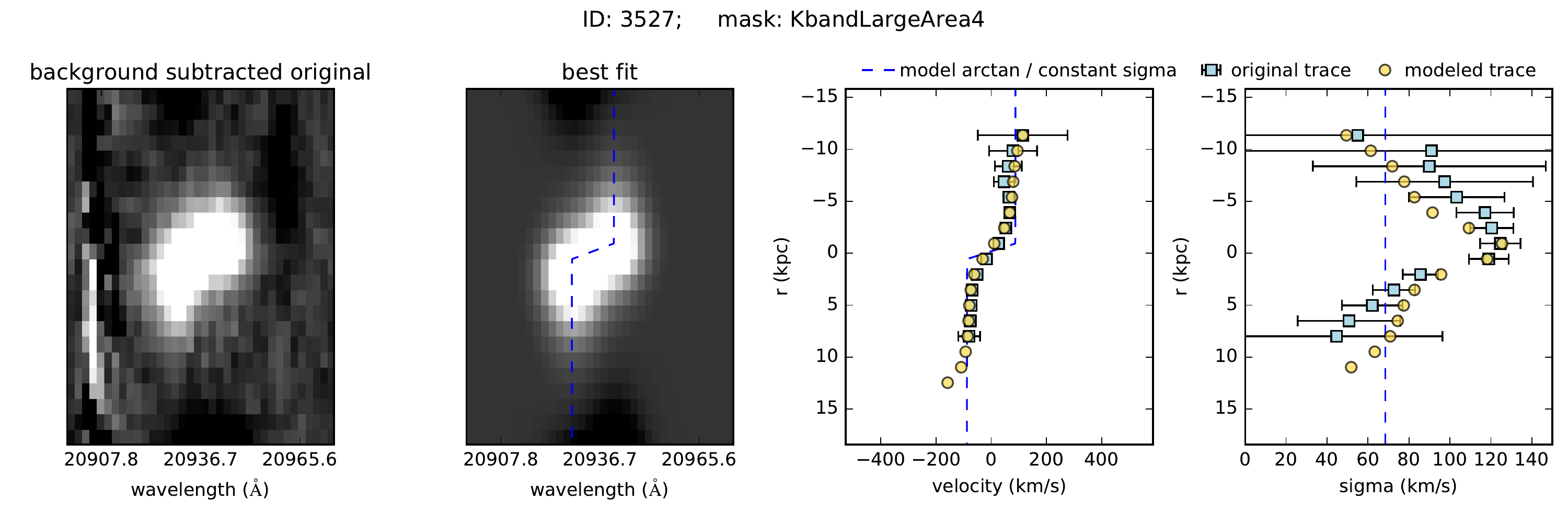}\\
\includegraphics[width=0.49\textwidth]{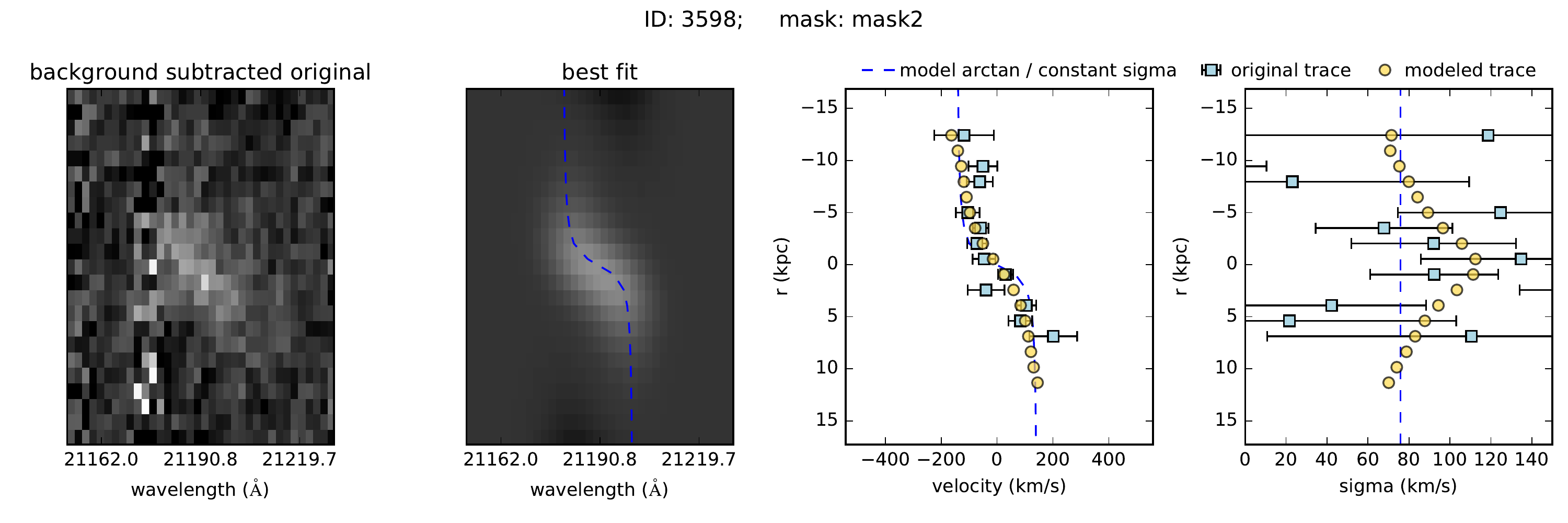}
\includegraphics[width=0.49\textwidth]{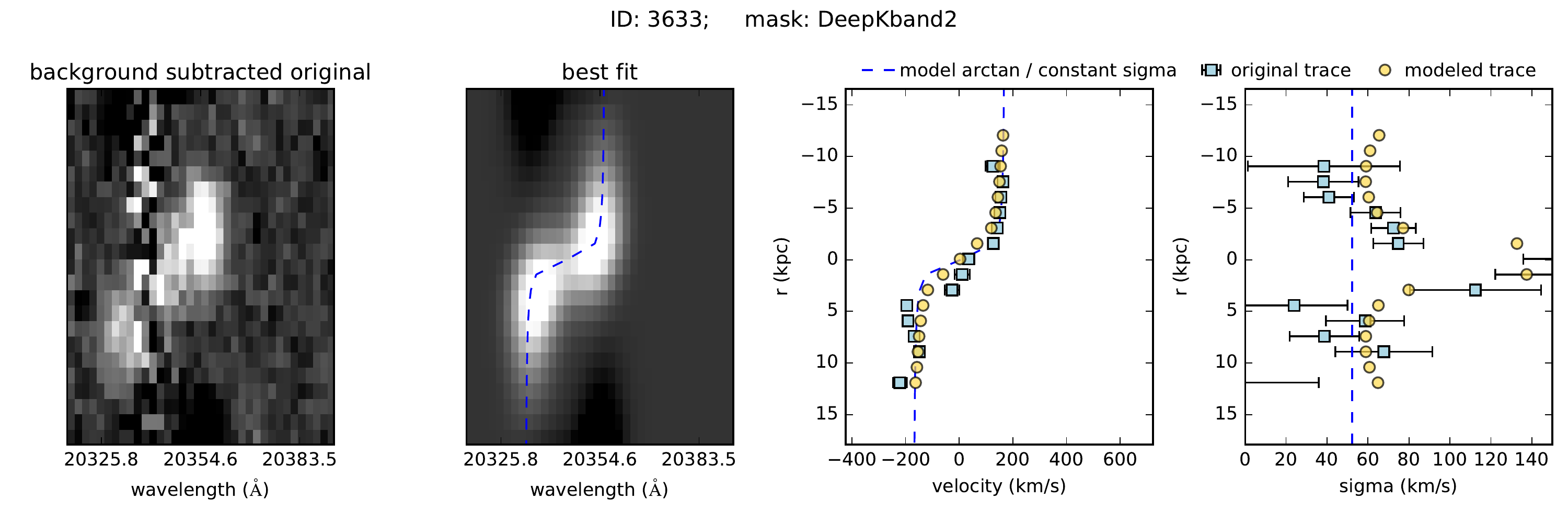}\\
\includegraphics[width=0.49\textwidth]{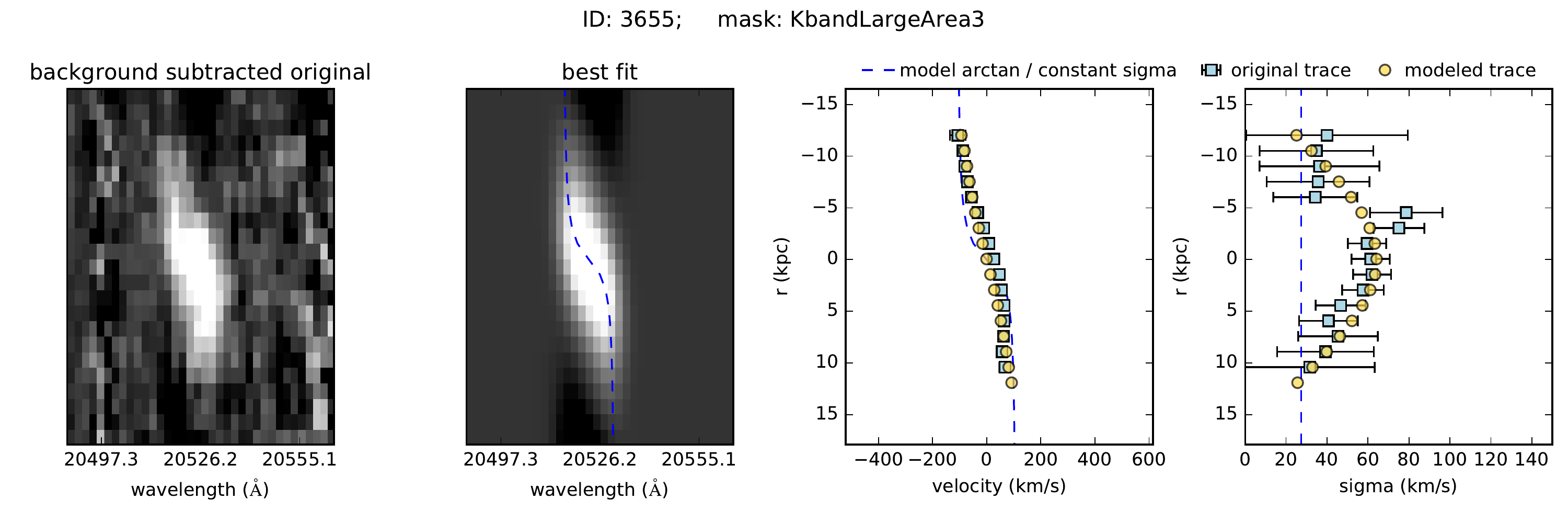}
\includegraphics[width=0.49\textwidth]{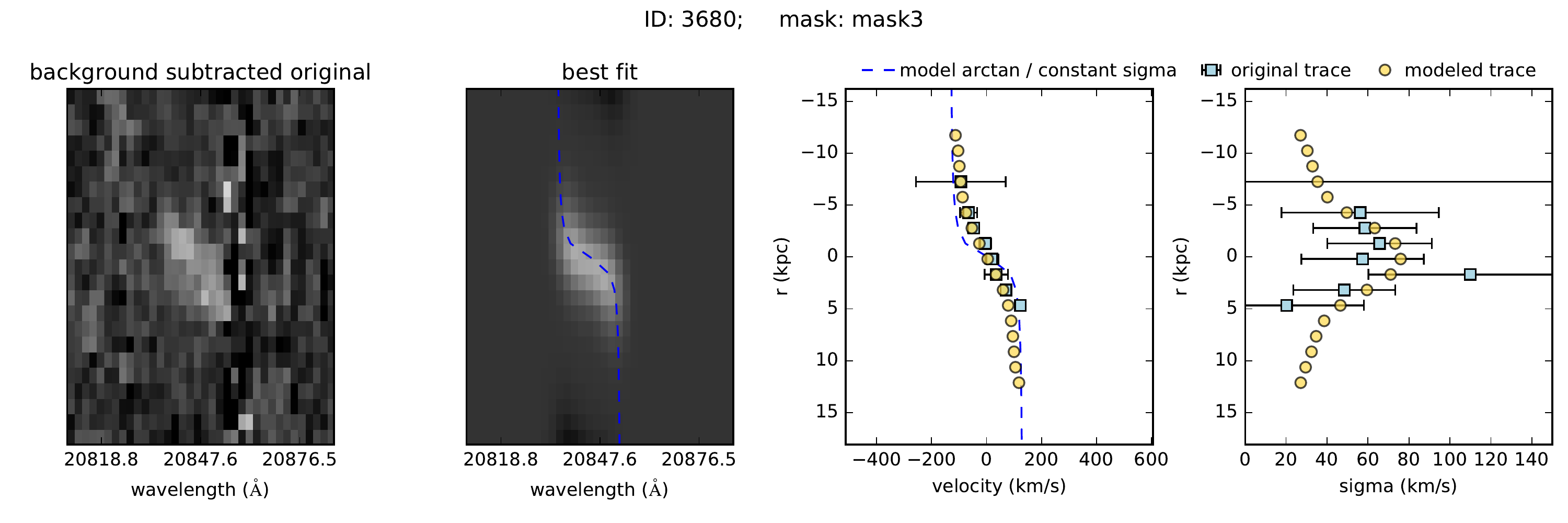}\\
\includegraphics[width=0.49\textwidth]{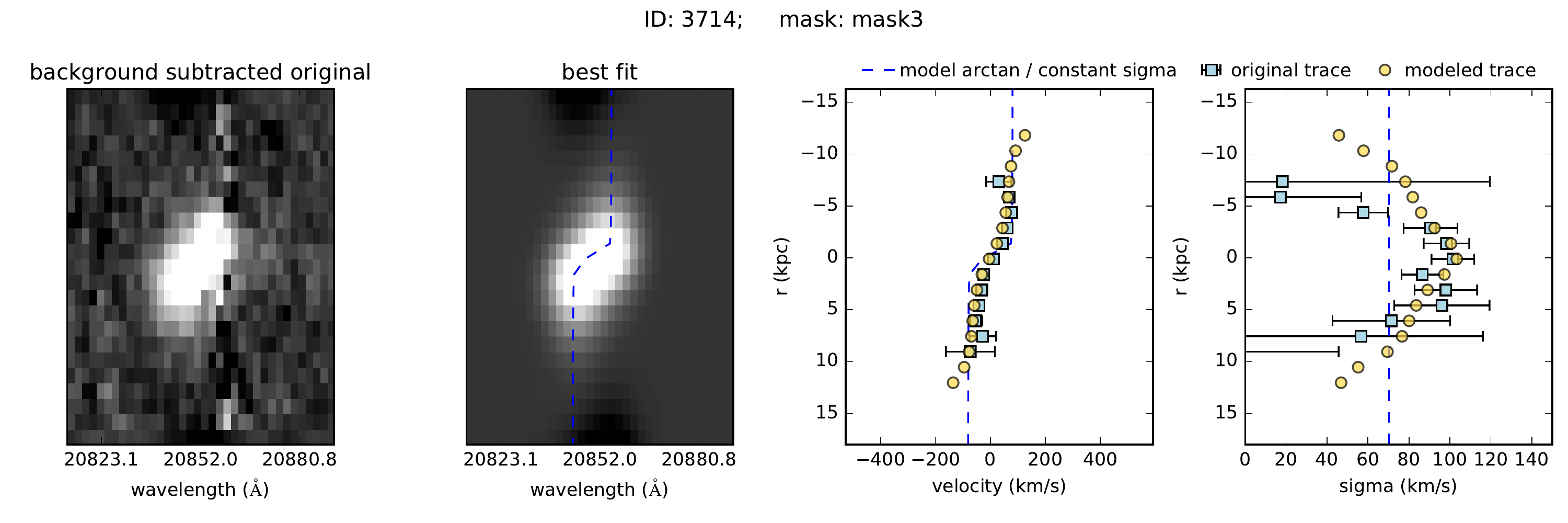}
\includegraphics[width=0.49\textwidth]{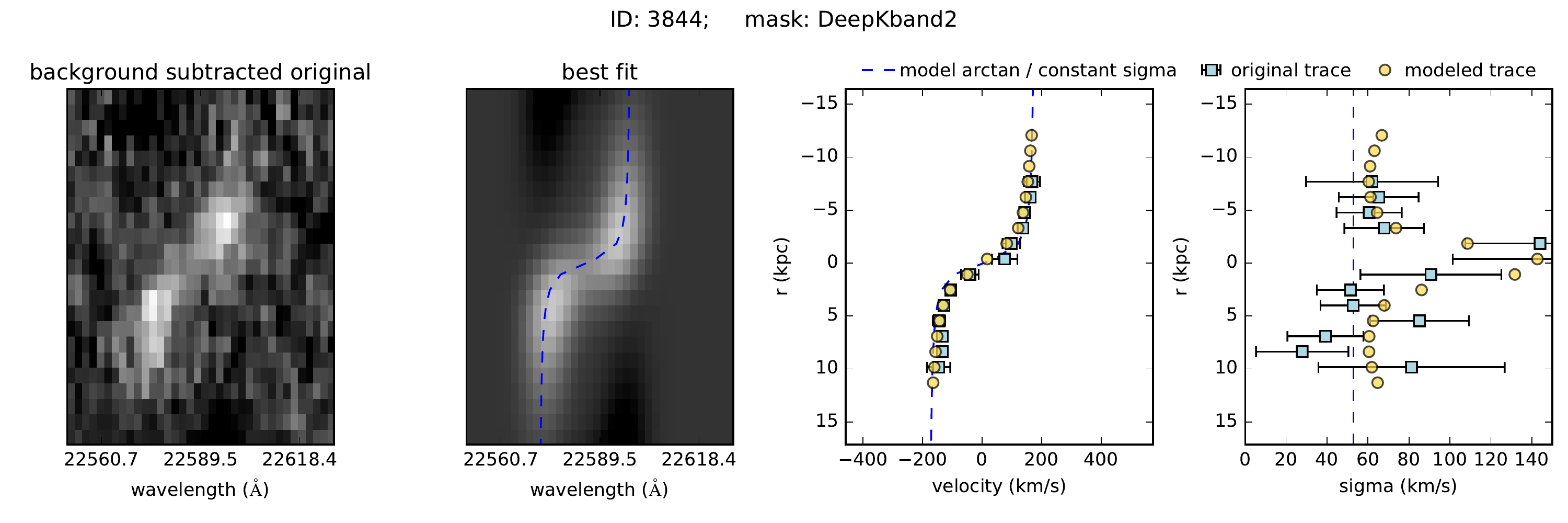}\\
\caption{{Individual fit results. For each source, from left to right: the original, sky-subtracted spectral image; the best-fit model; the radial velocity profiles of the data and the best-fit model; and the radial sigma profiles. The dashed lines in each panel are the intrinsic arctangents and velocity dispersions based on the fitted parameters. In the last two panels individual datapoints were obtained from gaussian fits to each row of the spectral image (blue squares) and best-fit model (yellow bullets). For the original data we only show rows with velocity and sigma errors $<500$ km/s and SNR$>1$. For the model profiles we show a spatial range up to 7 pixels to either side of the velocity center, where the dither pattern resulted in a positive signal. The velocities displayed here are not corrected for inclination and slit orientation.}} 
\label{fig:collage1}
\end{center}
\end{figure*}

\setcounter{figure}{13}    

\begin{figure*}
\begin{center}
\includegraphics[width=0.49\textwidth]{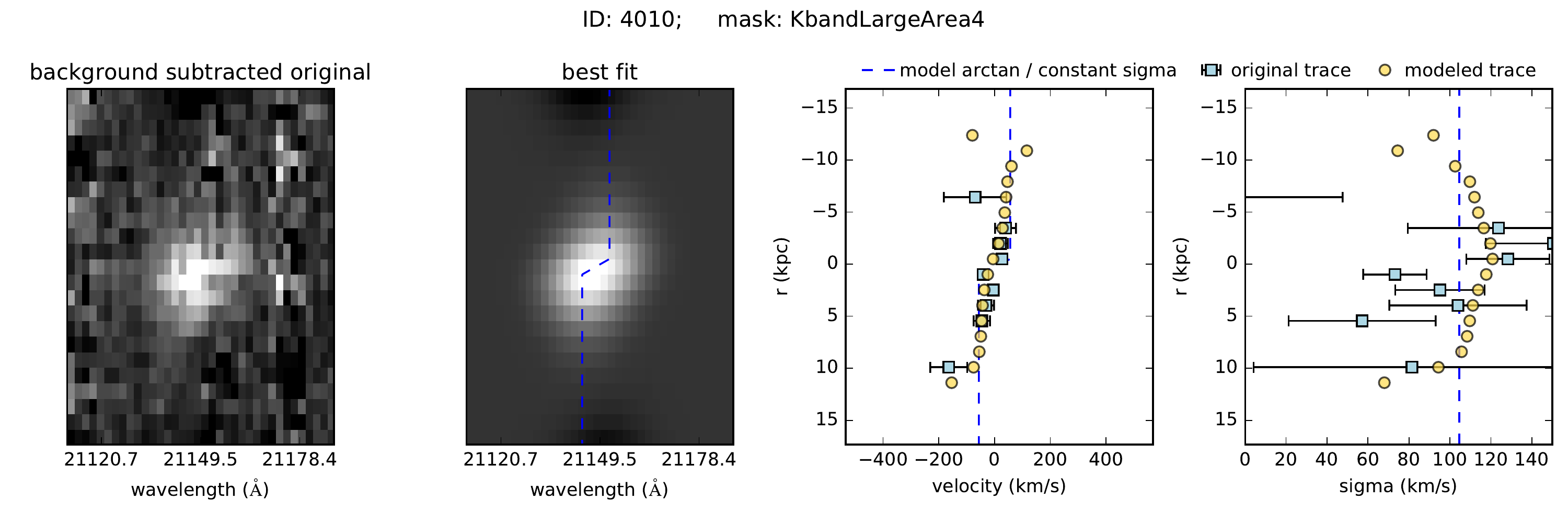}
\includegraphics[width=0.49\textwidth]{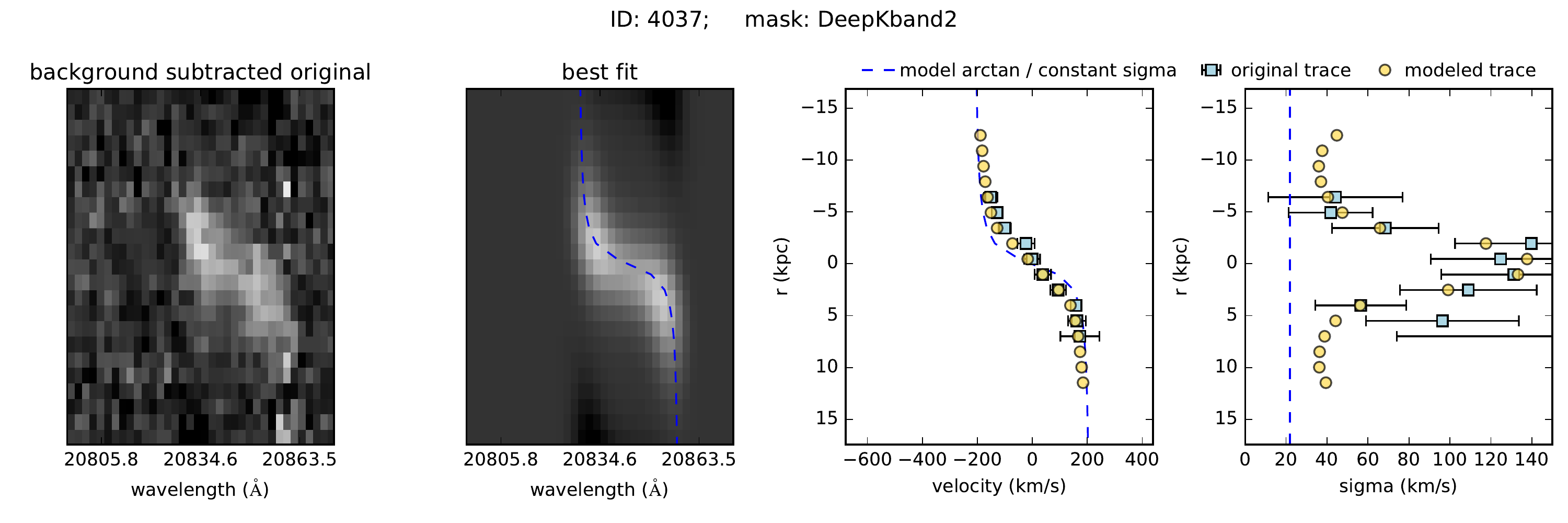}\\
\includegraphics[width=0.49\textwidth]{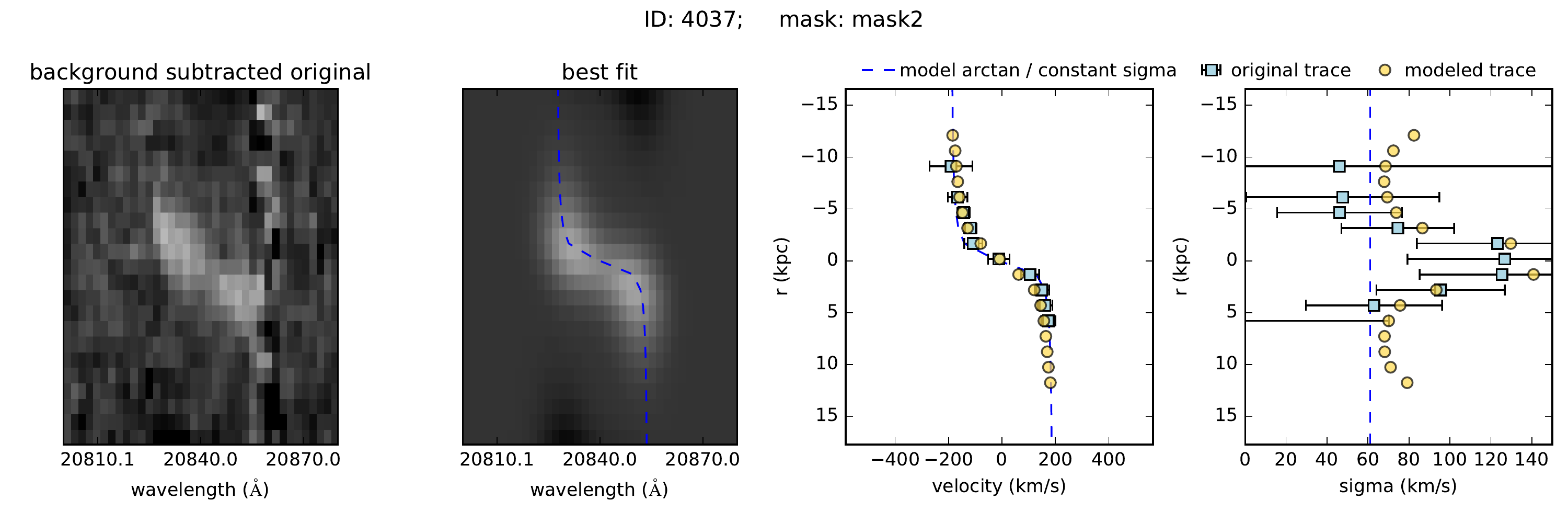}
\includegraphics[width=0.49\textwidth]{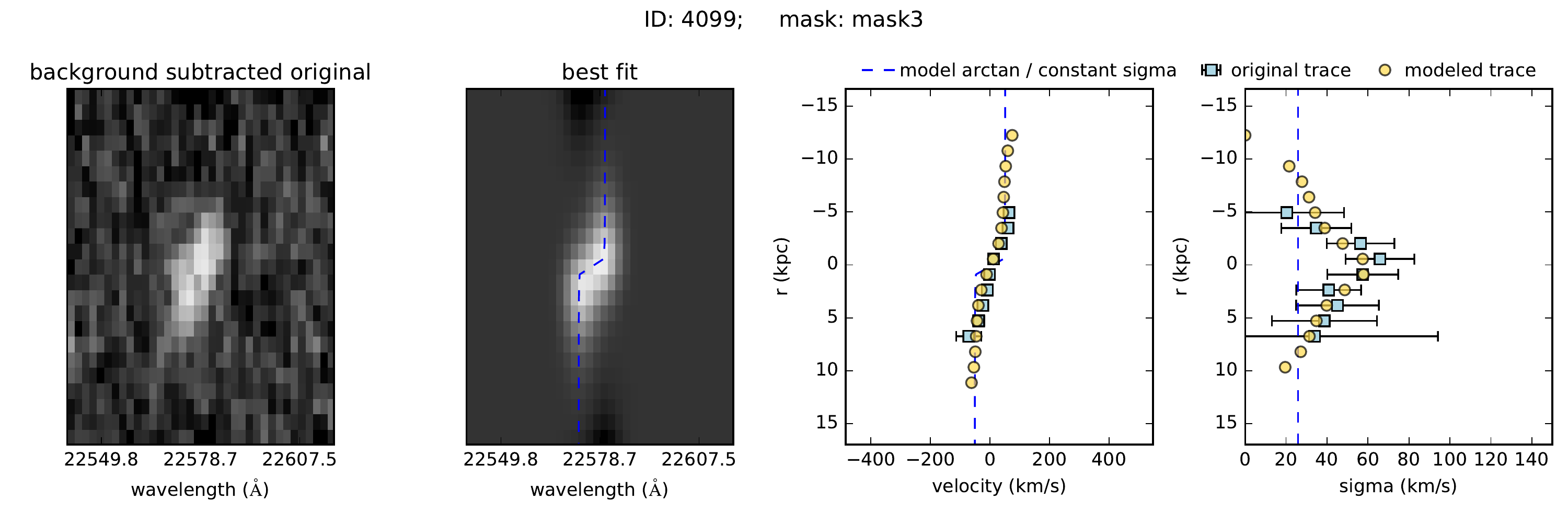}\\
\includegraphics[width=0.49\textwidth]{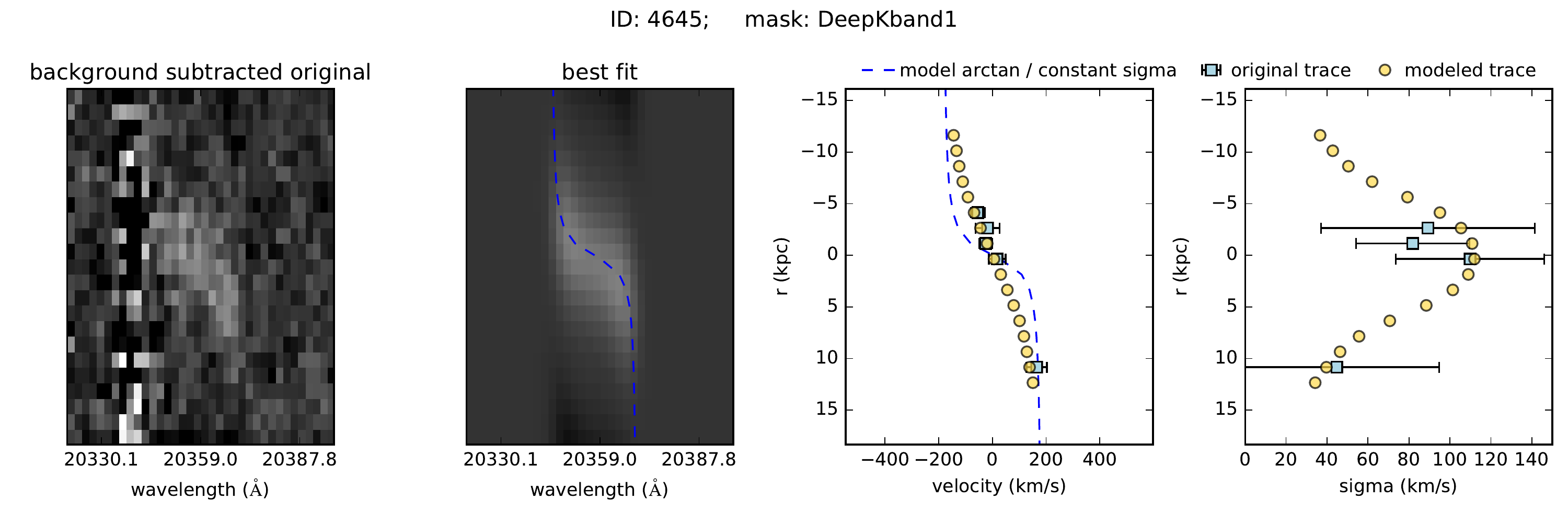}
\includegraphics[width=0.49\textwidth]{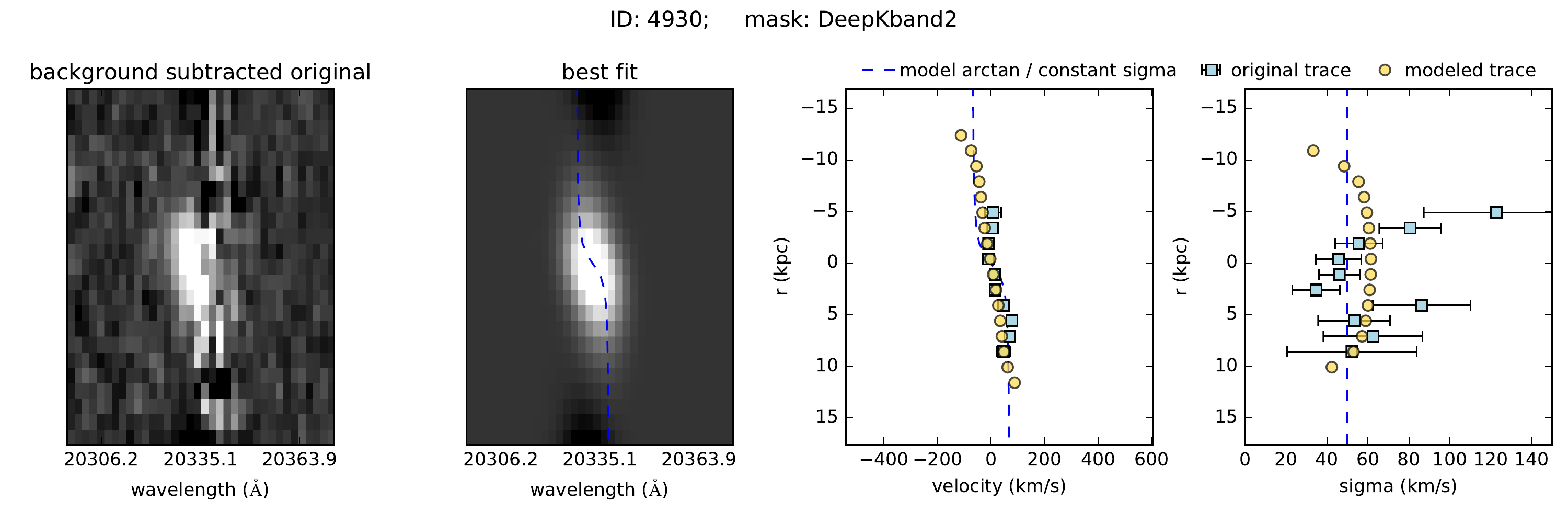}\\
\includegraphics[width=0.49\textwidth]{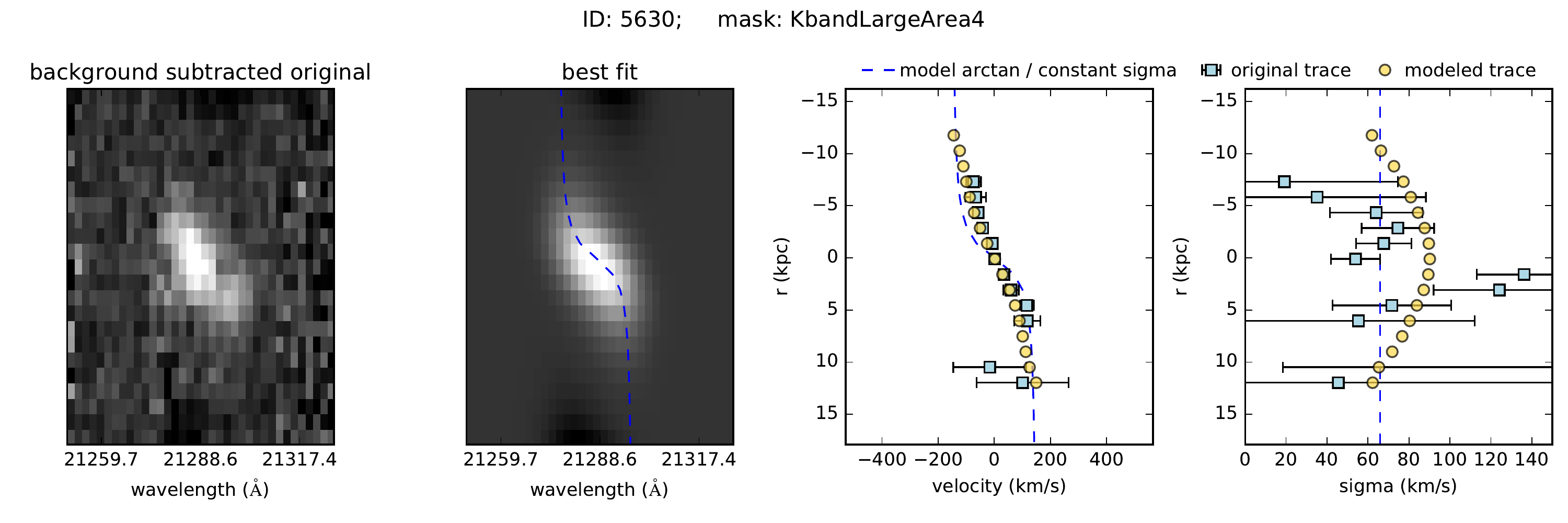}
\includegraphics[width=0.49\textwidth]{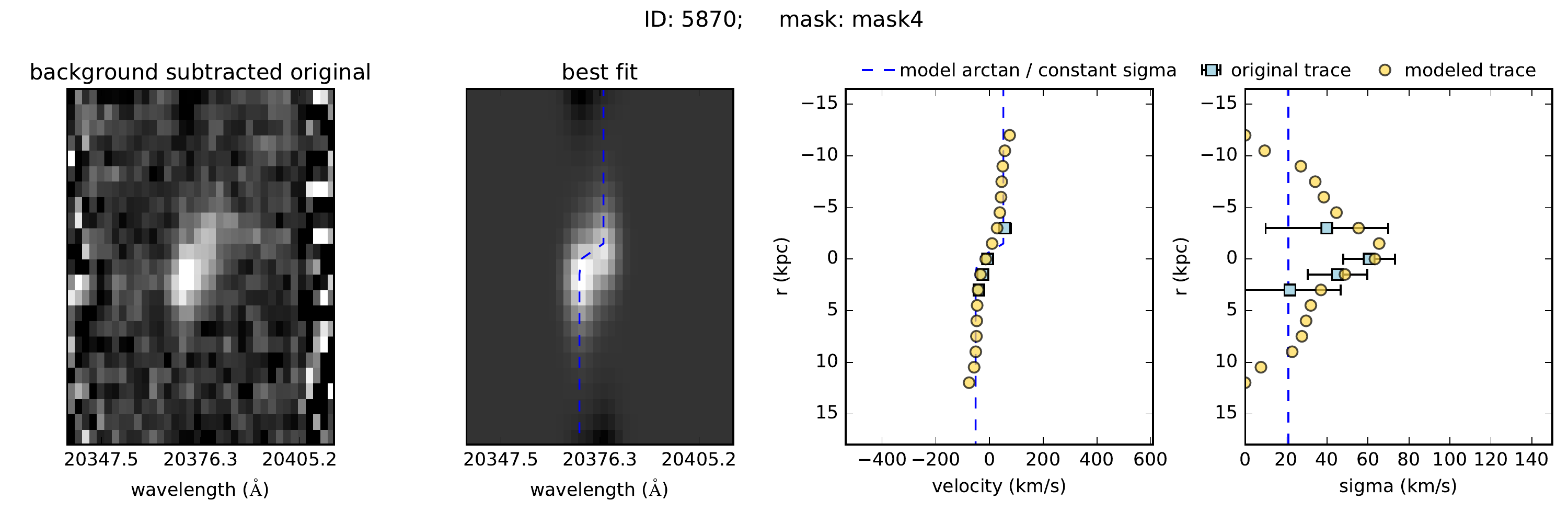}\\
\includegraphics[width=0.49\textwidth]{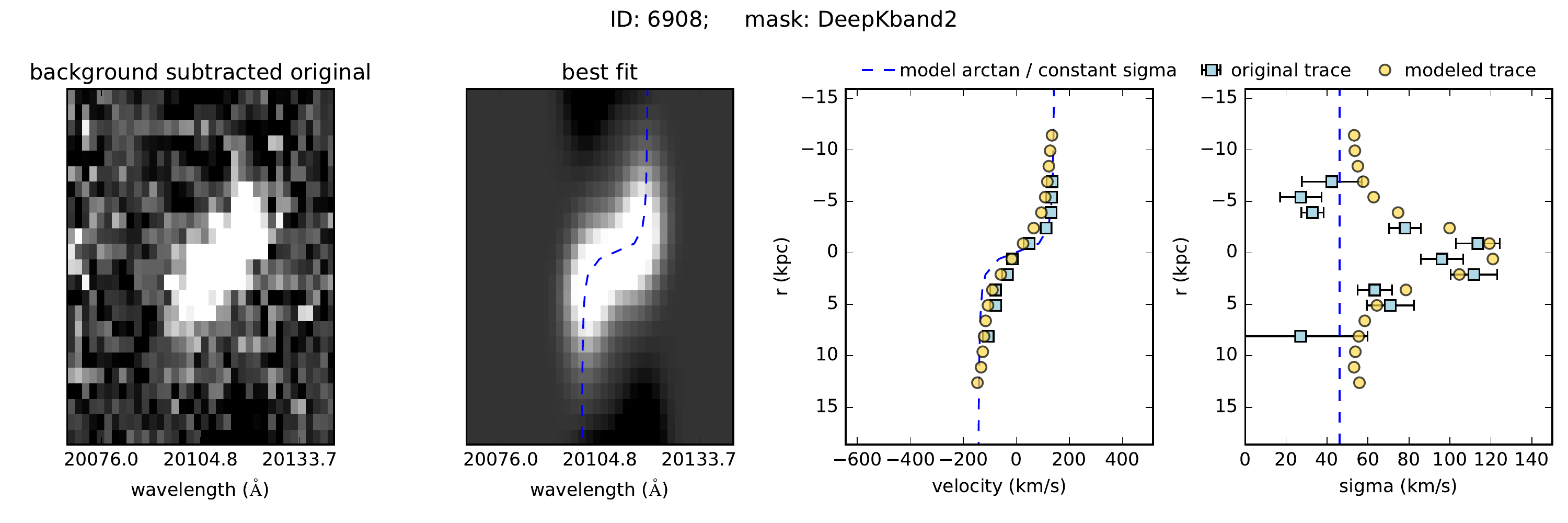}
\includegraphics[width=0.49\textwidth]{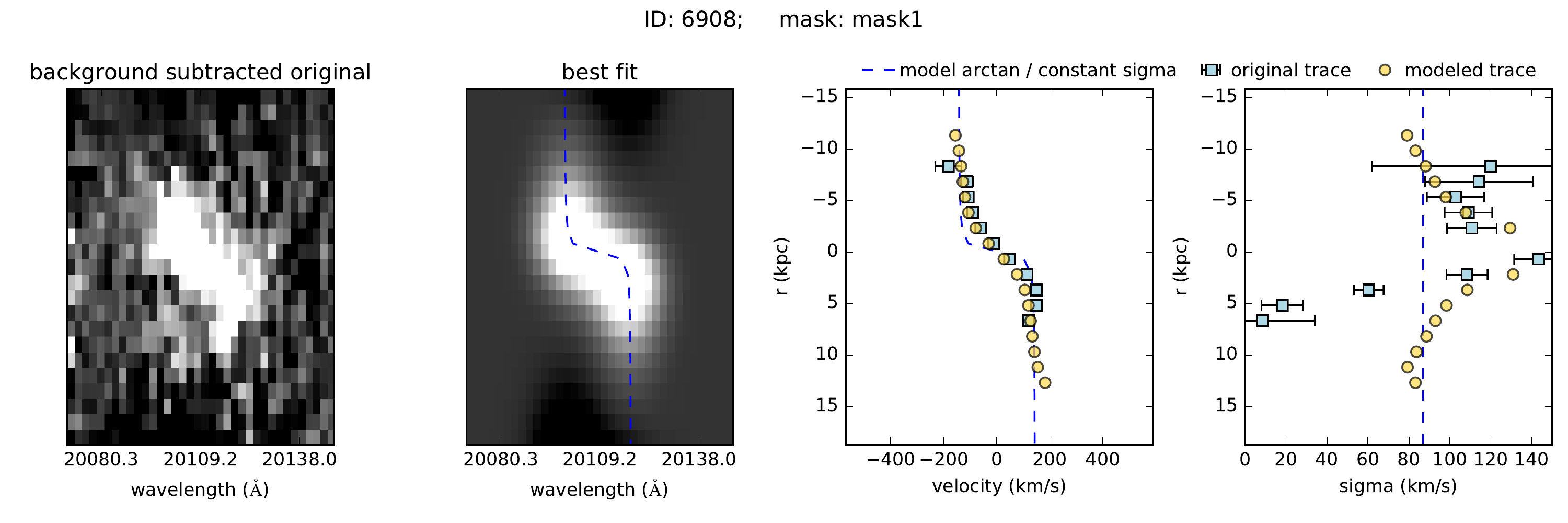}\\
\includegraphics[width=0.49\textwidth]{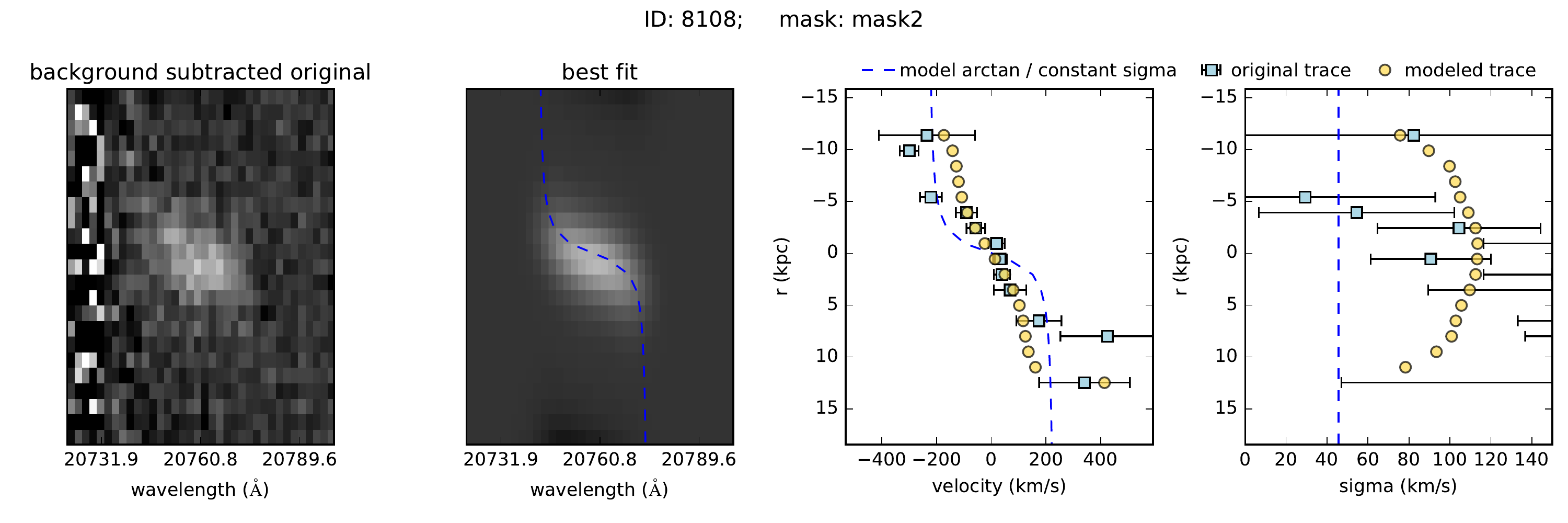}
\includegraphics[width=0.49\textwidth]{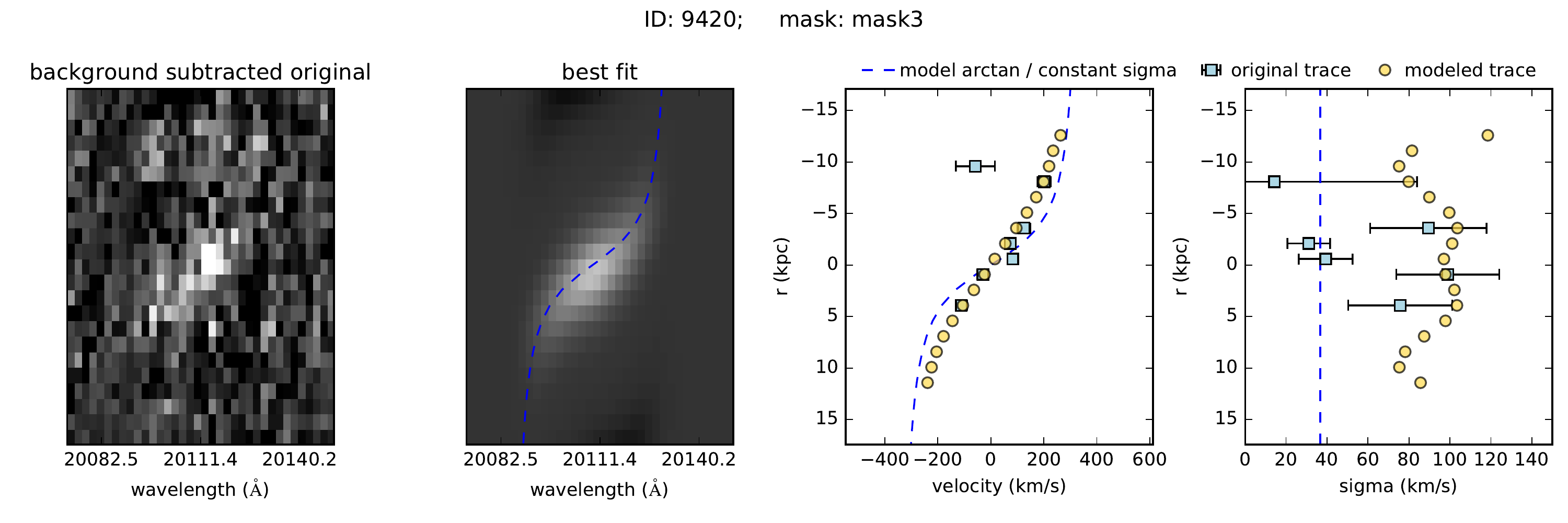}\\
\caption{continued.}
\label{fig:collage2}
\end{center}
\end{figure*}

{In this Section we present the spectral image stamps and best fits for 24 spectra (Figure \ref{fig:collage1}). In addition to the two-dimensional fits, we derived radial velocity and sigma profiles from gaussian fits to the flux intensity of each row of the spectral image stamp. The center of each gaussian traces the relative blue- or redshift from the center and can be converted to velocity using the kinematic center from the two-dimensional fits. In this way we can also derive the radial velocity dispersion profile, which is not constant due to the seeing (note that for the two-dimensional model we assumed an intrinsic constant velocity dispersion). For each spectrum we repeated this measurement for the modeled images. The errors on the trace were derived from the covariance matrix that was produced by the Python scipy {\tt optimize.leastsq} algorithm and give an indication of the goodness of the fits to each row.} 

{The radial velocity profile deviates in the center from the intrinsic arctangent, because it was measured on the convolved image. Overall, the original and modeled trace correspond within the uncertainties, indicating an arctangent profile is a reasonable approximation of the velocity curve. These figures further illustrate the advantage of a full two-dimensional analysis: we can accurately reproduce the velocity curves over a large radial range, whereas the trace measured from individual rows can have large uncertainties due to low SNR.}

\end{document}